\documentclass[11pt,a4paper]{article}
\usepackage{jheppub}

\usepackage[utf8]{inputenc}
\usepackage{amsmath}
\usepackage{amssymb}
\usepackage{mathrsfs}
\usepackage{graphicx}
\usepackage{bbold}
\usepackage{comment}
\usepackage{here}
\usepackage[dvipsnames]{xcolor}
\usepackage{braket}
\usepackage{bbm}
\usepackage{physics}
\usepackage{comment}
\usepackage{tikz}
\usepackage{multirow}
\numberwithin{equation}{section}

\newcommand{\be}{\begin{equation}}
\newcommand{\ee}{\end{equation}}
\newcommand{\nn}{\nonumber}
\newcommand{\del}{\partial}

\title{\boldmath Evidence of a CP broken deconfined phase 
in 4D SU(2) Yang-Mills theory at $\theta =\pi$ from imaginary $\theta$ simulations
}

\author[a,b]{Mitsuaki Hirasawa,}
\author[c,d,e]{Masazumi Honda,}
\author[c,d]{Akira Matsumoto,}
\author[f,g]{Jun Nishimura}
\author[h]{and Atis Yosprakob}

\affiliation[a]{
Department of Physics, University of Milano-Bicocca,\\
Piazza della Scienza 3, I-20126 Milano, Italy}

\affiliation[b]{
Istituto Nazionale di Fisica Nucleare (INFN), Sezione di Milano-Bicocca,\\
Piazza della Scienza 3, I-20126 Milano, Italy}

\affiliation[c]{
Yukawa Institute for Theoretical Physics, Kyoto University,\\
Kitashirakawa Oiwakecho, Sakyo-ku, Kyoto 606-8502 Japan}

\affiliation[d]{
Interdisciplinary Theoretical and Mathematical Sciences Program (iTHEMS), RIKEN,\\
2-1 Hirosawa, Wako, Saitama 351-0198 Japan}

\affiliation[e]{
Graduate School of Science and Engineering, Saitama University,\\
255 Shimo-Okubo, Sakura-ku, Saitama 338-8570, Japan}

\affiliation[f]{
KEK Theory Center, High Energy Accelerator Research Organization (KEK),\\
1-1 Oho, Tsukuba, Ibaraki 305-0801, Japan}

\affiliation[g]{
Graduate Institute for Advanced Studies, SOKENDAI,\\
1-1 Oho, Tsukuba, Ibaraki 305-0801 Japan}

\affiliation[h]{
Department of Physics, Niigata University,\\
8050 Ikarashi, 2-no-cho, Nishi-ku, Niigata 950-2181 Japan}

\emailAdd{mitsuaki.hirasawa(at)mib.infn.it}
\emailAdd{masazumi.honda(at)yukawa.kyoto-u.ac.jp}
\emailAdd{akira.matsumoto(at)yukawa.kyoto-u.ac.jp}
\emailAdd{jnishi(at)post.kek.jp}
\emailAdd{ayosp(at)phys.sc.niigata-u.ac.jp}

\preprint{
{
\begin{flushright}
YITP-24-166, KEK-TH-2669, RIKEN-iTHEMS-Report-24, STUPP-24-274
\end{flushright}
}
}

\abstract{
The spontaneous breaking of CP symmetry in
4D SU($N$)
pure Yang-Mills theory at $\theta=\pi$
has recently attracted much attention in the context of the higher-form symmetry
and the 't Hooft anomaly matching condition.
Here we use Monte Carlo simulations to study the $N=2$ case,
which is interesting
since it is the case opposite to the large-$N$ limit, where explicit
calculations are available.
In order to circumvent the severe sign problem due to the $\theta$ term for real $\theta$,
we first obtain results at imaginary $\theta$, where the sign problem is absent,
and make an analytic continuation to real $\theta$.
We use the stout smearing in defining the $\theta$ term
in the action to be used in our simulations.
Thus we obtain the expectation value of the topological charge 
and the deconfining temperature at $\theta=\pi$,
and provide evidence 
that the CP symmetry, which is spontaneously broken at low temperature,
gets restored \emph{strictly above} the deconfining temperature.
This conclusion is consistent with the anomaly matching condition
and yet differs from the prediction in the large-$N$ limit. 
}

\begin{document} 

%\begin{flushright}
%{\small YITP-22-xx, KEK-TH-xxxx, RIKEN-iTHEMS-Report-23}
%\end{flushright}

\maketitle

\flushbottom

%\tableofcontents

%%%%%%%%%%%%%%%%%
%%%%%%%%%%%%%%%%%
%%%%%%%%%%%%%%%%%
\section{Introduction} 
%%%%%%%%%%%%%%%%%
%%%%%%%%%%%%%%%%%
%%%%%%%%%%%%%%%%%

The topological aspect of quantum field theory has been one of the most
important subjects 
in both particle physics and condensed matter physics.
In particular, the dynamical
effect of the topological $\theta$ term,
%%parameter on the dynamics
which is purely non-perturbative, 
%%and its properties
remains elusive due to the notorious sign problem 
in
%% studying the theory using
standard Monte Carlo methods.

%%% condensed matter
%%For instance, in the context of
In condensed matter physics, 
the $\theta$ parameter has been discussed
intensively \cite{PhysRevB.36.5291, SHANKAR1990457, doi:10.1143/JPSJ.63.1277, Bietenholz:1995zk, doi:10.1143/JPSJ.65.1562, Alles:2007br, Bogli:2011aa, deForcrand:2012se, Alles:2014tta, Tang:2021uge} 
in 2D O(3) non-linear sigma model,
%% which is equivalent to 2D $\mathbb{CP}^1$ model, 
%%
which is known as an effective model of the anti-ferromagnetic spin system.
According to
Haldane’s conjecture \cite{PhysRevLett.50.1153, HALDANE1983464},
this
% It is expected that this
model is expected to be in a gapless phase at $\theta = \pi$,
while
%%although
it is in a gapped phase at $\theta = 0$.
%%which is well known as
%%
%%See Refs.~for various analytical and numerical works in this direction.

%%% particle physics
%%% strong CP ...
In particle physics,
%% quantum chromodynamics (QCD) accommodates
the topological $\theta$ term may appear
in 4D SU($N$) Yang-Mills (YM) theory including QCD ($N=3$),
and it breaks CP symmetry in general.
However, the experimental upper bound
%% observation of
on the neutron electric dipole moment
%%indicates
suggests
that the $\theta$ parameter is unnaturally small
as $|\theta| \lesssim 10^{-10}$ \cite{Baker:2006ts,Abel:2020pzs},
%%Thus, QCD somehow chooses $\theta = 0$,
%% considering that the CP symmetry is broken in the weak interaction
%% prefers
%% the CP-preserving choice $\theta = 0$,
%%This naturalness problem
which is a naturalness problem known as the strong-CP problem.
There have been many attempts
%% So far, many studies have been conducted
to explain this
%%the vanishing dipole moment 
either phenomenologically \cite{Peccei:1977hh, Peccei:1977ur, Weinberg:1977ma, Wilczek:1977pj, Nelson:1983zb, Barr:1984qx}
%%by reconsidering
%% the nature of topological charge or from a phenomenological perspective
or theoretically \cite{Ai:2020ptm, Ai:2024cnp, Schierholz:2024var}.
%Even though the effect of the $\theta$ term is not observed in our universe, 
%it is still theoretically important to reveal its nonperturbative aspect.

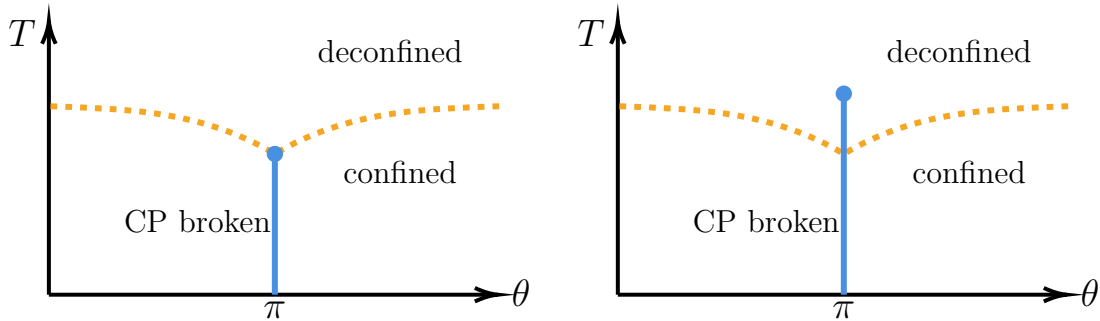
\begin{figure}[tb]
\centering
%%% possible phase diagrams
\tikzset{every picture/.style={line width=0.75pt}} %set default line width to 0.75pt        

\begin{tikzpicture}[x=0.75pt,y=0.75pt,yscale=-0.8,xscale=0.8]
%uncomment if require: \path (0,246); %set diagram left start at 0, and has height of 246

%Shape: Boxed Line [id:dp20977999097815325] 
\draw [line width=1.5]    (33,197) -- (310,197) ;
\draw [shift={(313,197)}, rotate = 180] [color={rgb, 255:red, 0; green, 0; blue, 0 }  ][line width=1.5]    (14.21,-4.28) .. controls (9.04,-1.82) and (4.3,-0.39) .. (0,0) .. controls (4.3,0.39) and (9.04,1.82) .. (14.21,4.28)   ;
%Shape: Boxed Line [id:dp9609166946377508] 
\draw [line width=1.5]    (33,197) -- (33,28) ;
\draw [shift={(33,25)}, rotate = 90] [color={rgb, 255:red, 0; green, 0; blue, 0 }  ][line width=1.5]    (14.21,-4.28) .. controls (9.04,-1.82) and (4.3,-0.39) .. (0,0) .. controls (4.3,0.39) and (9.04,1.82) .. (14.21,4.28)   ;
%Shape: Boxed Bezier Curve [id:dp9720328678795176] 
\draw [color={rgb, 255:red, 245; green, 166; blue, 35 }  ,draw opacity=1 ][line width=2.25]  [dash pattern={on 2.53pt off 3.02pt}]  (34,78.5) .. controls (78.03,80.29) and (127.33,80.29) .. (174,108.5) ;
%Shape: Boxed Line [id:dp7840035396976972] 
\draw [line width=1.5]    (388,197) -- (665,197) ;
\draw [shift={(668,197)}, rotate = 180] [color={rgb, 255:red, 0; green, 0; blue, 0 }  ][line width=1.5]    (14.21,-4.28) .. controls (9.04,-1.82) and (4.3,-0.39) .. (0,0) .. controls (4.3,0.39) and (9.04,1.82) .. (14.21,4.28)   ;
%Shape: Boxed Line [id:dp2253023204394431] 
\draw [line width=1.5]    (388,197) -- (388,28) ;
\draw [shift={(388,25)}, rotate = 90] [color={rgb, 255:red, 0; green, 0; blue, 0 }  ][line width=1.5]    (14.21,-4.28) .. controls (9.04,-1.82) and (4.3,-0.39) .. (0,0) .. controls (4.3,0.39) and (9.04,1.82) .. (14.21,4.28)   ;
%Shape: Boxed Bezier Curve [id:dp4654174137546858] 
\draw [color={rgb, 255:red, 245; green, 166; blue, 35 }  ,draw opacity=1 ][line width=2.25]  [dash pattern={on 2.53pt off 3.02pt}]  (389,78.5) .. controls (433.03,80.29) and (482.33,80.29) .. (529,108.5) ;
%Shape: Boxed Bezier Curve [id:dp6164487803381762] 
\draw [color={rgb, 255:red, 245; green, 166; blue, 35 }  ,draw opacity=1 ][line width=2.25]  [dash pattern={on 2.53pt off 3.02pt}]  (314,78.5) .. controls (269.97,80.29) and (220.67,80.29) .. (174,108.5) ;
%Shape: Boxed Bezier Curve [id:dp8475072789556974] 
\draw [color={rgb, 255:red, 245; green, 166; blue, 35 }  ,draw opacity=1 ][line width=2.25]  [dash pattern={on 2.53pt off 3.02pt}]  (669,78.5) .. controls (624.97,80.29) and (575.67,80.29) .. (529,108.5) ;
%Straight Lines [id:da9947499582119197] 
\draw [color={rgb, 255:red, 74; green, 144; blue, 226 }  ,draw opacity=1 ][line width=2.25]    (174,108.5) -- (174,197) ;
\draw [shift={(174,108.5)}, rotate = 90] [color={rgb, 255:red, 74; green, 144; blue, 226 }  ,draw opacity=1 ][fill={rgb, 255:red, 74; green, 144; blue, 226 }  ,fill opacity=1 ][line width=2.25]      (0, 0) circle [x radius= 3.22, y radius= 3.22]   ;
%Straight Lines [id:da09165655333223577] 
\draw [color={rgb, 255:red, 74; green, 144; blue, 226 }  ,draw opacity=1 ][line width=2.25]    (529,70.58) -- (529,197) ;
\draw [shift={(529,70.58)}, rotate = 90] [color={rgb, 255:red, 74; green, 144; blue, 226 }  ,draw opacity=1 ][fill={rgb, 255:red, 74; green, 144; blue, 226 }  ,fill opacity=1 ][line width=2.25]      (0, 0) circle [x radius= 3.22, y radius= 3.22]   ;

% Text Node
\draw (199,35) node [anchor=north west][inner sep=0.75pt]   [align=left] {{\large deconfined}};
% Text Node
\draw (215,111) node [anchor=north west][inner sep=0.75pt]   [align=left] {{\large confined}};
% Text Node
\draw (78,142) node [anchor=north west][inner sep=0.75pt]   [align=left] {{\large CP broken}};
% Text Node
\draw (6,22.4) node [anchor=north west][inner sep=0.75pt]  [font=\Large]  {$T$};
% Text Node
\draw (320,185.4) node [anchor=north west][inner sep=0.75pt]  [font=\Large]  {$\theta $};
% Text Node
\draw (165,199.4) node [anchor=north west][inner sep=0.75pt]  [font=\Large]  {$\pi $};
% Text Node
\draw (554,35) node [anchor=north west][inner sep=0.75pt]   [align=left] {{\large deconfined}};
% Text Node
\draw (570,111) node [anchor=north west][inner sep=0.75pt]   [align=left] {{\large confined}};
% Text Node
\draw (433,142) node [anchor=north west][inner sep=0.75pt]   [align=left] {{\large CP broken}};
% Text Node
\draw (361,22.4) node [anchor=north west][inner sep=0.75pt]  [font=\Large]  {$T$};
% Text Node
\draw (675,185.4) node [anchor=north west][inner sep=0.75pt]  [font=\Large]  {$\theta $};
% Text Node
\draw (520,199.4) node [anchor=north west][inner sep=0.75pt]  [font=\Large]  {$\pi $};

\end{tikzpicture}
\caption{Two scenarios for the $(\theta,T)$-phase diagram
of 4D pure SU($N$) Yang-Mills theory consistent with the anomaly matching condition.
The blue solid line represents
the first-order phase transition associated with the spontaneous breaking of CP symmetry.
The orange dotted curve represents
the deconfining transition that is believed to be of the first order 
for $N\geq 3$ and of the second order for $N=2$. 
The left panel represents the scenario $T_{\mathrm{CP}} = T_{\mathrm{dec}}(\pi) $,
while the right panel represents the scenario $T_{\mathrm{CP}} > T_{\mathrm{dec}}(\pi) $, 
which implies the existence of a deconfined phase with spontaneously broken CP
symmetry.}
\label{fig:phase_diag}
\end{figure}

%%% phase diagram
%% In this paper, we address another long-standing issue in particle physics, elucidating the phase diagram 
%% of the four-dimensional pure Yang-Mills (YM) theory for the temperature and the $\theta$-parameter.
In fact, 4D SU($N$) YM theory has CP symmetry not only at $\theta=0$ but also at $\theta=\pi$
due to the $2\pi$ periodicity in $\theta$.
%% the 2D $\mathbb{CP}^1$ model mentioned above,
%% $\theta=0$ and $\theta=\pi$ are special points in
%% SU($N$) YM theory
%% since the theory becomes CP symmetric only at these points.
%%
%% The CP symmetry in four-dimensional SU($N$) pure Yang-Mills theory at $\theta=\pi$
In particular, the CP symmetry
%%in SU($N$) pure YM theory at $\theta = \pi$,
at $\theta=\pi$ is considered to be spontaneously
broken at low temperature, while
%% it gets restored at high temperature.
%% Then, combined with the well-known result that
it is known to be unbroken at sufficiently
%%the theory is known to be in a CP-preserving deconfined phase at
high temperature \cite{Gross:1980br,Weiss:1980rj}.
Renewed interest in this issue has been triggered by recent developments 
in the higher-form symmetry \cite{Gaiotto:2014kfa} and
the 't Hooft anomaly matching \cite{tHooft:1979rat},
which predict
%, in SU($N$) pure YM theory
that, at $\theta = \pi$, either the CP
or the $\mathbb{Z}_N$ center symmetry 
should be
%is
spontaneously broken unless the theory becomes gapless.
Note here that the spontaneous breaking of
the $\mathbb{Z}_N$ center symmetry
corresponds to deconfinement.
%%is spontaneously broken in the deconfined phase, while it is preserved in the confined phase.
%% In other words,
Therefore the prediction implies that the theory
%%The upshot here is that
cannot be in a gapped confined phase
%%while preserving the
without breaking the
CP symmetry \cite{Gaiotto:2017yup,Cordova:2019bsd}.
%%at any temperature. 
%%
%% Therefore, the allowed situation is such that 
%% This condition is indeed satisfied by the known result
In other words, the anomaly-matching condition claims that 
\begin{equation}
  T_{\mathrm{CP}} \geq T_{\mathrm{dec}}(\pi) \ ,
  \label{inequality-CP-dec}
\end{equation}
where $T_{\mathrm{CP}}$ represents the temperature at which
the CP symmetry at $\theta=\pi$ gets restored, and
$T_{\mathrm{dec}}(\theta)$ represents the critical temperature 
of the deconfining transition, which depends on $\theta$ in general.
Thus, an interesting question is whether CP restoration coincides 
with the deconfining transition ($T_{\mathrm{CP}} = T_{\mathrm{dec}}(\pi)$)
or not ($T_{\mathrm{CP}} > T_{\mathrm{dec}}(\pi)$)
as depicted in the Left and Right panels of figure~\ref{fig:phase_diag}, respectively.
In particular, the second scenario implies the existence of a deconfined phase
with spontaneously broken CP symmetry.

It is known that the first scenario is realized
in the large-$N$ limit \cite{Witten:1980sp,Witten:1998uka}, 
where the CP symmetry at $\theta = \pi$ is spontaneously broken at low temperature,
and it gets restored precisely at the deconfining temperature,
%%\emph{as soon as}
at which the $\mathbb{Z}_N$ center symmetry gets spontaneously
broken.
%% While the situation is expected to be the same 
%% for a sufficiently large but finite $N$ as well,
%% whether the situation remains
%% the same all the way down to $N=2$ is
%% an interesting issue that has been addressed by various approaches.
%%
%%% analytical studies by SUSY
Given the belief that the deconfining transition 
is of the first order for $N\geq 3$ and of the second order for $N=2$,
it was argued in Ref.~\cite{Gaiotto:2017yup} that the first scenario is likely to be realized
for $N\geq 3$ but not necessarily for $N=2$.
This issue has also been investigated in the SU($N$)
supersymmetric Yang-Mills theory (SYM) 
deformed by the gaugino mass and compactified on $S^1$
with periodic boundary conditions \cite{Chen:2020syd},
%% Our result is also consistent with the analytic study
%% based on the SU($N$) deformed SYM \cite{Chen:2020syd},
%% which suggests the existence of a CP-broken deconfined phase,
%% i.e., $T_{{\rm dec}}(\pi)<T_{{\rm CP}}$ for $N=2$.
which flows into the pure (non-supersymmetric) YM theory in the IR limit.
Regarding the radius of $S^1$ as an analog of the inverse temperature, 
it was found that the phase diagram looks like figure~\ref{fig:phase_diag} (Left) for $N\geq 3$, 
whereas for $N=2$, it looks like figure~\ref{fig:phase_diag} (Right), 
namely, there exists a CP-broken deconfined phase\footnote{
It was pointed out \cite{Hanada:2021ksu} that this intermediate phase is analogous to
the partially deconfined phase in large-$N$ gauge theories,
where the transition from the completely deconfined phase to the partially deconfined phase 
should be accompanied with the spontaneous breaking of the global symmetry, 
which is the CP symmetry in the present case.}.
One should keep in mind, however, that the gaugino mass has to be small enough 
to make the analysis based on supersymmetry reliable.
Note also that the radius of $S^1$ with periodic boundary conditions 
cannot be regarded as the inverse temperature,
which actually requires anti-periodic boundary conditions for the gaugino field.
Thus there is a strong motivation
%% These situations motivate us
to investigate the phase diagram 
of 4D SU($N$) pure YM theory with small $N$ directly by numerical methods.

%%% numerical studies
There are various approaches that have been proposed
to overcome the sign problem caused by the $\theta$ term.
%% So far, various approaches have been proposed to study the gauge theory
%% with the $\theta$ term overcoming the sign problem. 
%%
%for investigating the phase diagram in the presence of the topological $\theta$ term.
%% subvolume
In particular,
%%the phase structure of the 4D SU(2) YM at $\theta = \pi$ has been investigated by
the so-called subvolume method was applied to
4D SU(2) YM theory, where it was suggested that
%%, which suggests that
the CP symmetry at $\theta=\pi$
is broken at zero temperature 
and gets restored at
%%$T_{{\rm CP}} \le 1.2T_{\mathrm{c}}$ \cite{Kitano:2020mfk, Kitano:2021jho, Yamada:2024vsk},
$T_{{\rm CP}} \le 1.2 \, T_{\mathrm{dec}} (0)$ \cite{Kitano:2020mfk, Kitano:2021jho, Yamada:2024vsk}.
%%where $T_{\mathrm{c}}=T_{\mathrm{dec}} (0)$
%%represents the deconfining temperature at $\theta = 0$.
%% in the continuum limit.
However, this does not tell us whether
%%it was not possible to obtain any information related to
the inequality \eqref{inequality-CP-dec}
is saturated or not.
%% has not been obtained
%%along this line.
%% CLM
On the other hand,
the density of states method (DOS) \cite{Gattringer:2020mbf} and
the complex Langevin method (CLM) \cite{Hirasawa:2020bnl} 
%%and the tensor renormalization group (TRG) \cite{Kuramashi:2019cgs,Hirasawa:2021qvh}
have been successfully applied to 
2D U(1) gauge theory with the $\theta$ term using an open boundary condition.
There are also some attempts to extend these works
%apply the DOS and CLM
%% extend these works
to 4D SU(2) YM theory \cite{Gattringer:2021xrb,Matsumoto:2021zjf}.
(See also Ref.~\cite{Bongiovanni:2014rna} for an earlier attempt to
apply the CLM to 4D SU(3) YM theory.)
%% It turned out, however, that
%% the smearing techniques that is needed to reduce the finite lattice spacing effects
%% in defining the topological charge properly in the 4D case requires
%% the periodic bounary condition, which leads to the topology freezing problem.
The tensor renormalization group (TRG) \cite{Kuramashi:2019cgs,Hirasawa:2021qvh}
has also been successfully applied to 
2D U(1) gauge theory with the $\theta$ term,
but treating a non-Abelian gauge group in TRG
%% internal degrees of freedom of SU($N$) for $N \ge 2$
seems to be still difficult
%%is still challenging
except in 2D \cite{Hirasawa:2021qvh,Fukuma:2021cni,Asaduzzaman:2023pyz}.
(See, however, Refs.~\cite{Kuwahara:2022ubg,Yosprakob:2023jgl,Yosprakob:2024sfd}
for recent developments.)

%% In these works, it was important to introduce either
%% an open boundary or a puncture for different purposes.
%% This makes the topological charge non-integer, but it was still possible to
%% obtain meaningful results since there was no need to use the smearing techniques
%% in the 2D U(1) case.
In our attempts \cite{Matsumoto:2021zjf}
to extend our CLM work \cite{Hirasawa:2020bnl} on the 2D U(1) case with
an open boundary to the 4D SU(2) case,
we have found that the topological information leaks out from the open boundary
by smearing, which was needed to define the topological charge
unlike in the 2D U(1) case.
Moreover, since the open boundary makes the topological charge non-integer,
the $2\pi$ periodicity in $\theta$ is lost and 
the CP symmetry at $\theta = \pi$ is explicitly broken.
In order to avoid these problems, we need to work with periodic boundary conditions,
which would then cause either the topology freezing problem or the
wrong convergence problem in the CLM
depending on the coupling constant \cite{Hirasawa:2020bnl}.
Thus simulating the theory directly at $\theta=\pi$ still remains a big challenge.

\begin{comment}
%% TN
Instead, it has been applied to the Schwinger model using a technique 
to handle Grassmann variables directly \cite{Shimizu:2014fsa}.
There are also some attempts to determine the phase structure of the 2D CP(1) model at $\theta = \pi$, 
which is interesting in the context of condensed matter physics mentioned earlier, 
using the bond-weighted TRG \cite{Nakayama:2021iyp} and the loop-TNR \cite{Kawauchi:2017dnj}. 
%%% Hamiltonian
In addition to these approaches based on Lagrangian formalism, 
there have been recent developments in numerical methods based on the Hamiltonian formalism, 
motivated by the application of quantum computing.
In particular, the Schwinger model has been investigated by using a quantum simulator 
\cite{Chakraborty:2020uhf,Honda:2021aum,Honda:2021ovk}
and the density-matrix renormalization group method 
\cite{Byrnes:2002nv,Buyens:2017crb,Funcke:2019zna,Honda:2022edn,Dempsey:2023gib}.
\end{comment}

%%% related works with imaginary theta
In this paper, we use yet another approach to the sign problem,
%% due to the $\theta$ term,
which is based on analytic continuation from the imaginary $\theta$ region \cite{Berni:2019bch},
where the sign problem is absent since
the $\theta$ term becomes real\footnote{This is analogous to
the imaginary chemical potential approach \cite{deForcrand:2002hgr,DElia:2002tig}
to finite density QCD.}.
%%
% instead of reproducing the distribution itself 
% by precise simulations with imaginary $\theta$,
% we concentrate on the change of the observable at $\theta = \pi$ 
% by assuming the similarity to the Gaussian model or instanton gas.
%%
This technique has been used to investigate
the deconfining temperature \cite{DElia:2012pvq, DElia:2013uaf, Bonanno:2023hhp},
the free energy \cite{Panagopoulos:2011rb, Bonati:2015sqt, Bonati:2016tvi},
the string tension, the glueball mass \cite{Bonanno:2024ggk}
and the electric dipole moment of the neutron \cite{Aoki:2008gv, Guo:2015tla}
as a function of $\theta$ in 4D SU($N$) gauge theory for $N\ge3$.
Similarly, we calculate the expectation value of the topological charge at
imaginary $\theta$, and by fitting the results to an appropriate holomorphic function,
we obtain $\langle Q\rangle_\theta$ at real $\theta$ through analytic
%%and obtain the results at real $\theta$ by analytic
continuation\footnote{In fact,
%% Instead of doing this, one can also think of
%% determining the topological charge distribution at $\theta = 0$
%% very precisely since
imaginary $\theta$ enhances configurations with
large topological charge, and hence it enables us to
probe
%%simulations at imaginary $\theta$ amount to probing
the tail of the topological charge distribution, which is difficult to determine otherwise.
%% at $\theta = 0$ very precisely.
Then, by making a Fourier transform of the topological charge distribution,
one can obtain the $\theta$ dependence of the partition function
and hence the expectation value of the topological charge
as well \cite{Azcoiti:2002vk}. (See also Ref.~\cite{Azcoiti:2003vv} for related work.)
In fact, this method was applied
to the Schwinger model \cite{Azcoiti:2017mxl} 
and the $\mathbb{CP}^{N-1}$
%%CP($N-1$)
model \cite{Azcoiti:2003qe,Imachi:2006qq,Azcoiti:2007cg}.
It would be interesting to apply the same method to the 4D SU(2) YM case.}.
In particular, if the result of $\langle Q\rangle_{\theta}$
at $\theta=\pi$ is nonzero, it tells us
that the CP symmetry is spontaneously broken.
%%at that temperature.
Unlike the previous imaginary $\theta$ simulations,
we use the stout smearing \cite{Morningstar:2003gk}
%%In this work For this purpose, it is important to define
in defining the $\theta$ term in the action to be used in our simulation.
%%based on the hybrid Monte Carlo algorithm.
This is of particular importance for our purpose since the CP symmetry
at $\theta=\pi$ assumes that the topological charge takes integer values,
which is not the case if one uses a naive definition.
%% One of the notable differences of our method from the previous studies is 
%% the implementation of the stout smearing 
%% to the molecular dynamics evolution in the hybrid Monte Carlo method.
%% Then the topological nature is properly encoded 
%% not only in the observables but also in the gauge configurations themselves.

%% In order to define the topological charge
%% properly on the lattice with finite lattice spacing,
%% we use the stout smearing method not only in the measurement but also
%% in the action for generating configurations \cite{Morningstar:2003gk}.
We find that the CP symmetry is indeed spontaneously broken at low temperature.
%%whereas it is restored at high temperature and exhibit the instanton-gas behavior.
As we increase the temperature,
%%Furthermore,
the order parameter $\langle Q\rangle_{\theta=\pi}$ decreases and vanishes
at some temperature $T_{\mathrm{CP}}$
%% , which is
close to $T_{\mathrm{dec}} (0)$.
%% the deconfining temperature
%% at $\theta=0$.
%%
%%$T_{\mathrm{c}}$.
%%indicating the CP restoration, 
We also estimate the deconfining temperature $T_{\mathrm{dec}}(\theta)$
at real $\theta$ by analytic continuation, and find that
%% we can safely conclude that
%%$T_{\mathrm{dec}}(\pi)  \lesssim 0.91 T_{\mathrm{dec}} (0)$,
$T_{\mathrm{dec}}(\pi)  < T_{\mathrm{dec}} (0)$,
%%$\theta=\pi$, 
%%and it seems to be lower than $T_{\mathrm{dec}} (0)$.
%%
%%even considering the uncertainty of analytic continuation, 
which is consistent with the general
expectation based on analytical studies \cite{Unsal:2012zj, Poppitz:2012nz, Anber:2013sga} and numerical studies \cite{DElia:2012pvq, DElia:2013uaf, Otake:2022bcq, Borsanyi:2022fub}.
Combining these results, we obtain
%%suggest
the relation
\begin{equation}
  %%  T_{\mathrm{CP}} \sim T_{\mathrm{c}} > T_{\mathrm{dec}}(\pi),
   T_{\mathrm{CP}} \sim T_{\mathrm{dec}} (0) > T_{\mathrm{dec}}(\pi) \ ,
\end{equation}
which suggests the existence of a CP-broken deconfined phase
as in figure~\ref{fig:phase_diag} (Right).
%%in an intermediate temperature regime.

%% In this paper, we investigate 4D SU(2) YM theory with the $\theta$ term numerically,
%% and provide some evidence for the existence of the deconfined phase
%% with spontaneously broken CP at $\theta = \pi$.
%% We avoid the sign problem
%% by performing Monte Carlo simulations at imaginary $\theta$,
%% which makes the action real.

%% As the order parameter for the spontaneous CP symmetry breaking,
%% we calculate the expectation value of the topological charge
%% $\langle Q\rangle_\theta$ at imaginary $\theta$,
%% and by fitting the results to an appropriate holomophic function,
%% we obtain $\langle Q\rangle_\theta$ at real $\theta$ by analytic continuation.

%%We evaluate the expectation value of the topological charge $\langle Q\rangle_\theta$ 
%%as an order parameter of the CP symmetry, 
%%
%% Then by analytic continuation,
%% we obtain the expectation value $\langle Q\rangle_\theta$ at real $\theta$.
%%
%%Here the topological charge is not only measured
%%as an observable but also is involved in the action.
%%Thus, we adopt the stout smearing method in the simulation \cite{Morningstar:2003gk} 
%%to incorporate the effect of smearing gauge configurations in the Monte Carlo update.

This paper is organized as follows.
In section \ref{sec:imaginary}, we explain our strategy
to probe the spontaneous CP symmetry breaking at $\theta=\pi$
by Monte Carlo simulations at imaginary $\theta$.
%%review the relationship between the topological charge distribution
%%and the $\theta$ dependence of the topological charge
%%(at real and imaginary $\theta$)
%%with an example of 2D U(1) gauge theory, which is exactly solvable.
In section \ref{sec:stout}, we discuss the definition of the topological charge
on the lattice based on the stout smearing,
which we use in our Monte Carlo simulations.
%%explain our simulation method
%%
%%including the stout smearing technique used in reducing the finite lattice spacing
%%effects in the topological charge.
In section~\ref{sec:result}, we present our simulation results, which provide evidence of
%%the existence of
a CP broken deconfined phase.
Section~\ref{sec:discussion} is devoted to a summary and discussions.
In appendix~\ref{app:stout}, we derive the explicit form of the drift term,
in the presence of the stout smearing,
which is used in our simulation based on the hybrid Monte Carlo algorithm.
%%when the stout smearing is implemented. 

\section{How to probe the spontaneous CP symmetry breaking at $\theta=\pi$}
  %%The SU($N$) gauge theory with the $\theta$ term}
\label{sec:imaginary}

The 4D SU($N$) gauge theory with the $\theta$ term is defined
by the partition function
\begin{equation}
  Z_{\theta}=\int \mathcal{D}A_\mu \, e^{-S_{\rm g} - i\theta Q} \ ,
  \label{partition-fn-theta}
\end{equation}
where $S_{\rm g}$ is the action for the gauge field $A_\mu$,
and $Q$ is the topological charge defined by
\begin{equation}
  Q=
  \frac{1}{32\pi^2} \epsilon_{\mu\nu\rho\sigma} \int d^4x 
       {\rm Tr}\left( F_{\mu\nu} F_{\rho\sigma} \right) \ .
\end{equation}
Under the CP transformation, the action $S_{\rm g}$ is invariant,
whereas the topological charge $Q$ flips its sign.
%% Therefore the CP transformation corresponds to a sign flip of the theta ($\theta \to -\theta$).
On a compact manifold such as a 4D torus,
the topological charge takes an integer value $Q\in\mathbb{Z}$,
which implies the $2\pi$ periodicity under $\theta \mapsto \theta + 2 \pi$.
Therefore the theory has CP symmetry not only at $\theta=0$ but also at $\theta=\pi$.
The order parameter of spontaneous CP symmetry breaking at $\theta=\pi$
can be defined as
%%we consider a quantity defined as
\begin{equation}
  O =
  \lim_{\epsilon\to0} \lim_{V \to \infty} \frac{\Braket{Q}_{\theta=\pi-\epsilon}}{V} \ ,
\end{equation}
where $V$ represents the space-time volume.
It should be nonzero in the CP broken phase while it vanishes in the CP restored phase.
%% Here and hence forth, we denote the spatial volume by $V_{\rm s}$
%% and the four-dimensional space-time volume by $V$.

Note that the expectation value $\Braket{Q}_{\theta}$
has the following property around $\theta = \pi$
\begin{equation}
  \Braket{Q}_{\theta = \pi-\epsilon}
  = - \Braket{Q}_{\theta = -(\pi-\epsilon)}=-\Braket{Q}_{\theta = \pi+\epsilon} \ ,
\end{equation}
%% \begin{equation}
%% \begin{split} 
%%     \Braket{Q}_{\theta = \pi-\epsilon} &= - \Braket{Q}_{\theta = -(\pi-\epsilon)}\\
%%     &=-\Braket{Q}_{\theta = \pi+\epsilon} \ ,
%% \end{split}
%% \end{equation}
where we use the CP symmetry and the $2\pi$ periodicity 
under $\theta \mapsto \theta + 2\pi$
in the first and second equalities, respectively.
This implies that
$\lim_{V \to \infty} \frac{\Braket{Q}_\theta}{V}$
changes discontinuously at $\theta = \pi$
in the CP broken phase while it is continuous at $\theta = \pi$ in the CP restored phase.
%%Thus the discontinuity of $ \Braket{Q}_\theta$ at $\theta=\pi$
%% that occurs when the CP is spontaneously broken
Thus, the spontaneous CP symmetry breaking at $\theta=\pi$
implies the existence of a first-order phase transition at $\theta=\pi$.

Unfortunately, the spontaneous CP symmetry breaking at $\theta = \pi$
is difficult to study by Monte Carlo simulations
due to the sign problem
as one can see from \eqref{partition-fn-theta}, where
the $\theta$ term appears as a pure phase factor in the integrand.
In order to
%%look for a possible way out,
motivate our strategy,
%%to circumvent this problem,
let us
%%pay attention to
consider the topological charge distribution
%%at $\theta=0$ defined by
\begin{equation}
  \rho(q) = \frac{1}{Z_0}\int
\mathcal{D}A_\mu \, 
%%  dU
\delta(q-Q) \, e^{-S_g} \ ,
  %= \frac{1}{Z_0}\int \frac{d\theta}{2\pi} e^{-i\theta q}Z_\theta.
\end{equation}
which can be calculated by Monte Carlo simulations at $\theta=0$
without the sign problem.
Note that $\rho(q)$ is an even function, \emph{i.e.,} $\rho(q)=\rho(-q)$
due to CP symmetry.
Using this distribution, the partition function
\eqref{partition-fn-theta}
can be written as
\begin{equation}
  Z_\theta
  %%= \int \mathcal{D}A_\mu \,   e^{-S_g+i\theta Q}
  = Z_0\int dq \, e^{-i\theta q} \, \rho(q) \ ,
  \label{Z-rho-Fourier-rel}
\end{equation}
which implies that $Z_\theta$ may be regarded as the Fourier transform
of the topological charge distribution $\rho(q)$ at $\theta=0$.
Note also that the expectation value of the topological charge
%%at arbitrary $\theta$ as
can be written as
\begin{equation}
  \Braket{Q}_\theta
  = i\frac{\partial}{\partial\theta}\log{Z_\theta} 
= \frac{\int dq \, q \, e^{-i\theta q}\, \rho(q)}{\int dq \, e^{-i\theta q} \, \rho(q)} \ .
\label{eq:Q_theta}
\end{equation}
Hence one can obtain $\Braket{Q}_\theta$ if $\rho(q)$ is known and vice versa.
%% Thus the discontinuity of $ \Braket{Q}_\theta$ at $\theta=\pi$
%% that occurs when the CP is spontaneously broken
%% implies the existence of a first order phase transition at $\theta=\pi$.

%%\section{$\theta$ dependence of $\Braket{Q}_\theta$ in two models}
%%In this section we

\begin{figure}
    \centering
    \includegraphics[width=0.45\hsize]{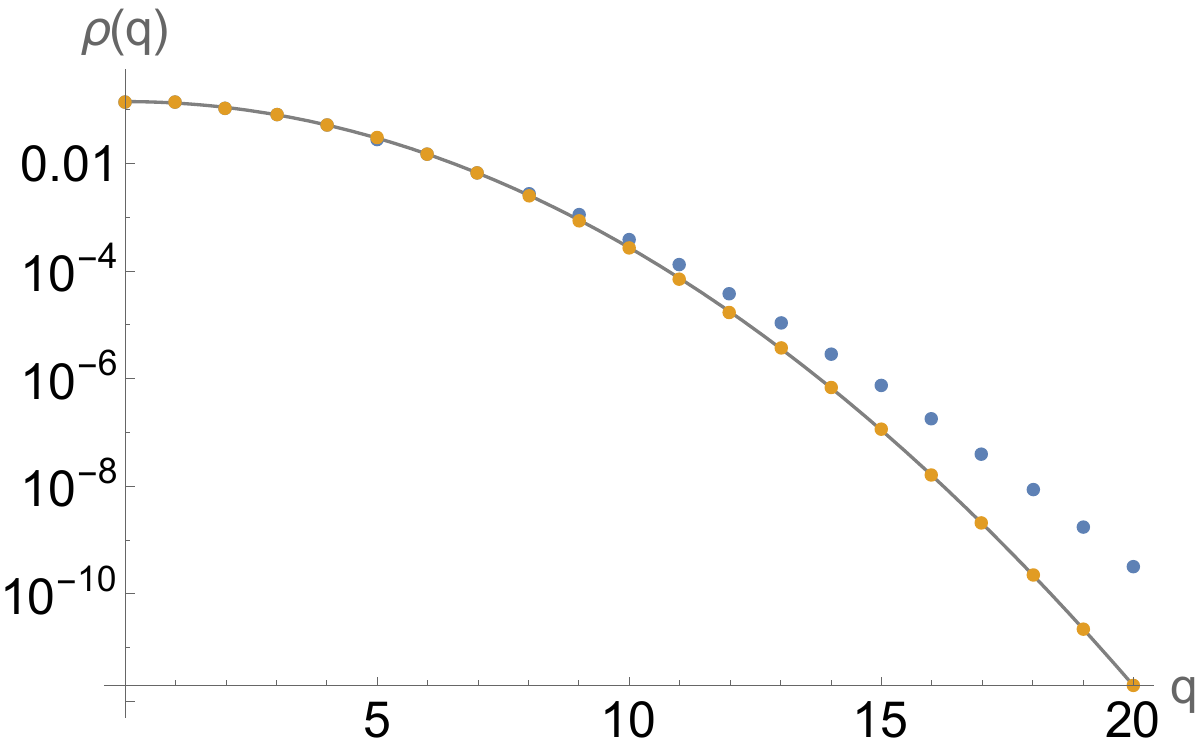}
    \caption{The topological charge distribution is plotted
      in the logarithmic scale
      for the result obtained by the instanton gas approximation (blue) and
      that obtained in the large-$N$ limit at low temperature (orange).
      The solid line represents a Gaussian behavior $\propto\exp\left(-q^2/2\chi_0V\right)$
      known for the latter case.}
%%      We can see a clear difference in the asymptotic behavior at large $q$.}
    \label{fig:rho_q}
\end{figure}

%%Let us discuss
%%the $\theta$ dependence of $\Braket{Q}_\theta$
There are actually known results in two simplified cases.
%% in which the behaviors of $\Braket{Q}_\theta$ are different.
One is the result obtained by the instanton gas approximation \cite{Gross:1980br,Weiss:1980rj},
which is considered to be valid
%%a good an effective model of the Yang-Mills theory
at sufficiently high temperature.
%%where the CP is restored.
The other is the result obtained in the large-$N$ limit \cite{Witten:1980sp,Witten:1998uka} at low temperature,
where the topological charge distribution is known to be Gaussian.
%%Gaussian model, which is an effective model
%% of the large-$N$ Yang-Mills theory at low temperature where the CP is broken.
%%
%%Using the Eq.~(\ref{eq:Q_theta}) and Eq.~(\ref{eq:q_dists}),
%%we obtain the $\theta$ dependence of
The quantity $\Braket{Q}_\theta$ in these two cases is given as
\begin{equation}
    \frac{\Braket{Q}_\theta}{V} = 
    \left\{
        \begin{array}{ll}
          i  \chi_0  \sin \theta &:\ \mbox{instanton gas approximation (high $T$)}\   ,\\
        i \chi_0  \, \theta &:\ \mbox{large-$N$ limit (low $T$)}  \ ,
        \end{array}
    \right.
\end{equation}
for $|\theta|<\pi$, where $\chi_0$ is the topological susceptibility at $\theta = 0$ defined by
\begin{align}
  \chi_0 &=  \left. \frac{i}{V} \frac{\del}{\del \theta}
  \langle Q \rangle _{\theta} \right|_{\theta=0} =
  \frac{1}{V}  \left\{  \langle Q^2 \rangle_{0} - (\langle Q \rangle_0)^2 \right\}  \ .
  \label{top-susceptibility}
\end{align}
This implies that the CP symmetry is spontaneously broken at low $T$,
whereas it gets restored at high $T$ in these simplified cases.
Using
%% \eqref{Z-rho-Fourier-rel} and
\eqref{eq:Q_theta},
we obtain the asymptotic behavior of the topological charge distribution $\rho(q)$
at large $|q|$ as
\begin{equation}
    \rho(q) \sim
    \left\{
        \begin{array}{ll}
          \exp\Big(-|q| \log{|q|}\Big)
          &:\ \mbox{instanton gas approximation (high $T$)} \ ,\\
          \exp\left(-\frac{q^2}{2\chi_0V}\right)
          &:\ \mbox{large-$N$ limit (low $T$)} \ .
        \end{array}
    \right.
\label{eq:q_dists}
\end{equation}
This suggests that the spontaneous breaking of the CP symmetry at $\theta=\pi$ 
is related to the asymptotic behavior of $\rho(q)$.
In figure~\ref{fig:rho_q}, we plot $\rho(q)$ against $q$
in the logarithmic scale for the two cases.
We can see a clear difference in the tail of the distribution,
which is difficult to probe by Monte Carlo simulations at $\theta=0$, however.
%%However, it is difficult to probe the tail of 

Let us here recall that Monte Carlo simulations can be performed
for imaginary $\theta$ since the sign problem is absent there.
Note, in particular, that the phase factor $e^{-i \theta Q}$ becomes
a weight $e^{\tilde{\theta}Q}$ with
$\tilde{\theta} = -i\theta \in \mathbb{R}$ in that case.
This makes it possible to probe the topological charge distribution $\rho(q)$
at large $|q|$.
In the above two simplified cases, the expectation value
$\Braket{Q}_{i\tilde{\theta}}$ becomes
%% of the topological charge 
%% by performing the analytic continuation $\tilde{\theta} = i\theta$,
%% we obtain the imaginary theta $\tilde{\theta}$ dependence
%% of $\Braket{Q}_{-i\tilde{\theta}}$ as
\begin{equation}
    \frac{\Braket{Q}_{i\tilde{\theta}}}{V} = 
    \left\{
        \begin{array}{ll}
          \chi_0  \sinh \tilde{\theta} &:\ \mbox{instanton gas approximation (high $T$)} \ ,\\
          \chi_0  \, \tilde{\theta} &:\ \mbox{large-$N$ limit (low $T$)} \ .
        \end{array}
    \right.
\label{eq:q_im_theta}
\end{equation}
The different asymptotic behaviors of the topological charge distribution
correspond here to the different growths of $\Braket{Q}_{i\tilde{\theta}}$
at large $\tilde{\theta}$.
In general, we expect exponential growth for the CP-restored phase
and power-law growth for the CP broken phase.
This is the basic strategy we adopt in this work.
%% Therefore, it is possible to observe the difference of the asymptotic behavior of the topological charge distribution by performing lattice simulations at $\tilde{\theta}$.

%%%%%%%%%%%%%%%%%%%%%%%%%%%%%%%%%%%%%%%%%
\section{Definition of the topological charge on the lattice}
%\section{Set up of our study}
\label{sec:stout}
%%%%%%%%%%%%%%%%%%%%%%%%%%%%%%%%%%%%%%%%%

In this section, we define the gauge theory on the lattice,
which we simulate by the standard hybrid Monte Carlo method.
In particular, we define the topological charge
carefully using the so-called stout smearing.
%%in the $\theta$ term
%%has to be defined carefully since a naive definition is known to have
%%severe lattice artifacts, which obscures the topological nature of this quantity.

%%present our simulation method.
%We study the following plaquette action:
The action for the gauge field $S_{\rm g}$ in \eqref{partition-fn-theta}
is represented on the lattice by
the standard Wilson plaquette action as
\begin{equation}
%S_{\beta}
S [U] 
=-\frac{\beta}{2N}
\sum_{n}\sum_{\mu\neq\nu}\mathrm{Tr}\left\{ P_{n}^{\mu\nu} (U) \right\} \ ,
\label{wilson-plaquette-action}
\end{equation}
where the plaquette $P_{n}^{\mu\nu}(U)$ is defined by
\begin{equation}
P_{n}^{\mu\nu}(U)
=U_{n,\mu}U_{n+\mu,\nu}U_{n+\nu,\mu}^{\dagger}U_{n,\nu}^{\dagger} \ .
\end{equation}
In this work, we use the
$L_{\mathrm{s}} \times L_{\mathrm{s}} \times L_{\mathrm{s}} \times L_{\mathrm{t}}$
lattice with
various $16 \le L_{\mathrm{s}}\le 64$ fixing $L_{\mathrm{t}}=5$.
%% in the spatial directions
%% $L_{\mathrm{t}}$ is the number of lattice sites in the temporal direction,

Since we use fixed $L_{\mathrm{t}}=5$,
we change the temperature $T=1/(aL_{\mathrm{t}})$
%%on a fixed lattice
by changing the lattice spacing $a$,
which amounts to changing $\beta$.
To be concrete, we use
%%using 
%%To be concrete, we use
the relationship between the temperature and $\beta$
given in Ref.~\cite{Engels:1994xj} as
\begin{equation}
  \frac{T}{
  T_{\rm c}}=\frac{1}{t_{\rm c} L_{\mathrm{t}}}
  \exp\left(\frac{51}{121}\log\frac{11}{6\pi^{2}\beta}+
  \frac{3\pi^{2}\beta}{11}-c_{3}\frac{18432\pi^{4}}{121\beta^{3}}\right) \ ,
  \label{T-Tc-beta-relation}
\end{equation}
where $T_{\rm c}$ is the deconfining temperature in the continuum limit, 
$t_{\rm c} = 21.45(14)$, and $c_{3}=5.529(63)\times10^{-4}$.
The value of $\beta$ corresponding to each $T/T_{\mathrm{c}}$
is given in table~\ref{tab:weight}.
%is the lattice cutoff scale,
%%$L_{\mathrm{t}}$ is the number of lattice sites in the temporal direction,
%The deconfining temperature is given by
%$T_{\text{c}}$/$\Lambda_{\mathrm{L}}=21.45(14)$.

%[Atis] The topological charge $Q[U]$ on the lattice is defined by Ref.~\cite{DiVecchia:1981aev}
A naive definition of the topological charge $Q(U)$ on the lattice is
%%defined naively
given by~\cite{DiVecchia:1981aev}
\begin{equation}
\label{eq: def_of_Q}
Q(U)
=-\frac{1}{32\pi^{2}}
\sum_{n}\frac{1}{2^{4}}\sum_{\mu,\nu,\rho,\sigma=\pm1}^{\pm4}
\tilde{\epsilon}_{\mu\nu\rho\sigma}
%%\epsilon_{\mu\nu\rho\sigma}
\mathrm{Tr} \left( P_{n}^{\mu\nu}P_{n}^{\rho\sigma} \right)  \ ,
\end{equation}
where $\tilde{\epsilon}_{\mu\nu\rho\sigma}$ is the totally anti-symmetric tensor satisfying
%% $\tilde{\epsilon}_{1234}=1$.
%%
$\tilde{\epsilon}_{(-\mu)\nu\rho\sigma}=-\epsilon_{\mu\nu\rho\sigma}.$
%% \begin{equation}
%% 1=\tilde{\epsilon}_{1234}=-\tilde{\epsilon}_{2134}=-\tilde{\epsilon}_{-1234}=\cdots \ .
%% \end{equation}
%[Atis] Note that while $Q[U]$ in \eqref{eq: def_of_Q} reproduces the continuum topological charge in the continuum limit,
While
%%$Q(U)$ in \eqref{eq: def_of_Q}
this quantity
reproduces
%%becomes
the topological charge in the continuum limit,
it typically has severe lattice artifacts, which obscure the topological nature.
%% is generically non-integer and not topological due to the discretization effect.
%%
%The topological charge \eqref{eq: def_of_Q} becomes non-integer valued due to the discretization effect.
In order to recover the topological nature even at finite lattice spacing,
%%property of the gauge theory with the $\theta$ term,
we remove the high momentum modes in each configuration
by applying the stout smearing 
%, which was first proposed in Ref.~
\cite{Morningstar:2003gk}.
%In the following, we briefly review how to apply the stout smearing to our simulation.

Here we briefly review the procedure.
%%how we apply the stout smearing in our simulation.
The original configuration $U_{n,\mu}$ is replaced
%%are updated to
with the smeared configuration $\tilde{U}_{n,\mu}$ obtained
by $N_\rho$ steps of the stout smearing as
\begin{equation}
(U_{n,\mu} \equiv) U_{n,\mu}^{(0)} \to U_{n,\mu}^{(1)} \to \cdots
  \to (\tilde{U}_{n,\mu} \equiv ) U_{n,\mu}^{(N_\rho)}  \ ,
  \end{equation}
%By the following formulae,
where each step is defined by
%%We obtain the $(k+1)$-times smeared link
%%variables $U_{n,\mu}^{(k+1)}\ (0 \leq k \leq N_\rho-1)$ as follows:
\begin{align}
  U_{n,\mu}^{(k+1)}
  &= \exp \qty(iY_{n,\mu}^{(k)}) \, U_{n,\mu}^{(k)} \ ,
\label{exp-Y}
  \\
  iY_{n,\mu}^{(k)}
  &= -\frac{\rho}{2} \Tr \qty(J_{n,\mu}^{(k)} \tau^a) \tau^a
=-\frac{\rho}{2} \qty{J_{n,\mu}^{(k)} - \frac{1}{2}\Tr \qty(J_{n,\mu}^{(k)}) \mathbbm{1}}
\ ,
\label{smearing-param}
\\
J_{n,\mu}^{(k)}
&= U_{n,\mu}^{(k)} \, \Omega_{n,\mu}^{(k)}
%% - \bar{\Omega}_{n,\mu}^{(k)} \, \qty(U_{n,\mu}^{(k)})^\dagger \ , \\
- {\Omega_{n,\mu}^{(k)}}^\dagger \, {U_{n,\mu}^{(k)}}^\dagger \ , \\
\Omega_{n,\mu}^{(k)} 
&= \sum_{\sigma(\neq \mu)}
\qty(
U_{n+\hat{\mu},\sigma}^{(k)} 
%%\qty(U_{n+\hat{\sigma},\mu}^{(k)})^\dagger
U_{n+\hat{\sigma},\mu}^{(k)\dagger}
%%\qty(U_{n,\sigma}^{(k)})^\dagger +
U_{n,\sigma}^{(k)\dagger} +
%%\qty(U_{n+\hat{\mu}-\hat{\sigma},\sigma}^{(k)})^\dagger
U_{n+\hat{\mu}-\hat{\sigma},\sigma}^{(k)\dagger}
%%\qty(U_{n-\hat{\sigma},\mu}^{(k)})^\dagger U_{n-\hat{\sigma},\sigma}^{(k)} } \ , \\
U_{n-\hat{\sigma},\mu}^{(k)\dagger} U_{n-\hat{\sigma},\sigma}^{(k)} ) \ .
% \bar{\Omega}_{n,\mu}^{(k)} 
% &= \sum_{\sigma(\neq \mu)} 
% \qty( U_{n,\sigma}^{(k)} U_{n+\hat{\sigma},\mu}^{(k)}
%   U_{n+\hat{\mu},\sigma}^{(k)\dagger}
%   +U_{n-\hat{\sigma},\sigma}^{(k)\dagger} U_{n-\hat{\sigma},\mu}^{(k)}
%   U_{n+\hat{\mu}-\hat{\sigma},\sigma}^{(k)} ) \ .
\end{align}
We have defined the SU(2) generators $\tau^a$ with the normalization
$\tr (\tau^a \tau^b) = \frac{1}{2} \delta_{ab}$.
%% $\tau^a=\sigma^a/2$
%% in terms of the Pauli matrices $\sigma^a$.
%[Atis] The smearing parameter $\rho$ is positive and should be chosen appropriately, which depends on the system.
The smearing parameter $\rho>0$ in \eqref{smearing-param}
%%is positive and
should be chosen appropriately as we discuss shortly.
%% depending on the system.
Note that the stout smearing is nothing but
%%can be regarded as
the discretized version
of the gradient flow method \cite{Luscher:2010iy}
since $J_{n,\mu}^{(k)}$
%%is nothing but
represents
the derivative of the Wilson plaquette action
with respect to the link variable $U_{n,\mu}^{(k)}$.
One of the biggest advantages of the stout
smearing among various smearing techniques
is that the smeared link is differentiable with respect to the original link, 
which is crucial in applying the HMC algorithm to the action including a term
defined with the smeared link.

In the SU(2) case, since the eigenvalues of the traceless Hermitian matrix $Y_{n,\mu}$ 
are given by $\pm \kappa_{n,\mu}$ with 
\begin{equation}
  \kappa_{n,\mu}
  %% = \sqrt{-\det Y_{n,\mu}}
  = \sqrt{\frac{1}{2} \Tr(Y_{n,\mu})^2} \ ,
\label{def-kappa}
\end{equation}
the matrix $\exp(iY_{n,\mu})$ in \eqref{exp-Y} can be obtained easily by
\begin{equation}
  \exp(iY_{n,\mu}) =(\cos \kappa_{n,\mu}) \mathbbm{1}
  + \frac{\sin \kappa_{n,\mu}}{\kappa_{n,\mu}} \, iY_{n,\mu} \ .
  \label{def-exp-Y}
\end{equation}

%%%
\begin{figure}[tb]
\centering
\includegraphics[scale=0.4]{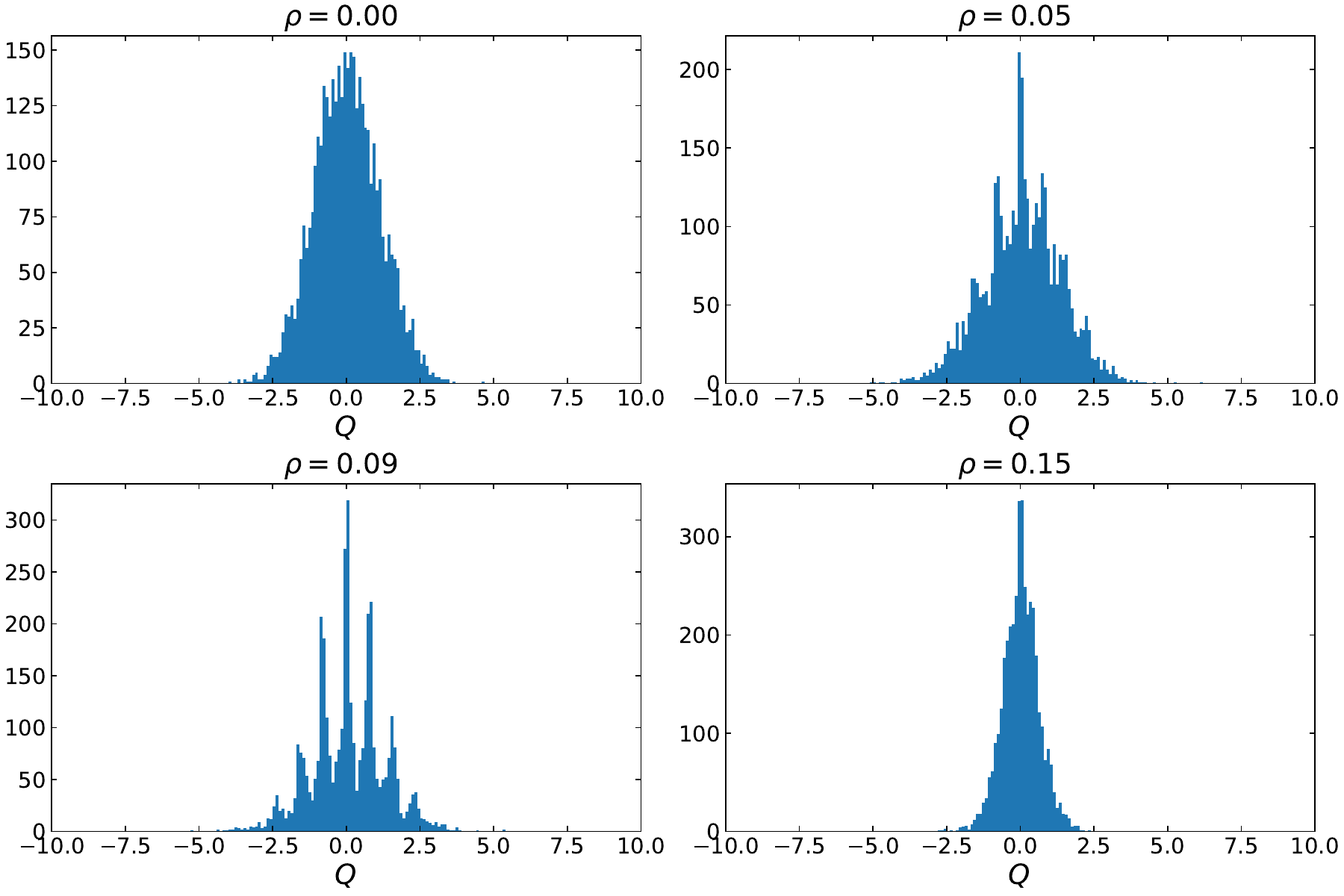}
\caption{The histogram of the topological charge after the stout smearing
for $\rho=0$, 0.05, 0.09, and 0.15 with the fixed number of
smearing steps $N_{\rho}=40$.
%%The lattice volume is $V_{\rm L}=20^3\times 5$, and
The lattice volume is $V_{\rm L}=20^3\times 5$, and
the temperature is $T=1.2 \, T_{\mathrm{c}}$.}
\label{fig:Q_hist}
\end{figure}
%%%

%We choose the parameters $N_\rho$ and $\rho $ such that the topological charge distribution exhibits comb-like structures.
In figure~\ref{fig:Q_hist}, 
we plot the distribution of the topological charge $Q[\tilde{U}]$ 
computed from the smeared link variables $\tilde{U}_{n,\mu}$ 
%%the topological charge distribution
at $\theta=0$
%[Atis] for various values of the smearing parameter $\rho$ while the number of smearing steps is fixed to $N_{\rho}=40$.
for various $\rho$
%% smearing parameter 
with the fixed number of smearing steps $N_{\rho}=40$.
%[Atis] We easily see that the topological charge distribution for $\rho=0.09$ shows the comb-like structure.
We find that a comb-like structure is most clearly seen for $\rho=0.09$.
Therefore, we use $N_\rho=40$ and $\rho=0.09$ in our simulation.
%[Atis] Note that the positions of peaks are slightly deviate from integer values due to the lattice artifacts. Therefore we need to renormalize the parameter $\theta$. To determine the renormalization factor, we use a method that is explained in appendix~\ref{app:ren_theta}.
%The results we show bellow are obtained after the renormalization.

Note that the positions of the peaks are slightly shifted from integer values
due to the lattice artifacts.
It is well known that
one can actually make the position of the peaks closer to integers
by rescaling $Q\rightarrow wQ$ with some parameter $w>0$.
(See, for instance, Ref.~\cite{DelDebbio:2002xa}.)
%% In this work, 
%% we consider
%% the renormalization of the $\theta$ parameter, which has been done
%% in the literature. (See, for instance, Ref.~\cite{DelDebbio:2002xa}.)
In this work, we numerically determine the value of $w$ from
$Q[\tilde{U}]$ obtained for each configuration 
in such a way that the cost function
%%so that $wQ[\tilde{U}]$ is closed to integers as much as possible.
%%The determination of $w$ is performed by minimizing the cost function,
\begin{equation}
  F(w)=\Braket{1-\cos(2\pi wQ[\tilde{U}])}
  %%\ ,
\end{equation}
is minimized, where the symbol $\Braket{\cdots}$
represents the ensemble average.
%%of the Monte Carlo data.
(A similar cost function has been used in Ref.~\cite{Bonati:2015sqt}, for instance.)
%% If $wQ[\tilde{U}]$ is exactly integer, the cost function becomes zero.
In practice, we calculate $F(w)$ as a function of $w$,
and determine the optimal value $w_{\mathrm{opt}}$ by fitting the result to
%%$F(w)=A(w-w_{\mathrm{opt}})^2+C$ near the minimum.
$F(w)=A(w-w_{\mathrm{opt}})^2+B$ near the minimum.

%% As we have seen in figure~\ref{fig:Q_hist},
%% the topological charge $Q[\tilde{U}]$ defined with the smeared link $\tilde{U}$
%% has a comb-like distribution for appropriate choice of the smearing parameters.
%% However, one finds that the position of the peaks is slightly shifted from integers.

%Note that we compute values of ww at L=16L=16 and use them in Sec.?????????????????????\ref{sec:result} without taking the infinite volume extrapolations because it is practically impossible to find the minima at larger lattice volume due to the unclear comb-like structure in the topological charge distribution, especially in lower temperature region.

In table~\ref{tab:weight}, we present
the optimal value of $w$ obtained in this way
for $L=16$ at each temperature.
For a larger lattice at low temperature,
it turns out to be difficult to find the minimum 
since the comb-like structure becomes unclear.
On the other hand, at high temperature,
we find that
%%the volume dependence of
the optimal value of
$w$ is actually almost independent of the volume.
Therefore, we use the values in table~\ref{tab:weight}
for all $L_{\rm s}$.
%%in Section \ref{sec:result}.
%% without taking the infinite volume extrapolations.
%%since the volume dependence of $w$ turns out to be very small. 

In our simulation, the topological charge
is defined by $\tilde{Q} \equiv w Q[\tilde{U}]$
with the smeared link variables $\tilde{U}_{n,\mu}$,
whereas the plaquette action is
defined with the original link variables $U_{n,\mu}$.
%% \red{Since the smearing factor \eqref{def-exp-Y} is differentiable with respect to the original link, 
%% the topological charge $\tilde{Q}$ can be employed in the action for the molecular dynamics evolution.}
We will denote $\tilde{Q}$ simply as $Q$ in what follows.
In appendix~\ref{app:stout}, we review how to calculate
the force term in the hybrid Monte Carlo algorithm
with the stout smearing.
%%for thesmeared link variables.
%% for the reader's convenience.
%% Accordingly, we have to calculate the force term \eqref{drift-after-each-smearing}

\begin{table}[t]
  \centering
 \begin{minipage}[b]{0.48\columnwidth} 
  \centering
   \begin{tabular}{|c||c|c|}\hline
        $T/T_{\mathrm{c}}$ & $\beta$ & $w$ \tabularnewline\hline 
        \hline 
        0.90 & 2.33381 & 1.2234(3) \tabularnewline\hline 
        0.96 & 2.35320 &1.2176(3) \tabularnewline\hline 
        0.98 & 2.35943 &1.2167(1) \tabularnewline\hline 
        0.99 & 2.36250 &1.2183(4) \tabularnewline\hline
        1.00 & 1.36554 &1.2183(1) \tabularnewline\hline 
    \end{tabular}
 \end{minipage}
 \hspace{-0.1\columnwidth}
%%   \hspace
 \begin{minipage}[b]{0.48\columnwidth}
  \centering
    \begin{tabular}{|c||c|c|}\hline
        $T/T_{\mathrm{c}}$ & $\beta$ & $w$ \tabularnewline\hline 
        \hline 
        1.01 & 2.36856 &1.2200(3) \tabularnewline\hline 
        1.02 & 2.37155 &1.2217(3) \tabularnewline\hline 
        1.03 & 2.37452 &1.2233(2) \tabularnewline\hline 
        1.04 & 2.37745 &1.2247(2) \tabularnewline\hline 
        1.10 & 2.39458 &1.2431(1) \tabularnewline\hline 
        %1.20 & 1.2650(1) \tabularnewline\hline 
    \end{tabular}
\end{minipage}
 \caption{The value of $\beta$ in the Wilson plaquette action
   \eqref{wilson-plaquette-action} used in our simulations
   is given at each temperature based on the relation \eqref{T-Tc-beta-relation}.
   We also present the rescaling factor $w$ that makes the topological charge
   close to integers at each temperature for $L_{\rm s}=16$.}
    \label{tab:weight}
\end{table}

%Therefore, we expect the above treatment does not occur any significant problem.
%% Therefore, we expect the above treatment does not lead to any significant problem.
%% We also expect that we can estimate $w$ at a larger lattice by approaching the continuum limit or by eliminating $\mathcal{O}(a)$ contribution using an improved action.

%% Therefore we need to renormalize
%% the parameter $\theta$. To determine the renormalization factor,
%% we use a method that is explained in appendix~\ref{app:ren_theta}.
%% The results we show below are obtained after the renormalization.

%For the parameters regarding the stout smearing, we take $N_\rho=40,\ \rho=0.09$ in our simulation.
%This is because we can see the comb-like structures in the topological charge distribution at $\theta=0$ in this parameter setting as shown in figure~\ref{fig:Q_hist}.

%%%%%%%%%%%%%%%%%%%%%%%%%%%%%%%%
%%%%%%%%%%%%%%%%%%%%%%%%%%%%%%%%
%%%%%%%%%%%%%%%%%%%%%%%%%%%%%%%%
\section{Simulation results}
%\section{Results}
\label{sec:result}
%%%%%%%%%%%%%%%%%%%%%%%%%%%%%%%%
%%%%%%%%%%%%%%%%%%%%%%%%%%%%%%%%
%%%%%%%%%%%%%%%%%%%%%%%%%%%%%%%%

In this section, we present our results of the Monte Carlo simulation.
%% In order to investigate the behavior near the phase transition
%% associated with the spontaneous CP symmetry breaking,
%% we make an infinite volume extrapolation.
%%
%% for the topological charge expectation value $\Braket{Q}_{-i\tilde{\theta}}$
%% and the topological charge susceptibility $\chi_0$.
We compute the topological charge $\Braket{Q}_{i\tilde{\theta}}$ 
%%extrapolated values of
%%either to a polynomial or to a series of hyperbolic sine functions,
and fit the result to some holomorphic function of $\tilde{\theta}$.
%% where some linear combination of the coefficients
%% is fixed by $\chi_0$ determined by the infinite volume extrapolation.
%%We compare the quality of the two fits based on the $\chi^2$ values.
By making an analytic continuation of the fitting functions,
we obtain $\Braket{Q}_{\theta}$ for real $\theta$
%[Atis] we observe the CP breaking or restoration.
and determine whether the CP symmetry is spontaneously broken or not
at each temperature,
which tells us the critical temperature $T_{\rm CP}$
for the spontaneous CP symmetry breaking.
We also determine $T_{\rm dec}(\theta)$ by analytical continuation
and show that $T_{\rm CP}> T_{\rm dec}(\pi)$.

In order to justify the application of analytic continuation, $\Braket{Q}_{\theta}$ has to be an analytic function of $\theta$, which cannot be true in the rigorous sense since there is a deconfining phase transition as one increases $\theta$ at some temperature fixed in a certain range. 
(See figure~\ref{fig:phase_diag}.)  
However, in the present case of the SU(2) YM theory, the deconfining phase transition is of the second order\footnote{
In the case of the SU(3) YM theory, the deconfining phase transition is of the first order, which implies that $\Braket{Q}_{\theta}$ is no longer continuous across the transition.
%%the free energy is no longer analytic across the transition.
Thus, for the temperature at which the deconfining phase transition occurs at some $\theta$, the analytic continuation with respect to $\theta$ cannot be extended further.
The spontaneous CP symmetry breaking can still be investigated outside this temperature region.
} and hence $\Braket{Q}_{\theta}$ is continuous across the transition as a function of $\theta$.
Here we assume that it is not only continuous but also analytic across the transition
within small errors given that the non-analytic behavior at a second
order transition is observed only in a detailed analysis of finite volume effects
close to the transition.
Indeed this assumption seems to be valid
%% to good accuracy
at least in the imaginary $\theta$
region that can be directly explored by our simulations.
(See the last paragraph of Section \ref{sec:deconf-temp-theta-dep}.)
Under this assumption, it is expected that the analytic continuation is valid until one hits
the first-order phase transition line at $\theta=\pi$ if it exists.

\subsection{$\theta$ dependence of the topological charge density}
%%\texorpdfstring{$\Braket{Q}_{-i\tilde{\theta}}/V_{\rm s}$}{Q/V}}

\begin{figure}[H]
    \centering
    \includegraphics[width=0.475\hsize]{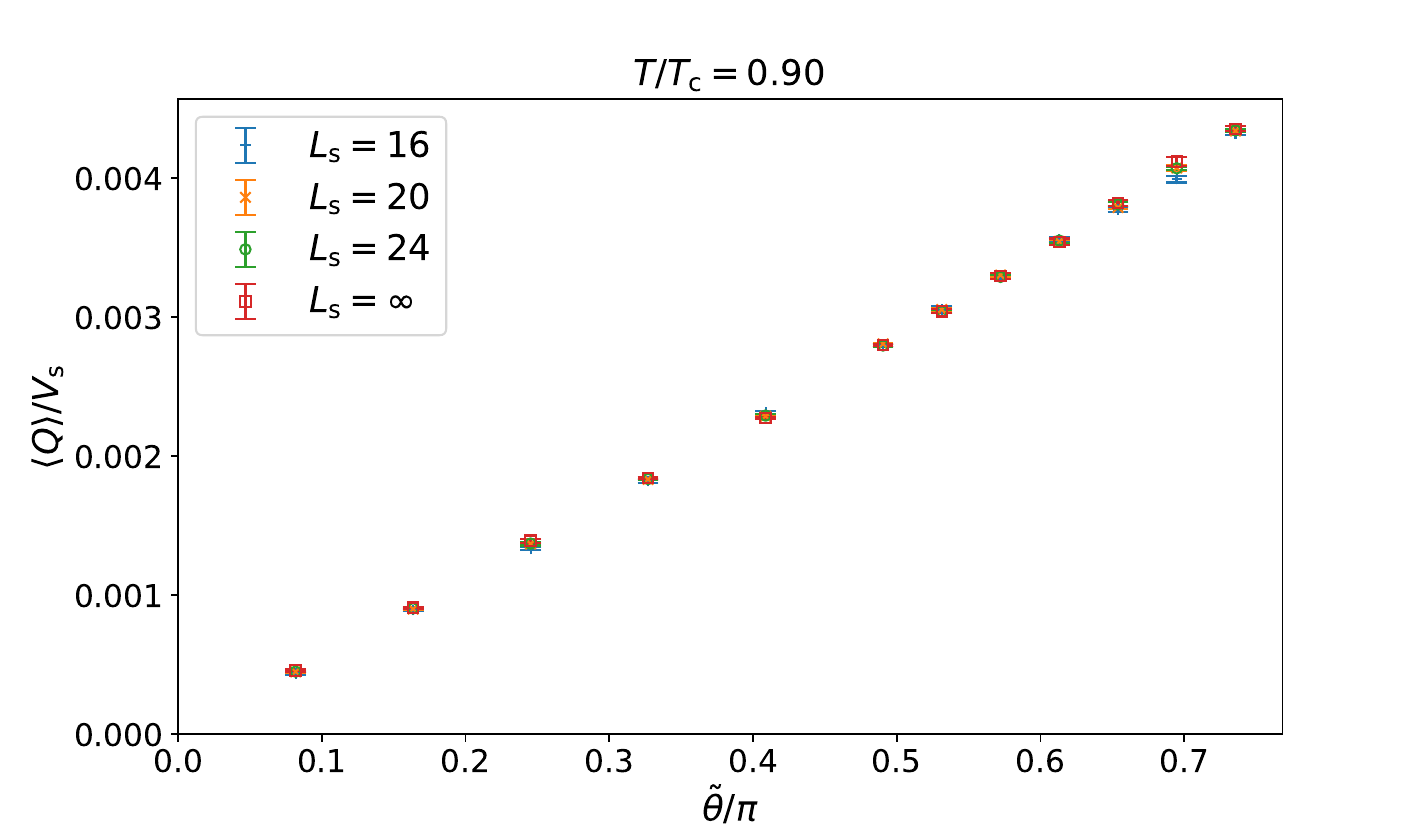}
    \includegraphics[width=0.475\hsize]{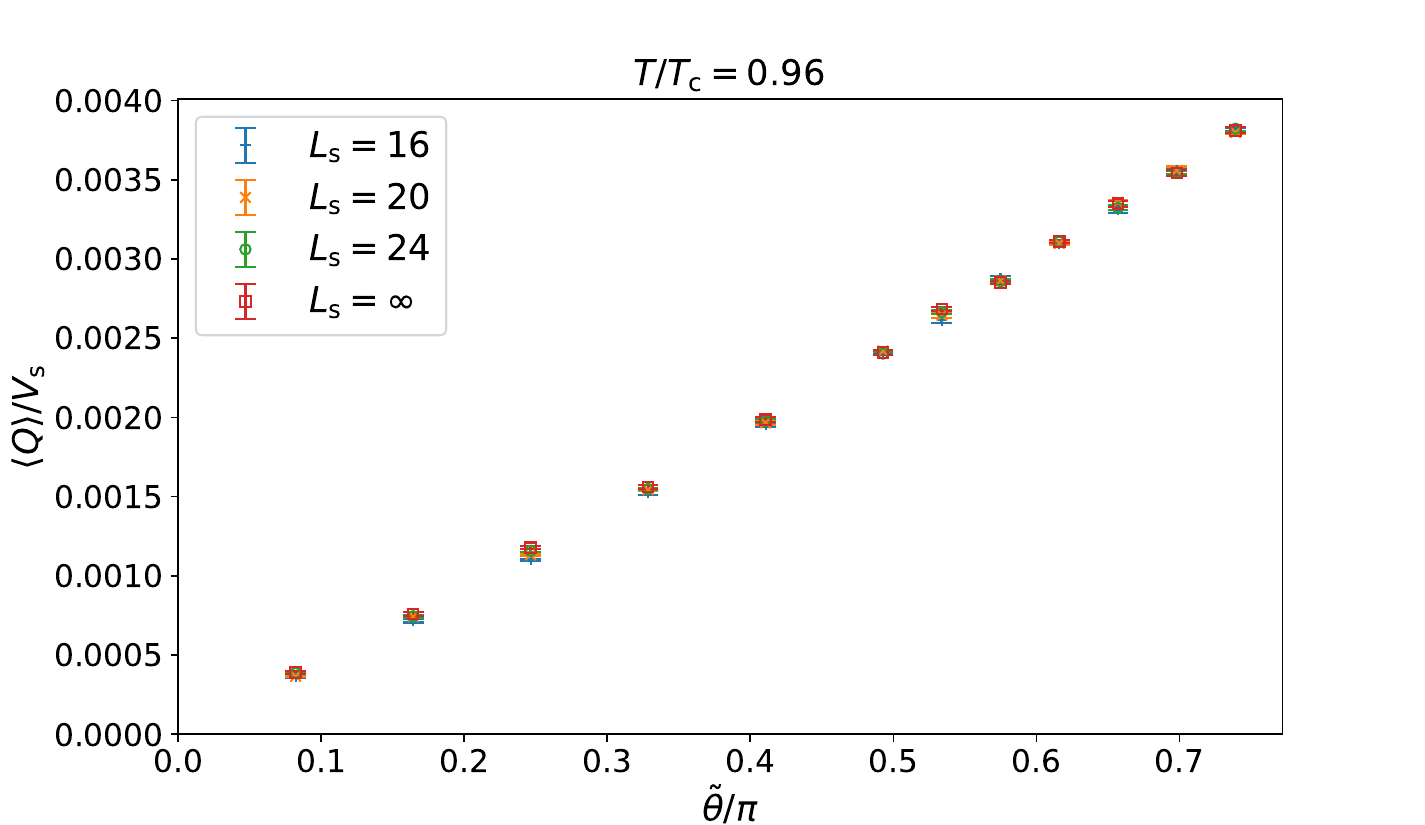}\\
    \includegraphics[width=0.475\hsize]{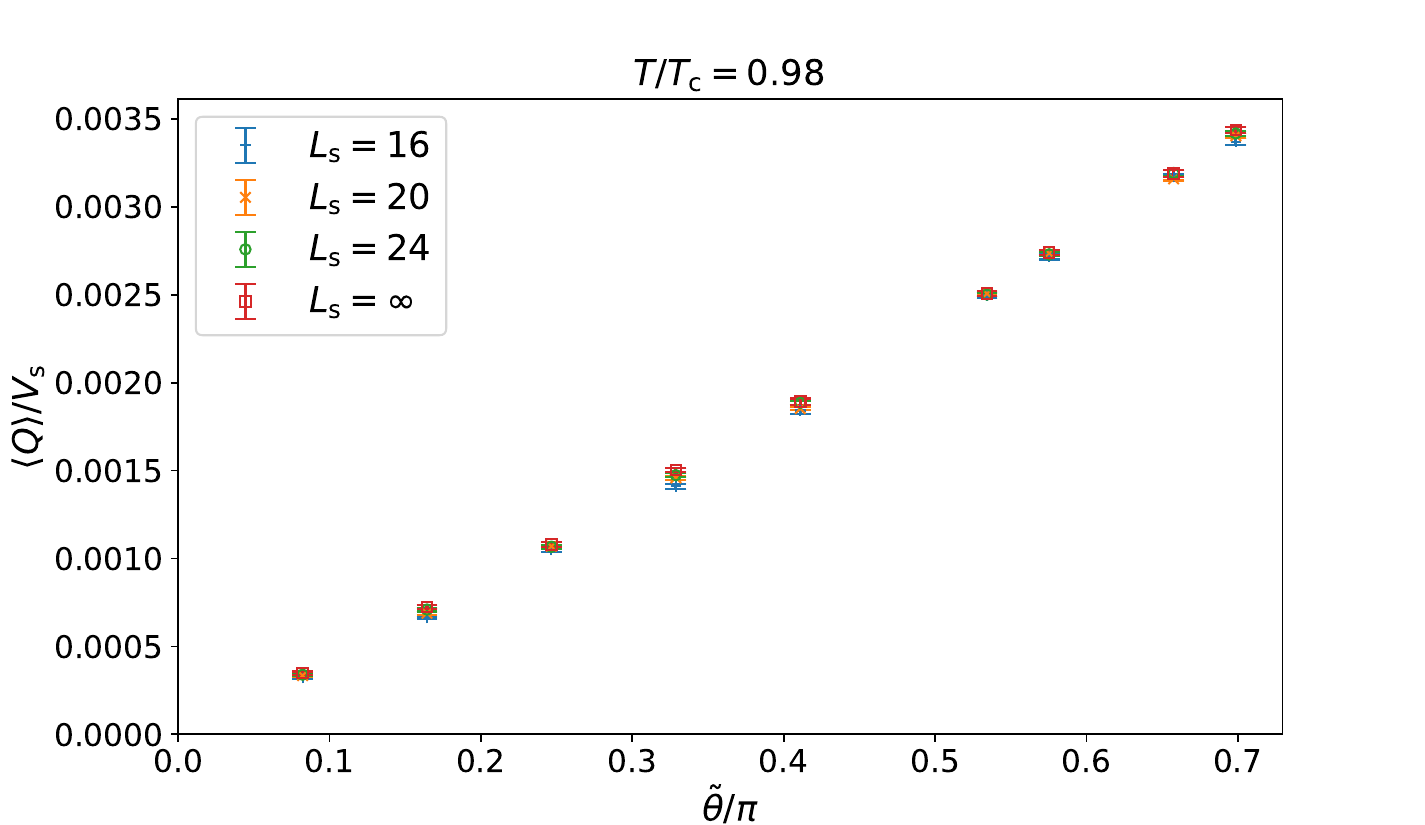}
    \includegraphics[width=0.475\hsize]{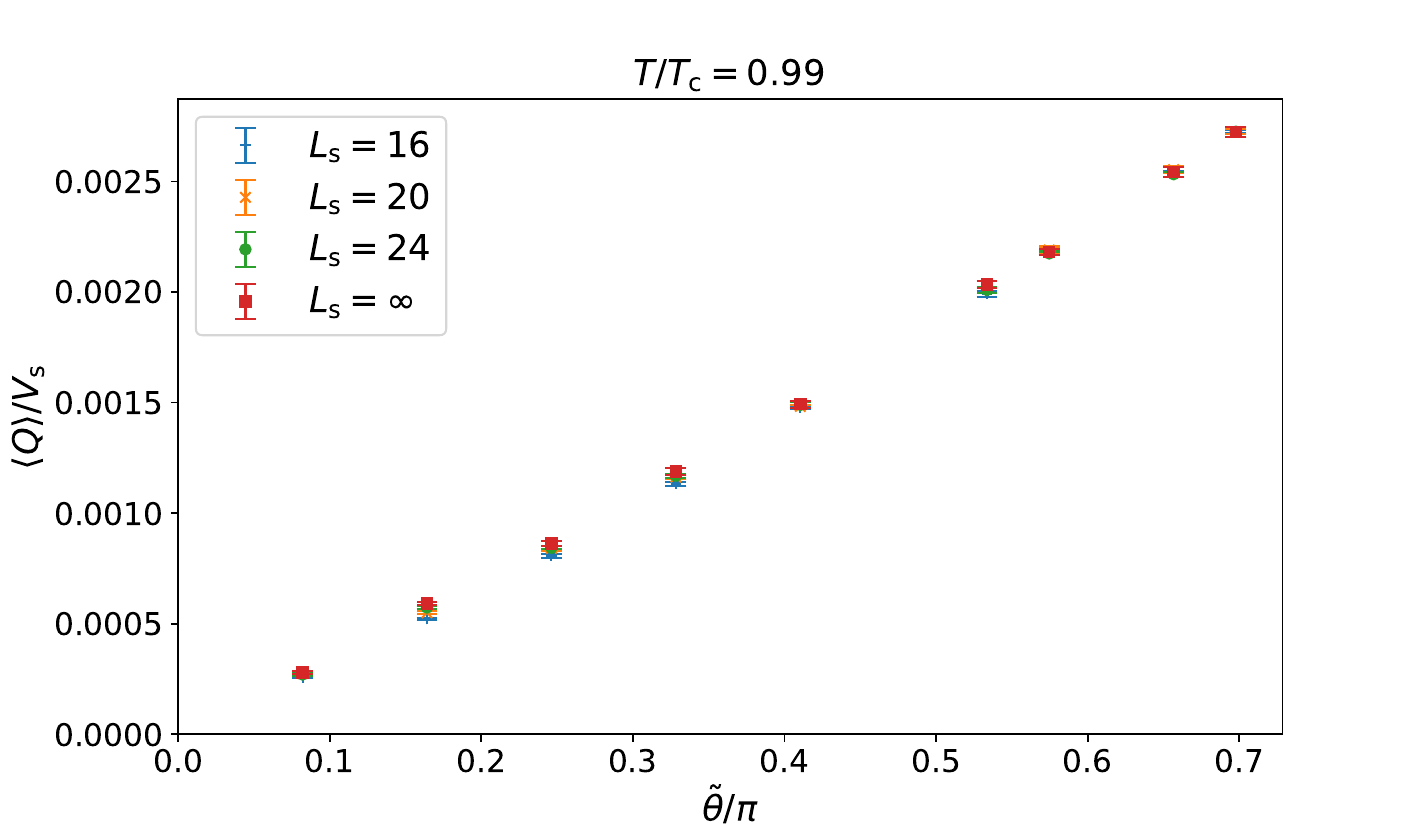}\\
    \includegraphics[width=0.475\hsize]{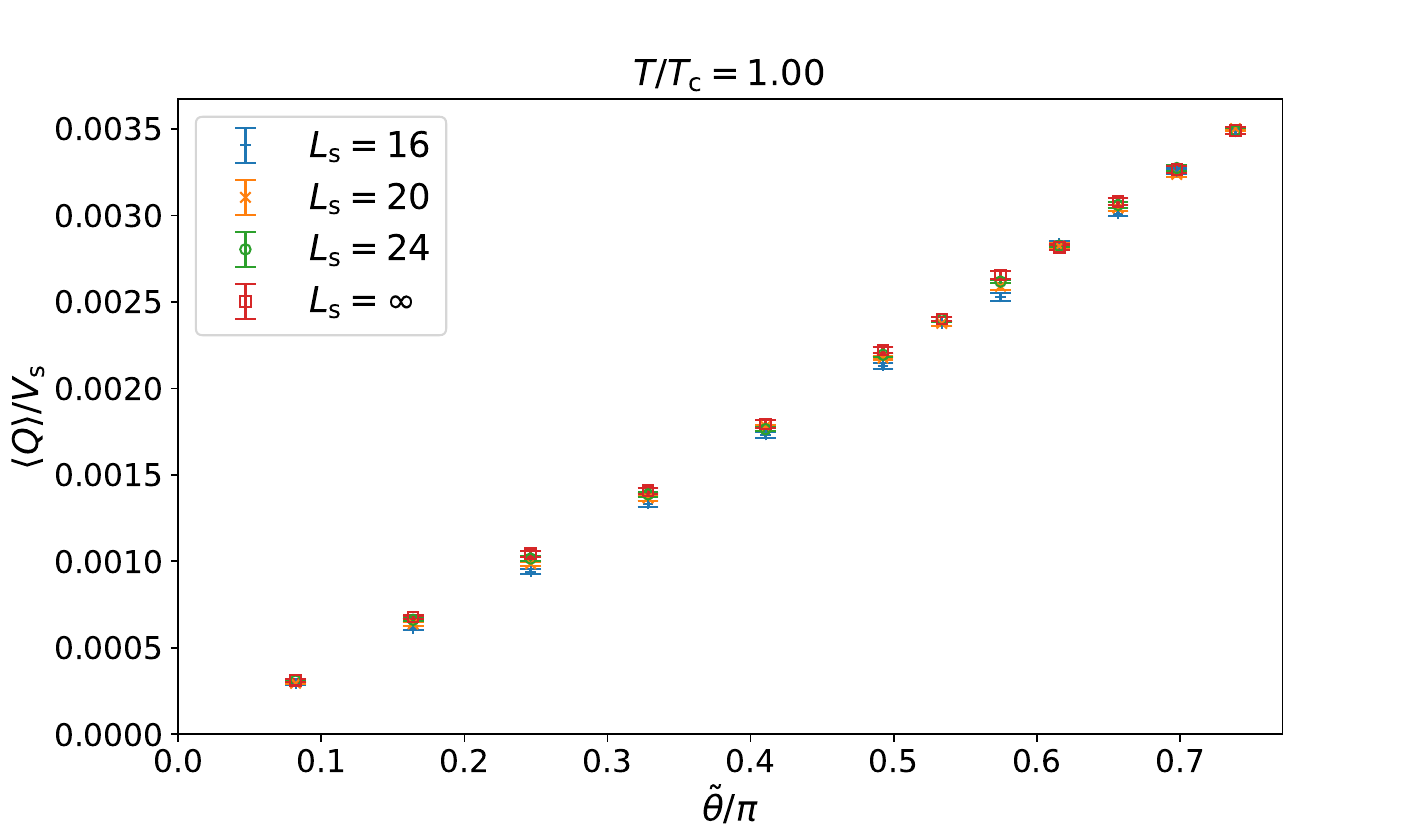}
    \includegraphics[width=0.475\hsize]{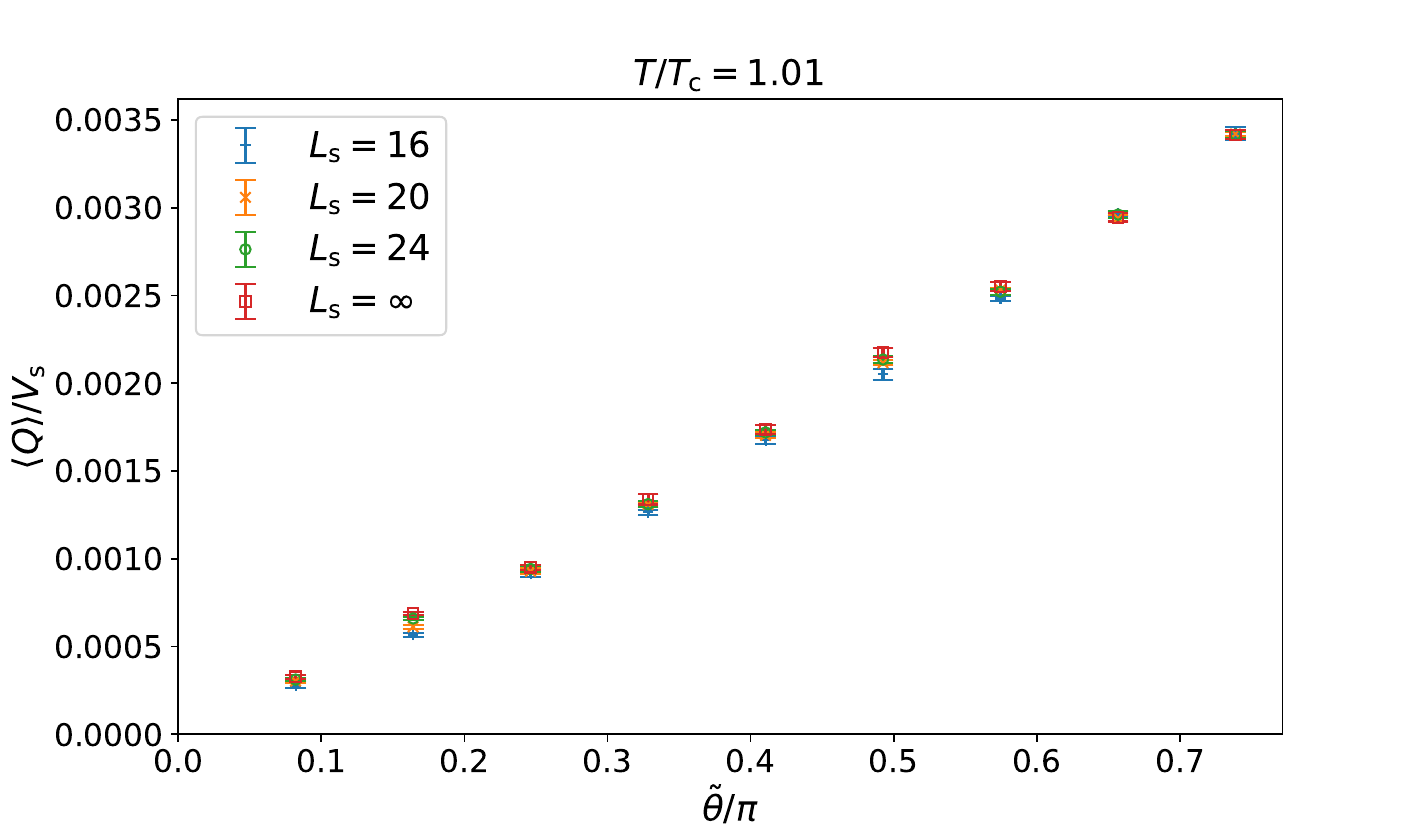}\\
    \includegraphics[width=0.475\hsize]{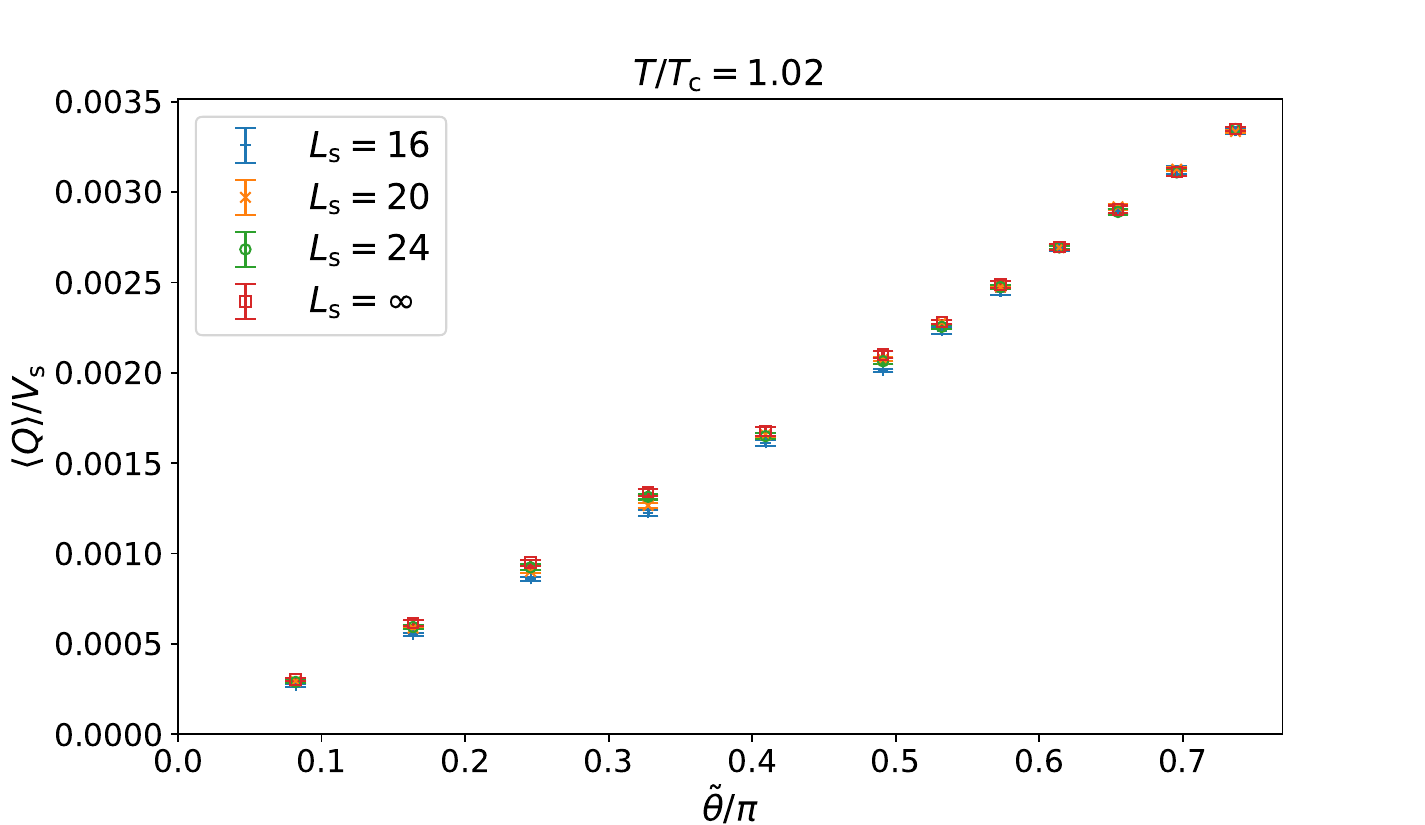}
    \includegraphics[width=0.475\hsize]{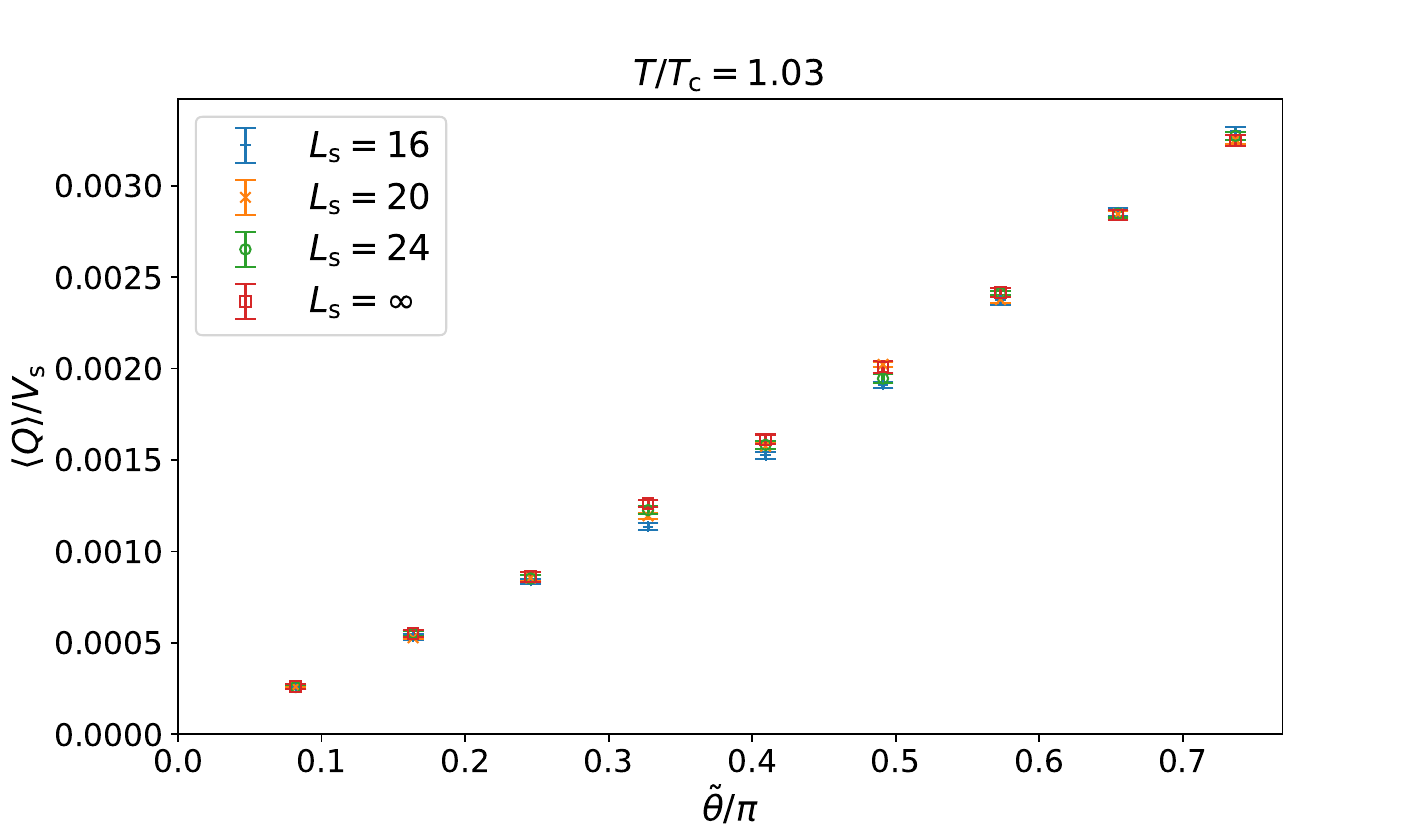}\\
    \includegraphics[width=0.475\hsize]{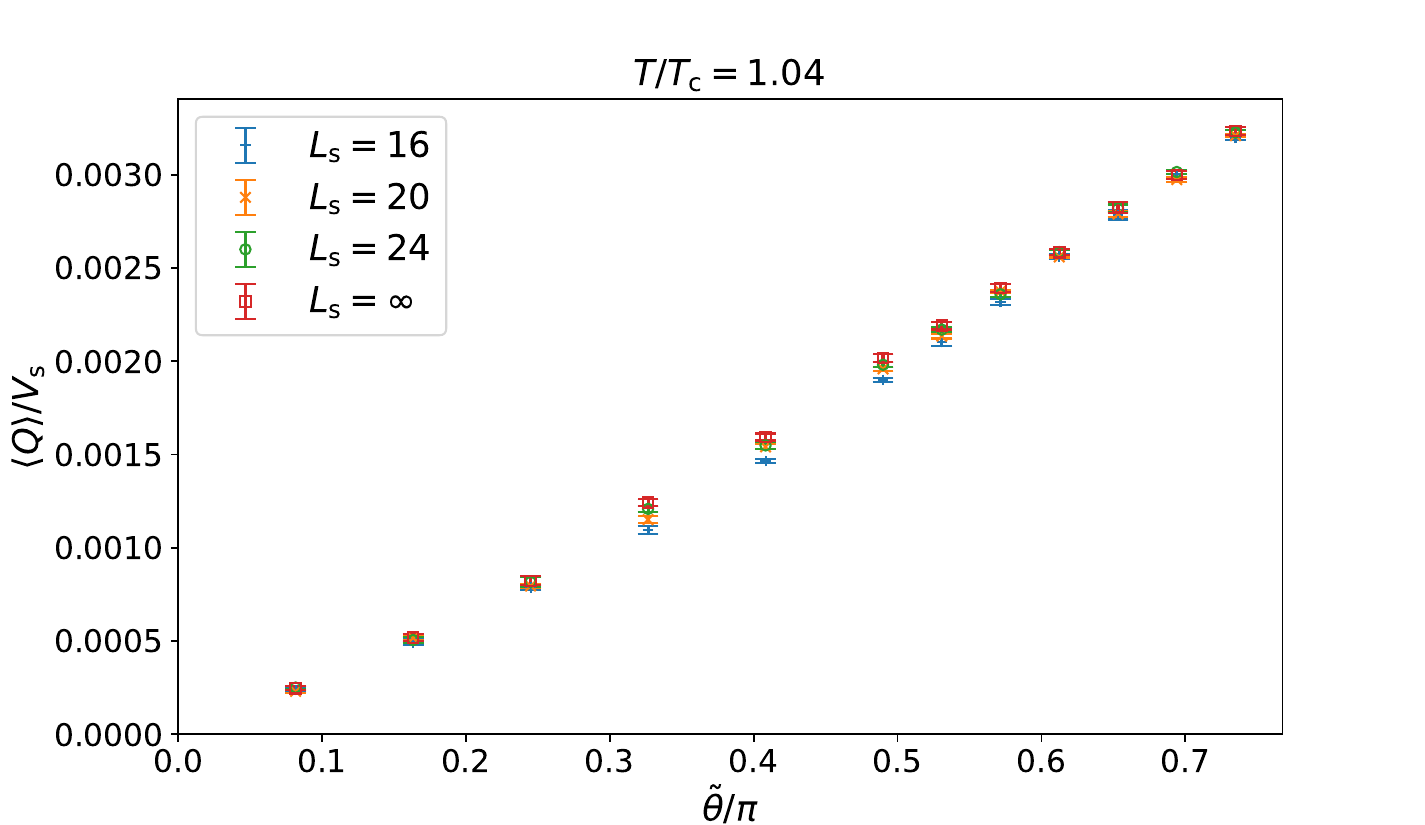}
    \includegraphics[width=0.475\hsize]{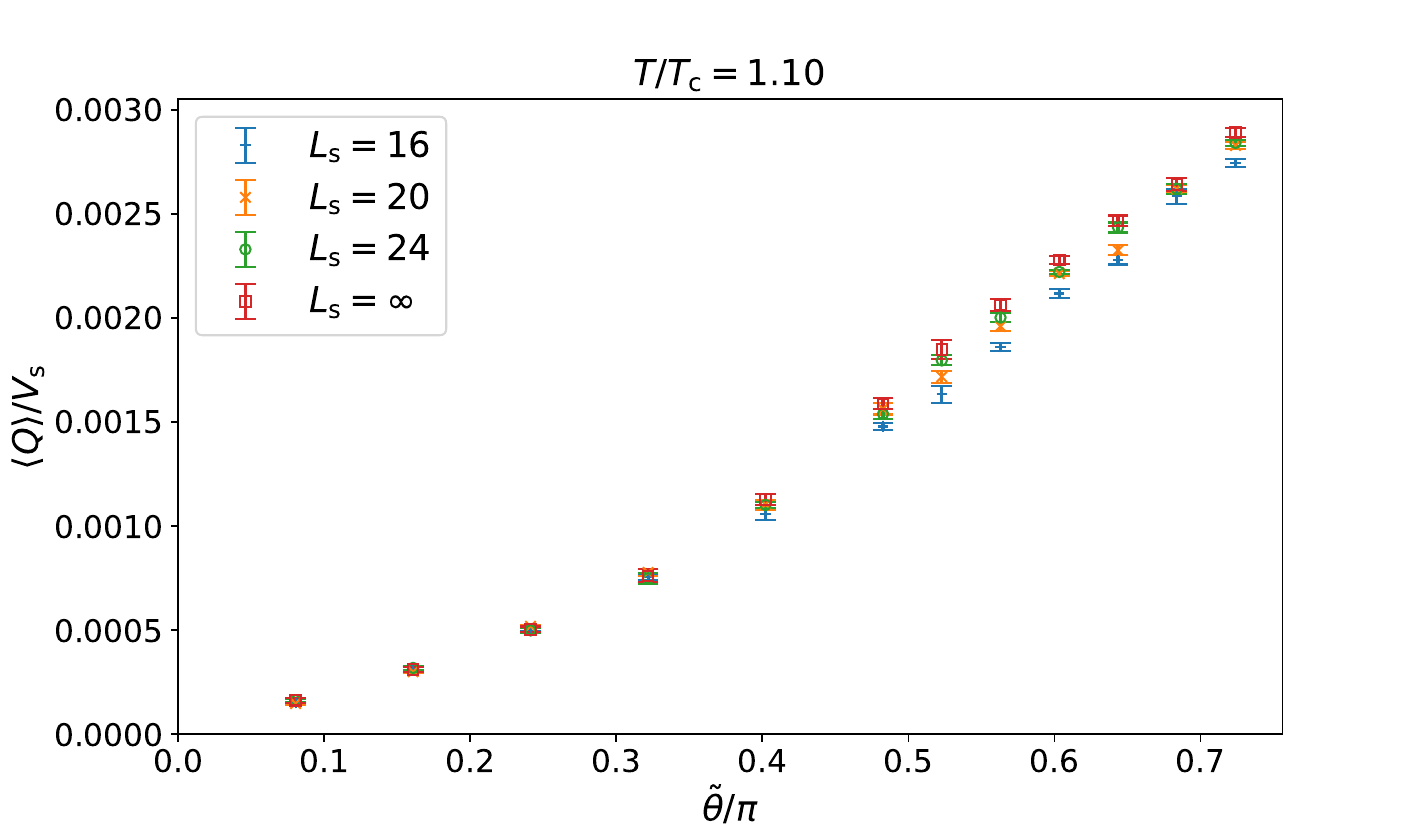}\\
    \caption{The topological charge density $\Braket{Q}_{i\tilde{\theta}}/V_{\rm s}$ is plotted
    against $\tilde{\theta}/\pi$ for $L_{\rm s}=16, 20, 24$
    at various temperature within $0.9 \le T/T_{\mathrm{c}} \le 1.1$.
    The results of the infinite volume extrapolation are plotted as well.} 
    \label{fig:infinite_V}
\end{figure}

In figure~\ref{fig:infinite_V}, we plot the topological charge density
$\Braket{Q}_{i\tilde{\theta}}/V_{\rm s}$
against $\tilde{\theta}$ for $L_{\rm s} = 16, 20, 24$
with fixed $L_{\rm t}=5$.
%%Let us recall that
%%the lattice size in the temporal direction is fixed to $L_{\rm t}=5$ in this work.
%% We do not observe a significant volume dependence.
For each $\tilde{\theta}$, we fit the data points to
the linear function $f(V_{\rm s}) = c + b/V_{\rm s}$, 
where $c$ and $b$ are the fitting parameters
and $V_{\rm s} = L_{\rm s}^3$ is the spatial lattice volume.
The results obtained by the $L_{\rm s}\rightarrow \infty$ extrapolation
are shown in the same figure.

%%\subsection{Infinite volume extrapolation for \texorpdfstring{$\chi_0$}{chi0}}

\begin{figure}[H]
    \centering
    \includegraphics[width=0.475\hsize]{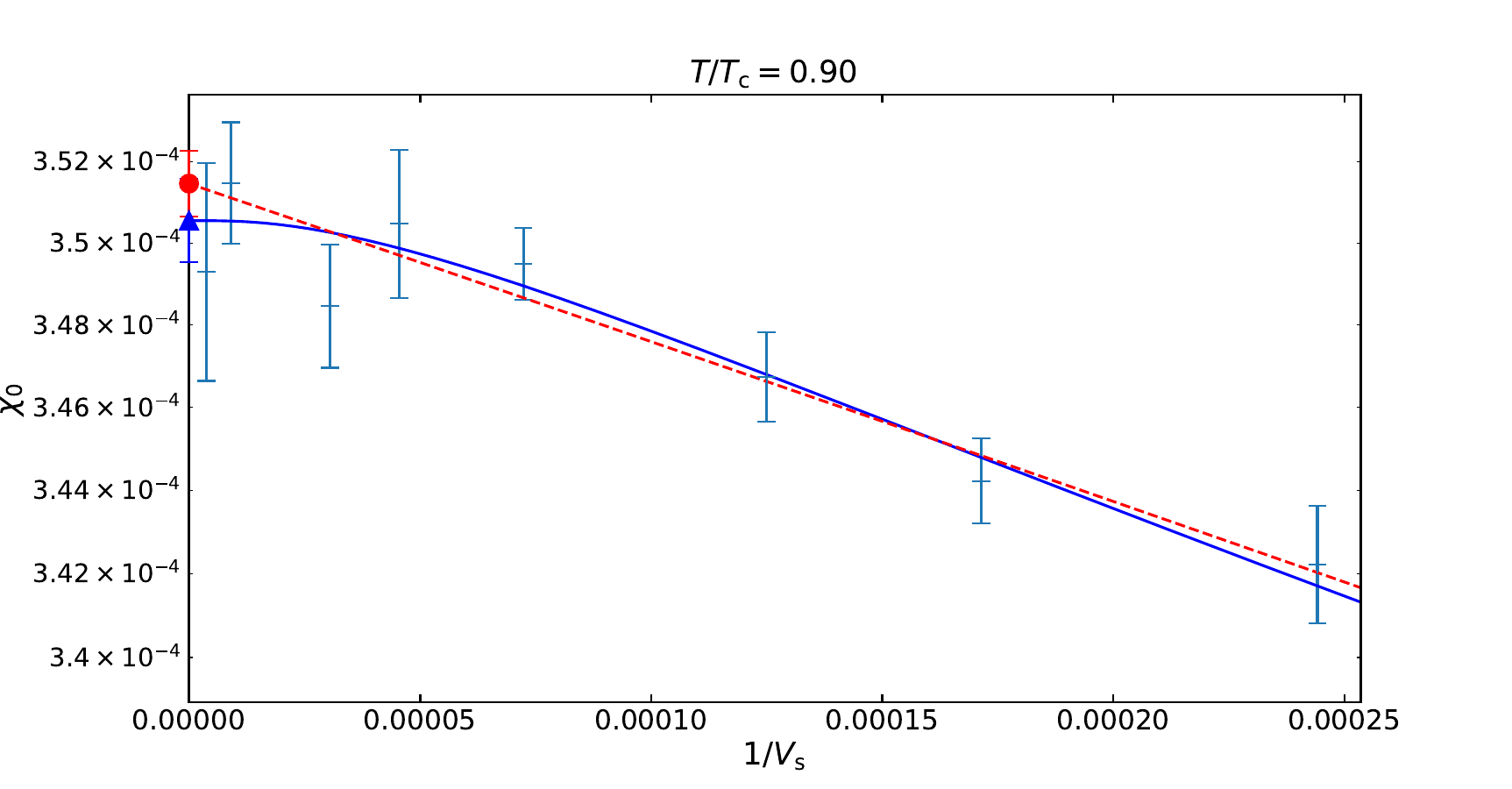}
    \includegraphics[width=0.475\hsize]{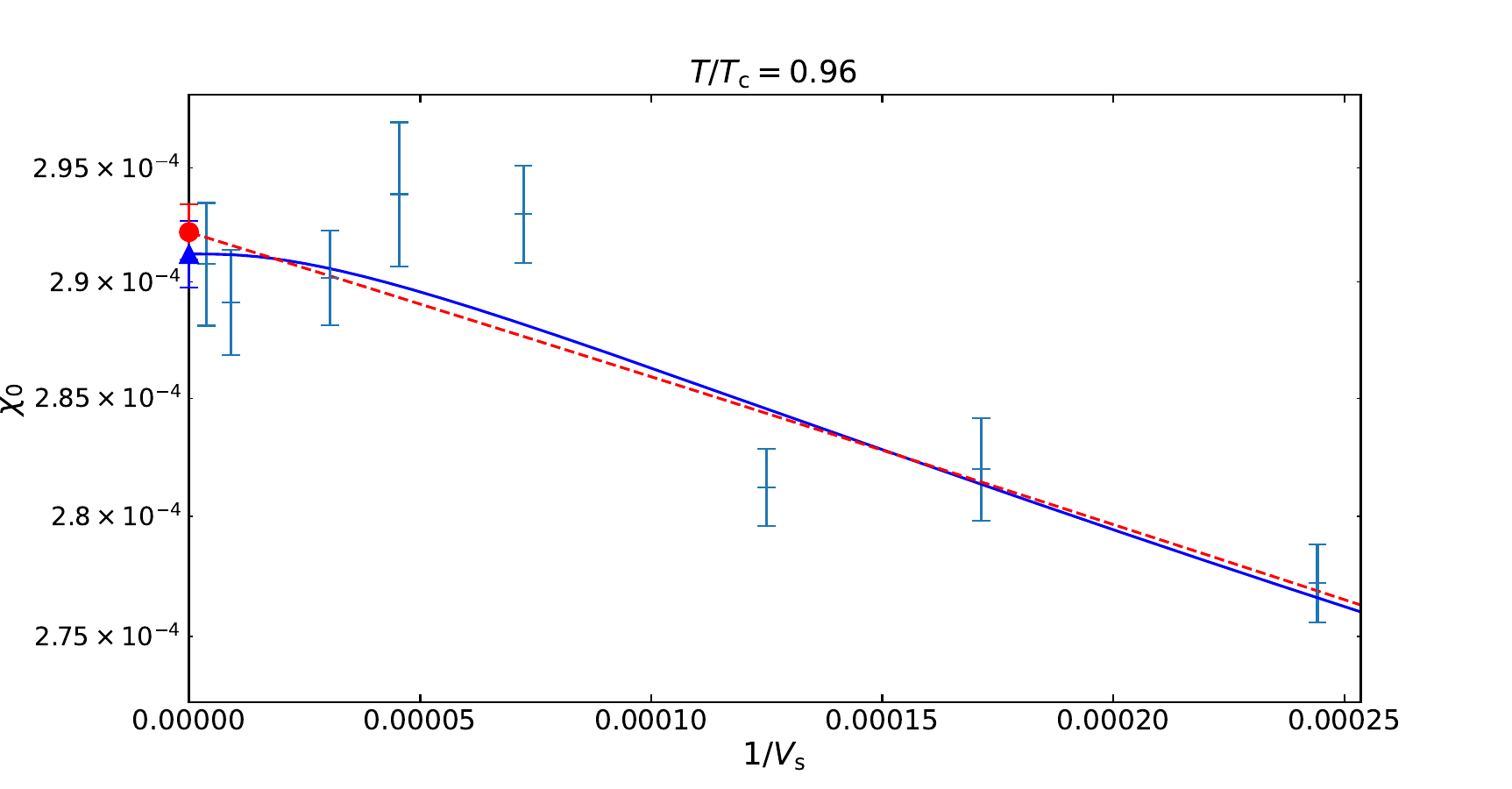}\\
    \includegraphics[width=0.475\hsize]{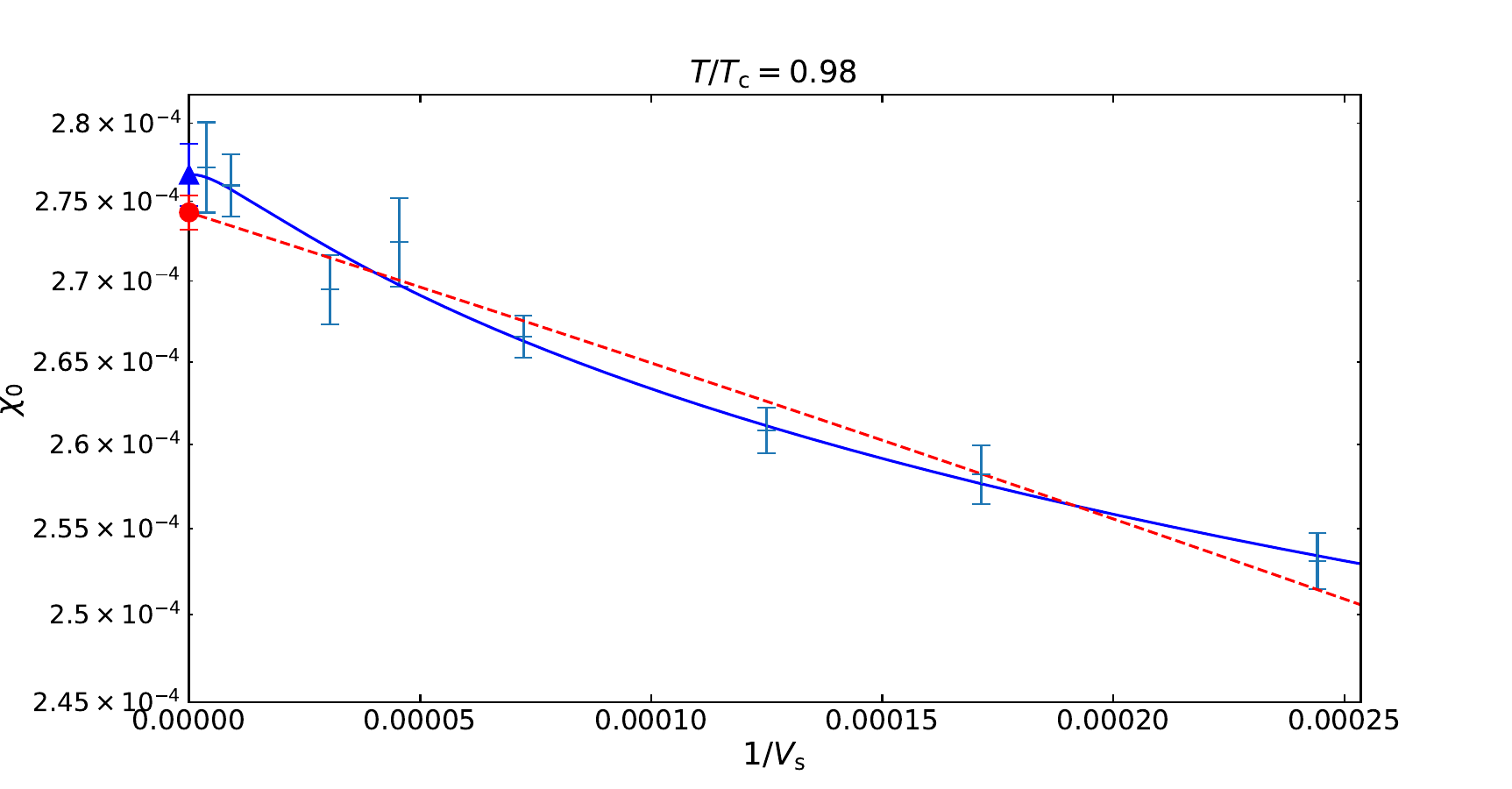}
    \includegraphics[width=0.475\hsize]{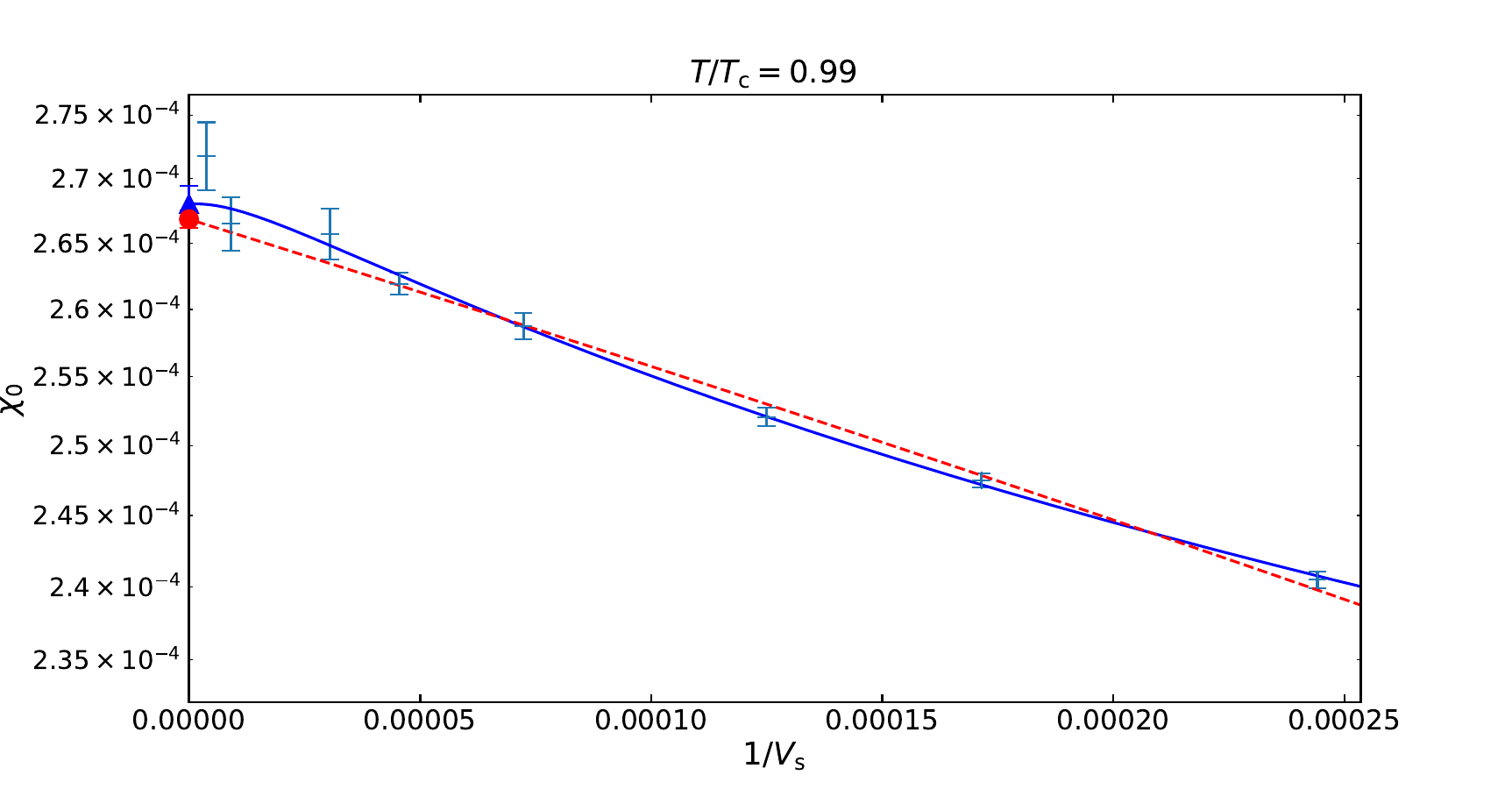}\\
    \includegraphics[width=0.475\hsize]{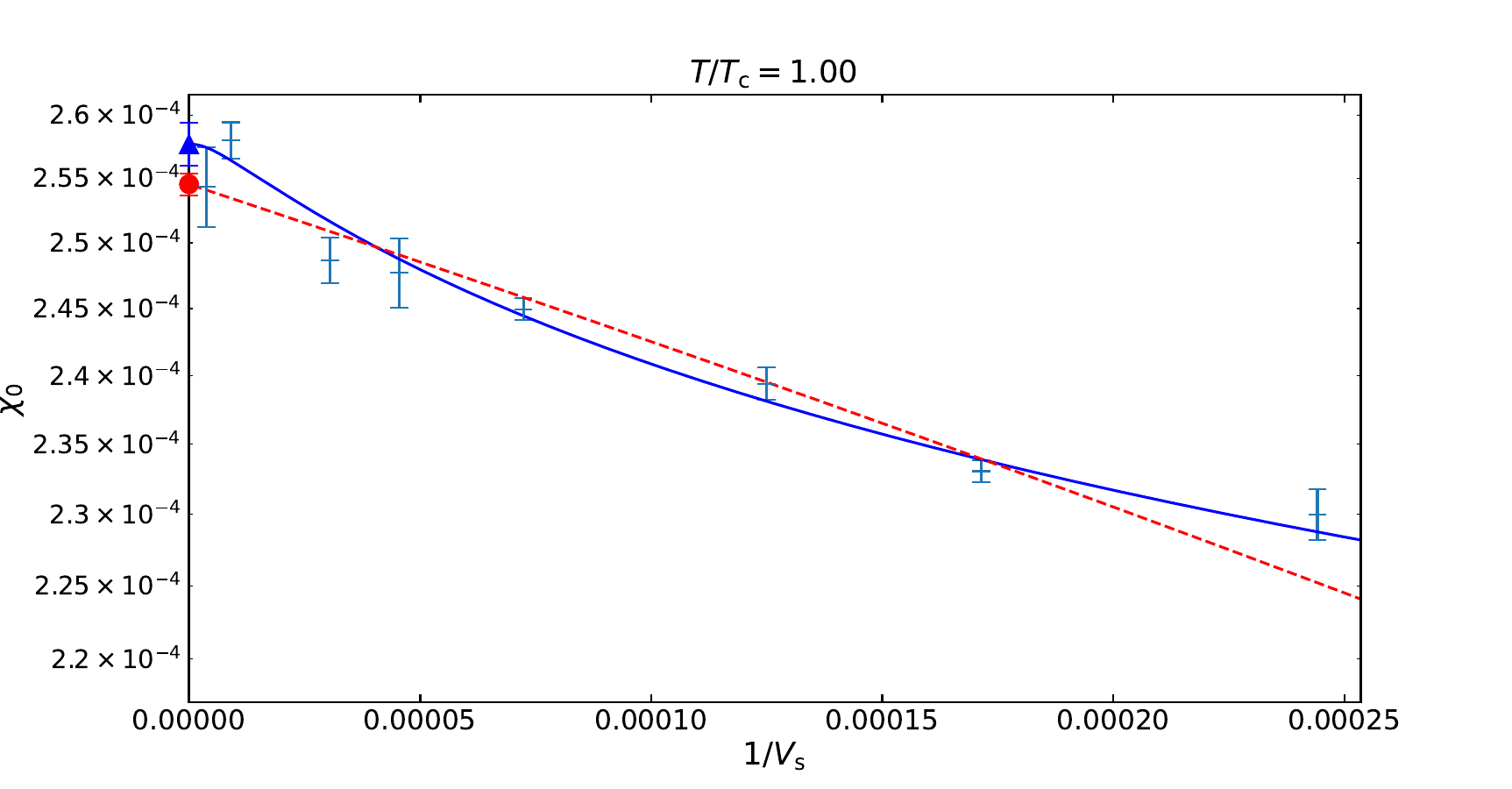}
    \includegraphics[width=0.475\hsize]{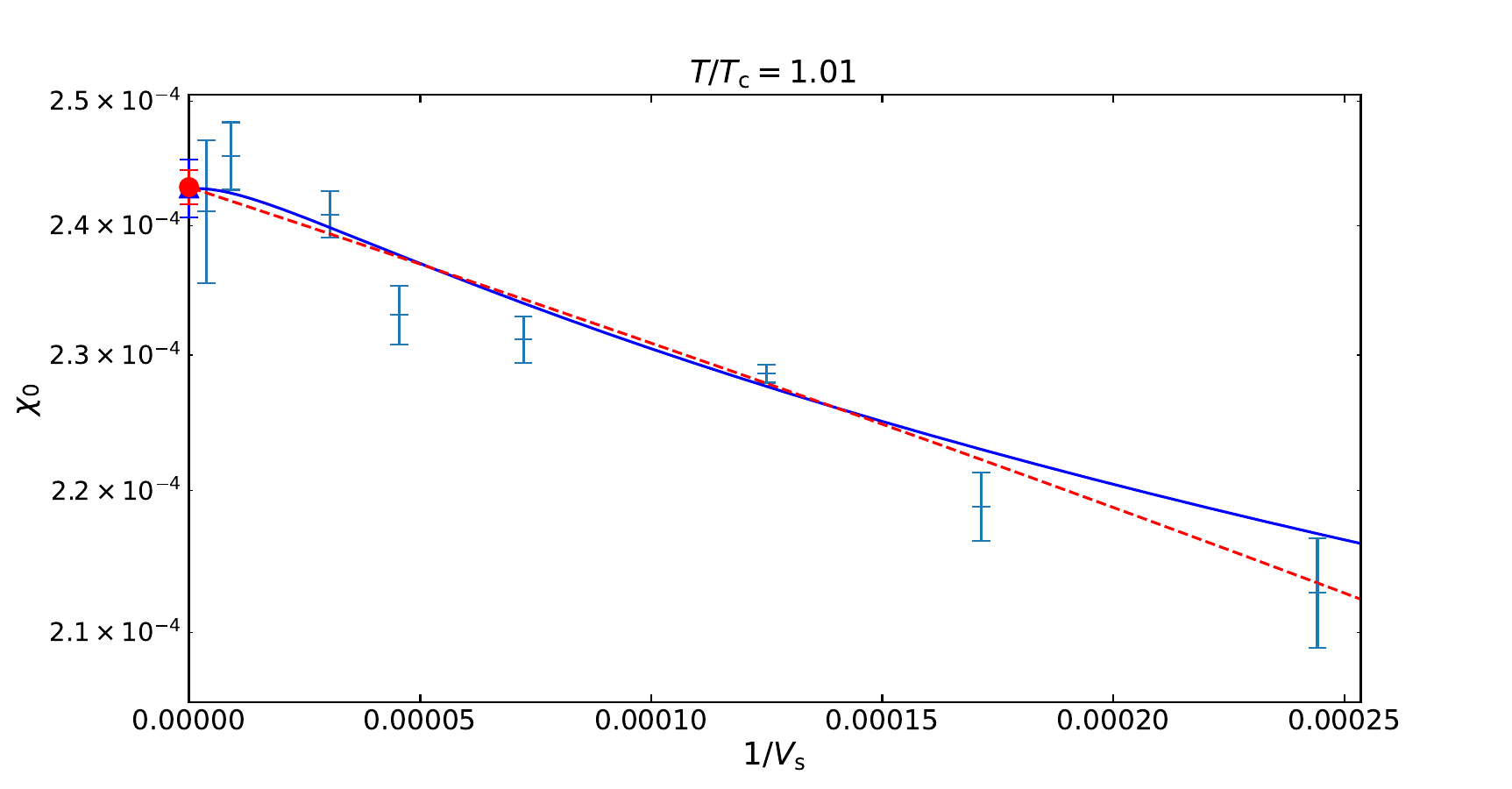}\\
    \includegraphics[width=0.475\hsize]{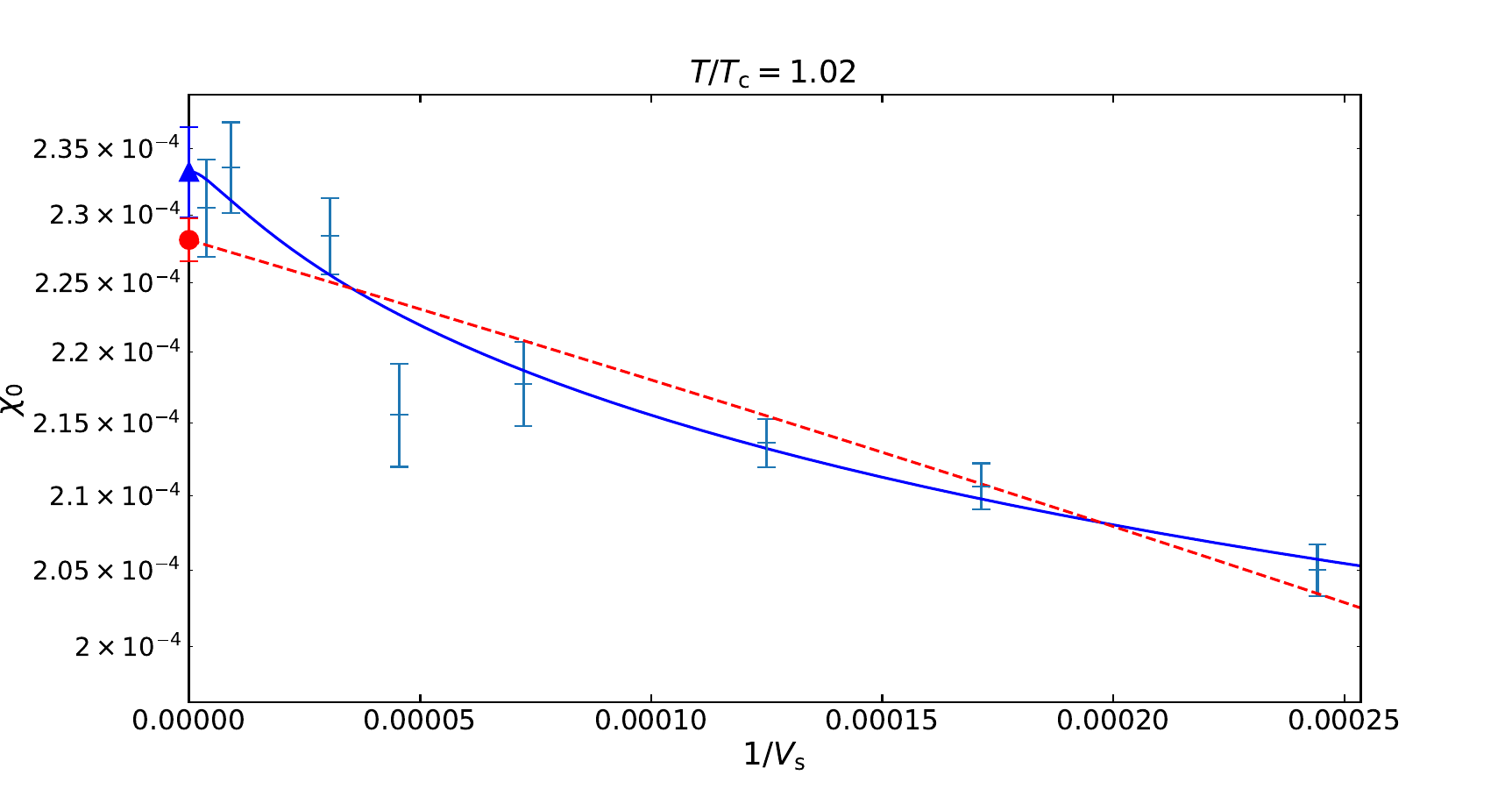}
    \includegraphics[width=0.475\hsize]{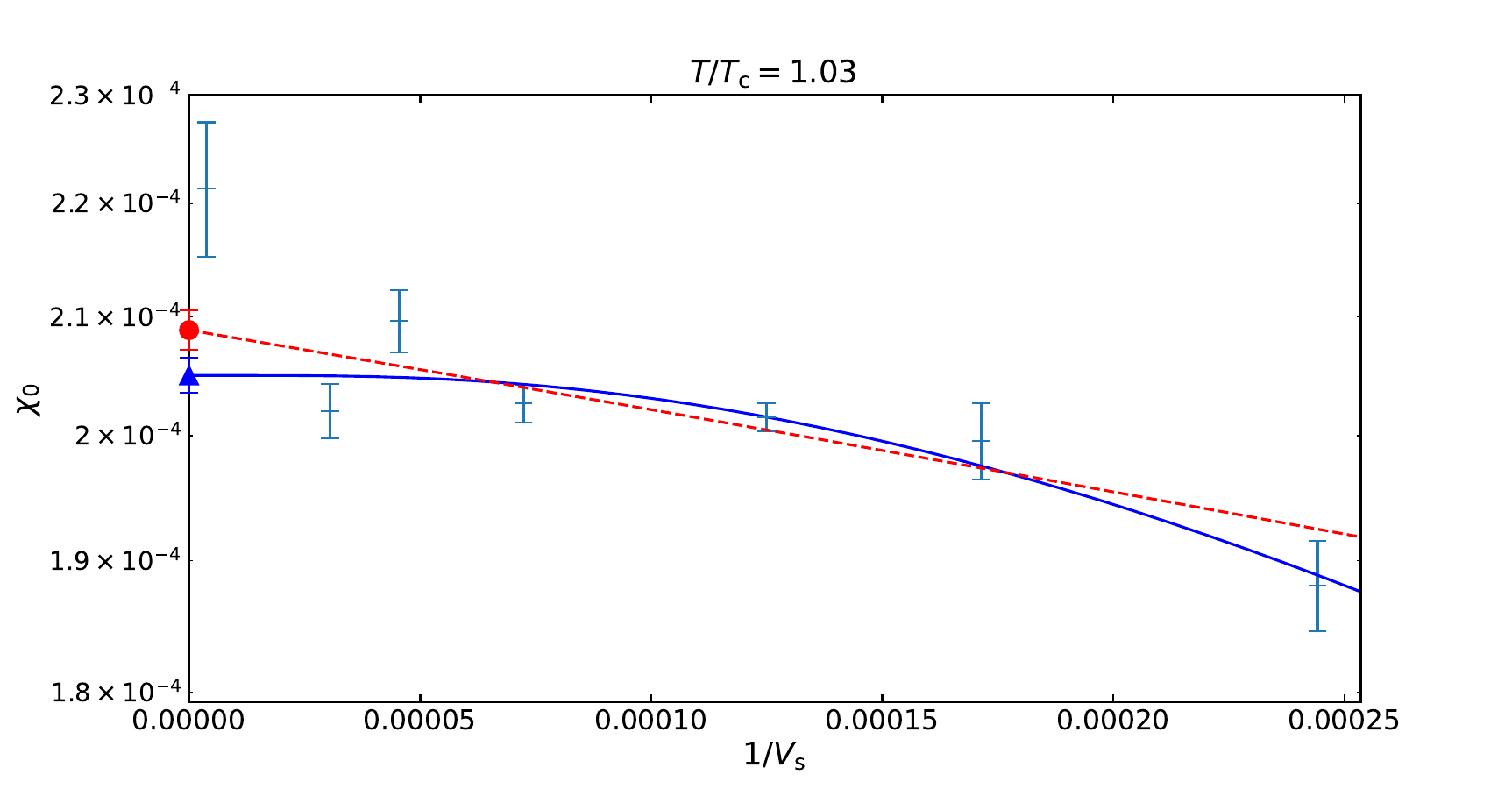}\\
    \includegraphics[width=0.475\hsize]{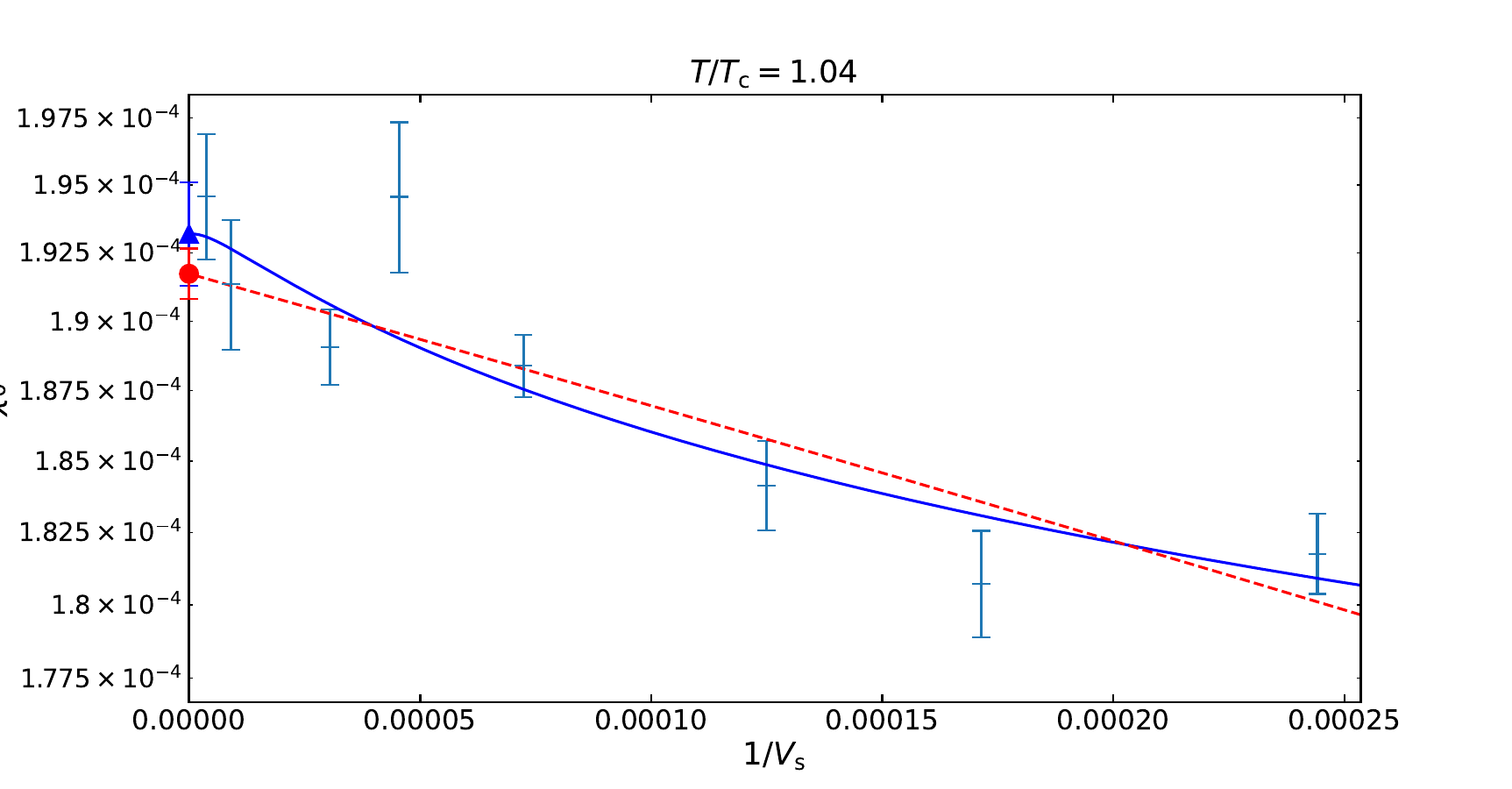}
    \includegraphics[width=0.475\hsize]{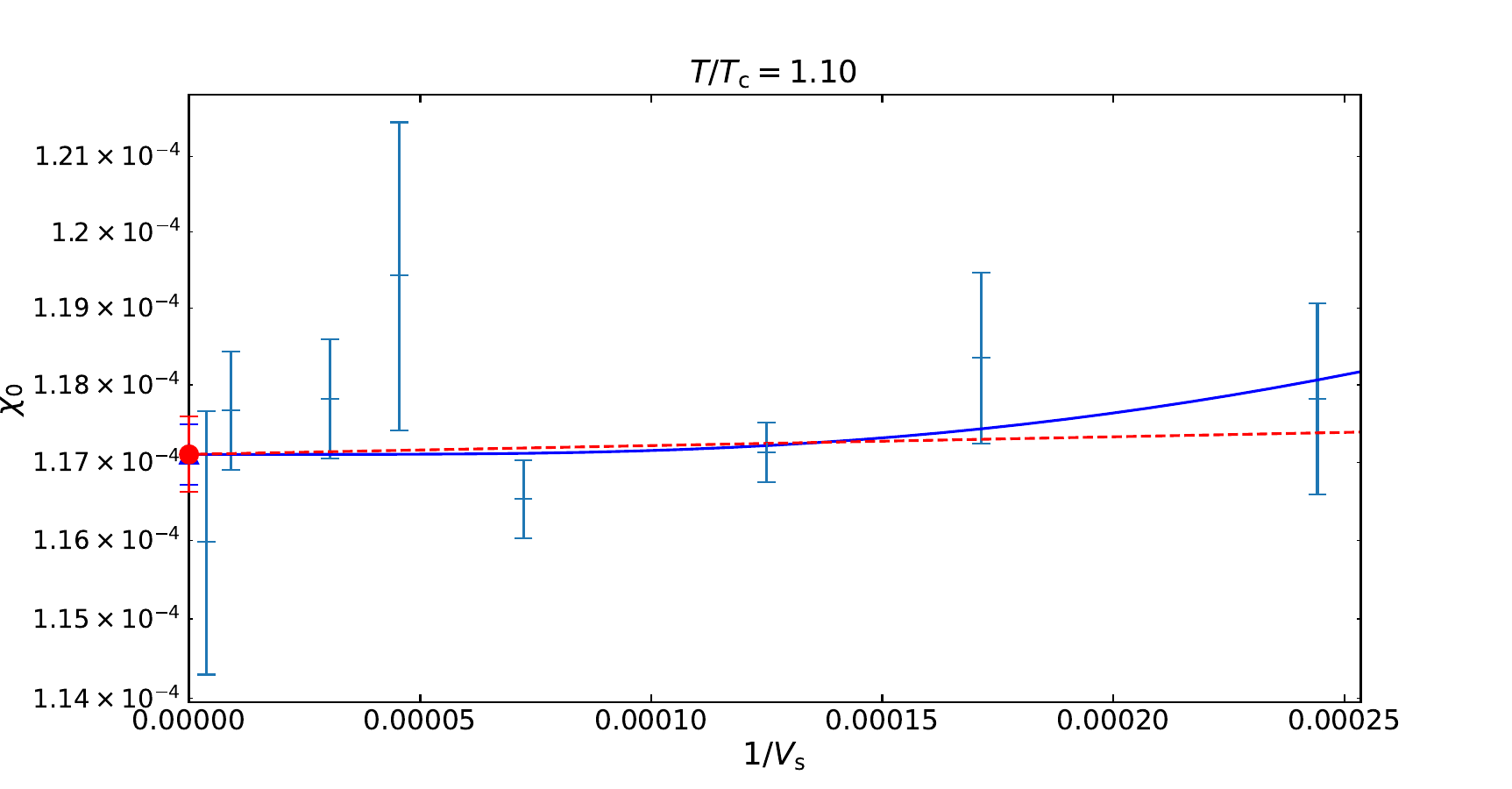}\\
    \caption{The topological susceptibility $\chi_0$ in the lattice unit
      is plotted against $1/V_{\rm s}$.
      The blue solid lines represent fits to the function \eqref{fitting-chi0-low},
      while the red dashed lines represent fits to the function \eqref{fitting-chi0-high}.
      The blue triangles and the red circles represent 
      the corresponding extrapolated values of
      $\chi_0$ at each temperature.}
    %%, which are given in table~\ref{tab:chi0}.}
    \label{fig:chi0_con}
\end{figure}

%% is defined as follows depending on the temperature: 
Next, we show our results\footnote{
Since this only requires simulations at $\tilde{\theta}=0$,
which is much cheaper than simulations at $\tilde{\theta}\neq 0$,
we were able to obtain many data points within $16 \le L_{\rm s} \le 64$.}
for the
%% discuss the infinite volume extrapolation of the
topological susceptibility $\chi_0$ at $\theta=0$, which is defined by
\eqref{top-susceptibility}.
From now on, we assume that $\chi_0$ is made dimensionless by using the lattice unit.
%%to extrapolate $\chi_0$.
In figure~\ref{fig:chi0_con}, 
the topological susceptibility
$\chi_0$ is plotted
%% against $L_{\rm s}^{-1}$ for $T<T_{\rm c}$, and
against $1/V_{\rm s}$ for $16 \le L_{\rm s} \le 64$.
%%for $T \ge T_{\rm c}$, respectively.
The natural fitting functions are\footnote{The reason for this choice
is that the nonzero glueball mass leads to an exponentially fast approach
to the infinite volume limit in the confined phase \cite{Durr:2006ky}.
%%(QUOTE PAPERS!!!)
On the other hand, the glueball contribution is expected to be suppressed
in the deconfined phase, and therefore the leading
%% order of the
finite volume effect is expected to be proportional to $1/V_{\rm s}$ there.
In the temperature region investigated in this work, however,
the extrapolation does not depend much on this choice.} 
\begin{align}
%%  \begin{array}{rll}
    \label{fitting-chi0-low}
    f_{\rm low}(L_{\rm s}) &= c_0 + c_1 e^{-c_2 L_{\rm s}}  &  \mbox{for $T\le T_{\rm c}$}  \ ,\\
    f_{\rm high}(V_{\rm s}) &= d_0 + d_1/V_{\rm s}   &  \mbox{for $T> T_{\rm c}$} \ ,
    %% f_{T<T_{\rm c}}(L_{\rm s}) &= c_0 + c_1 e^{-c_2 L_{\rm s}} \ , \\
    %% f_{T\ge T_{\rm c}}(V_{\rm s}) &= d_0 + d_1/V_{\rm s} \ ,
%%\end{array}
\label{fitting-chi0-high}
\end{align}
where $c_0,\ c_1,\ c_2,\ d_0$ and $d_1$ are the fitting parameters.
The values of $\chi_0$ after the infinite volume
%%$L_{\rm s}\rightarrow \infty$
extrapolation are shown in table~\ref{tab:chi0}.
In figure~\ref{fig:chi0_con}, we show the fitting curves for both functions
\eqref{fitting-chi0-low} and \eqref{fitting-chi0-high} at each temperature for comparison.
We find that the choice of the fitting function
does not affect the extrapolated values beyond the fitting errors.
%% the dependence of the extrapolated values on the fitting functions
%% is within the fitting errors.

\begin{table}[t]
  \centering
 \begin{minipage}[b]{0.48\columnwidth} 
    \centering
    \begin{tabular}{|c||c|c|}\hline
        $T/T_{\mathrm{c}}$ & $\chi_0$ & $\chi^2/ N_{\rm DF}$  \tabularnewline\hline 
        \hline 
        0.90 & 0.000351(1) & 0.60 \tabularnewline\hline 
        0.96 & 0.000291(1) & 2.37 \tabularnewline\hline 
        0.98 & 0.000277(2) & 0.53 \tabularnewline\hline 
        0.99 & 0.000268(1) & 0.75 \tabularnewline\hline 
        1.00 & 0.000258(2) & 1.67 \tabularnewline\hline 
    \end{tabular}
 \end{minipage}
 \hspace{-0.1\columnwidth}
%%   \hspace
 \begin{minipage}[b]{0.48\columnwidth}
  \centering
    \begin{tabular}{|c||c|c|}\hline
        $T/T_{\mathrm{c}}$ & $\chi_0$ & $\chi^2/ N_{\rm DF}$  \tabularnewline\hline 
        \hline 
        1.01 & 0.000243(1) & 2.12 \tabularnewline\hline 
        1.02 & 0.000228(2) & 2.25 \tabularnewline\hline 
        1.03 & 0.000209(2) & 2.91 \tabularnewline\hline 
        1.04 & 0.0001917(9) & 1.78 \tabularnewline\hline 
        1.10 & 0.0001171(5) & 0.98 \tabularnewline\hline 
    \end{tabular}
\end{minipage}
    \caption{The topological susceptibility $\chi_0$ in the lattice unit
      after the infinite volume extrapolation for various
      temperature within $0.9 \le T/T_{\mathrm{c}} \le 1.1$.}
    \label{tab:chi0}
\end{table}

\begin{figure}[H]
    \centering
    \includegraphics[width=0.475\hsize]{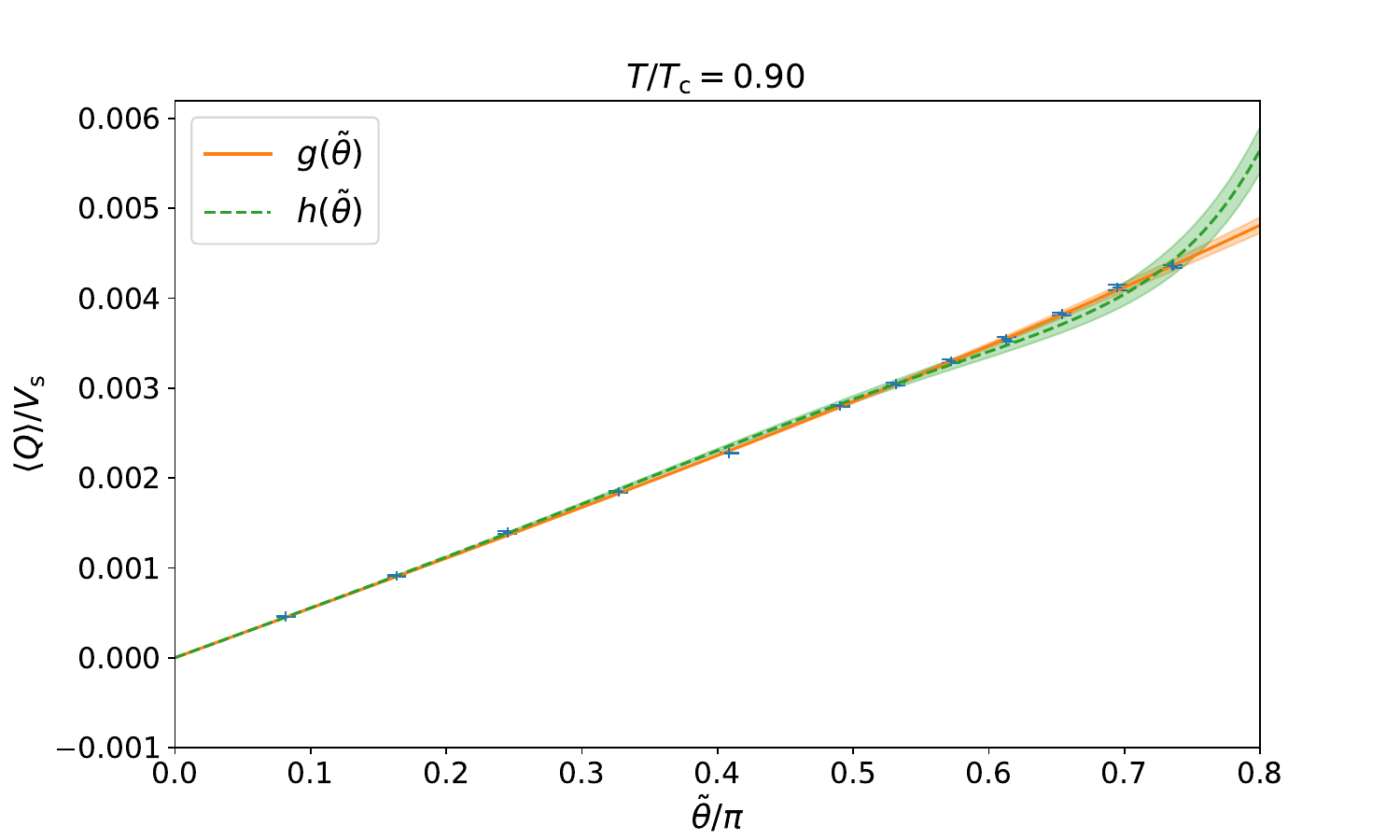}
    \includegraphics[width=0.475\hsize]{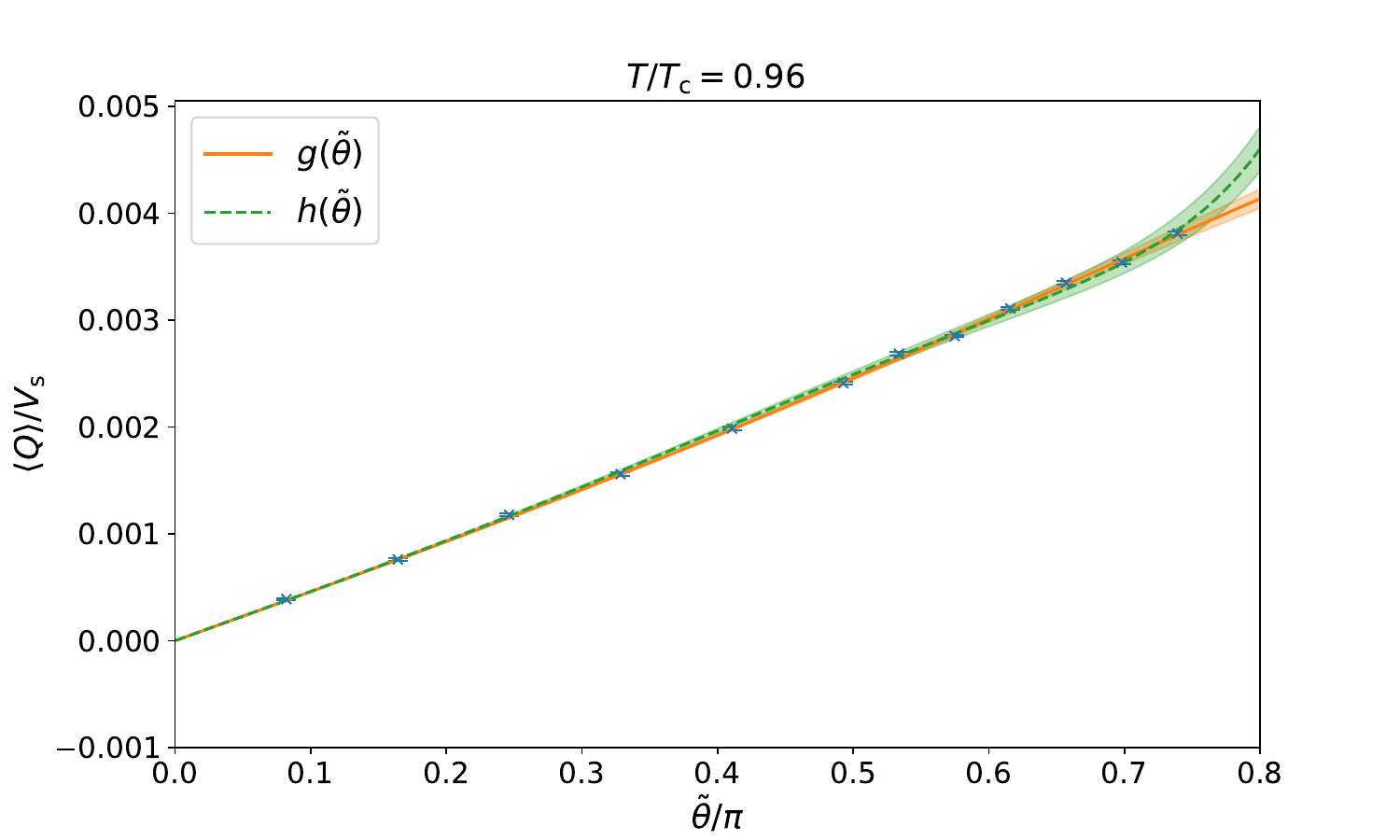}\\
    \includegraphics[width=0.475\hsize]{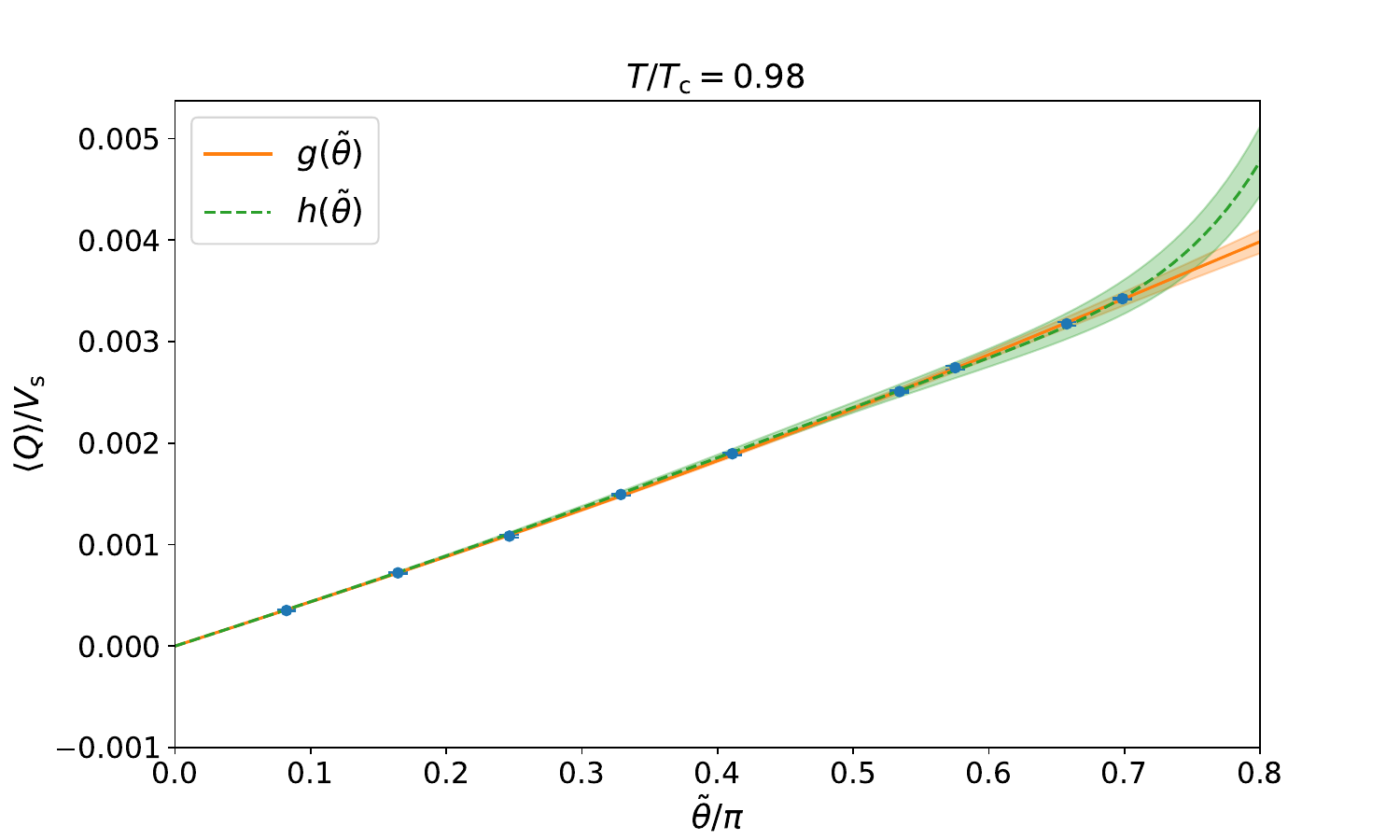}
    \includegraphics[width=0.475\hsize]{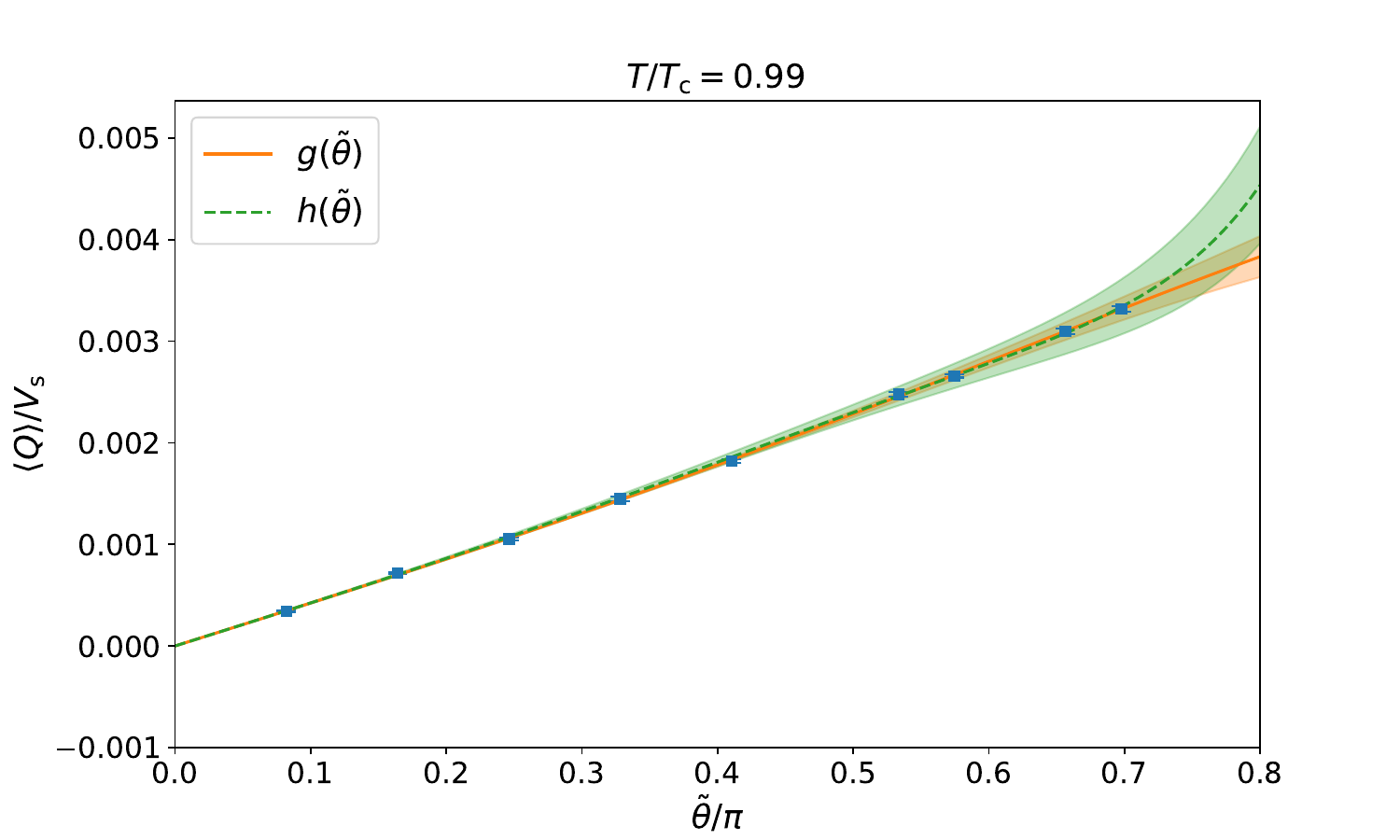}\\
    \includegraphics[width=0.475\hsize]{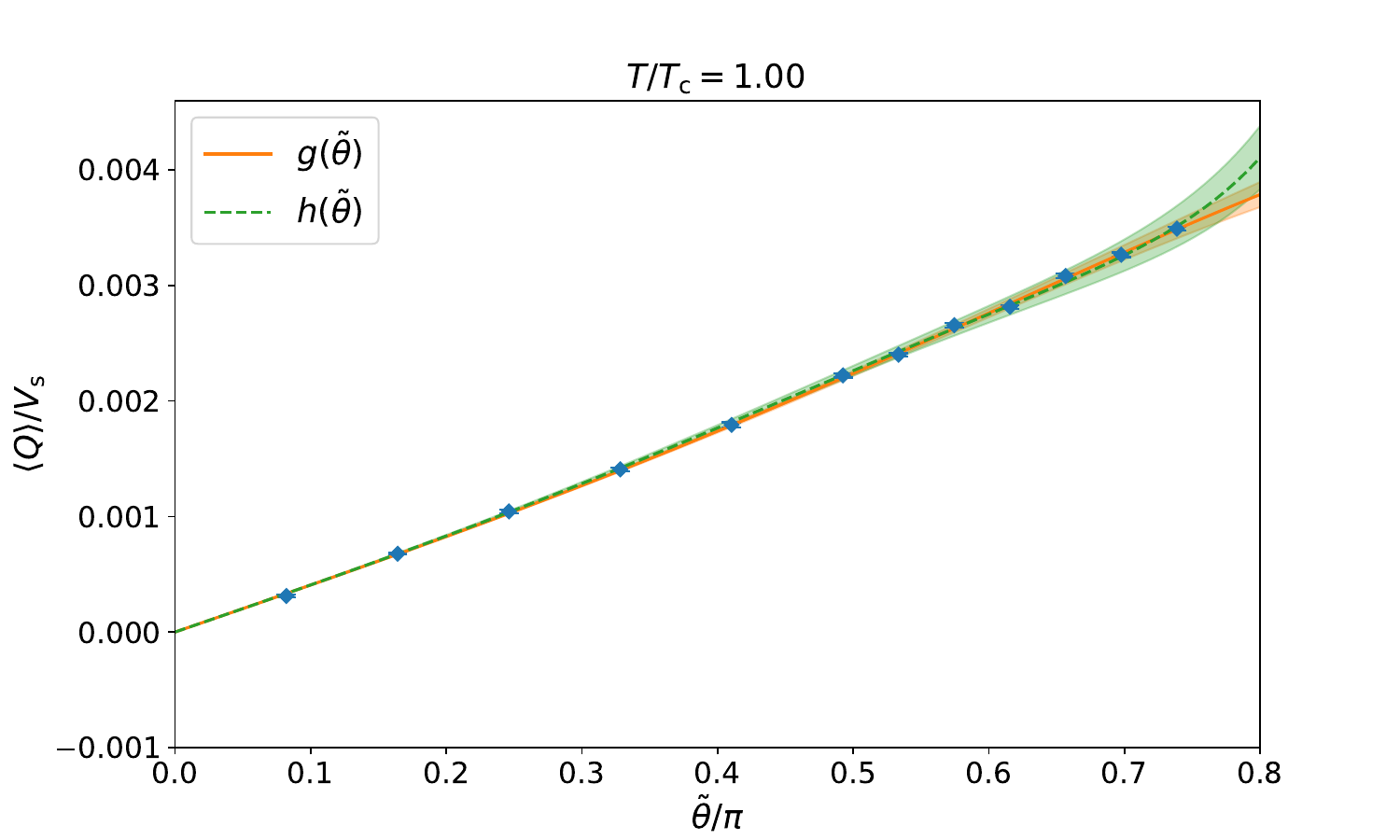}
    \includegraphics[width=0.475\hsize]{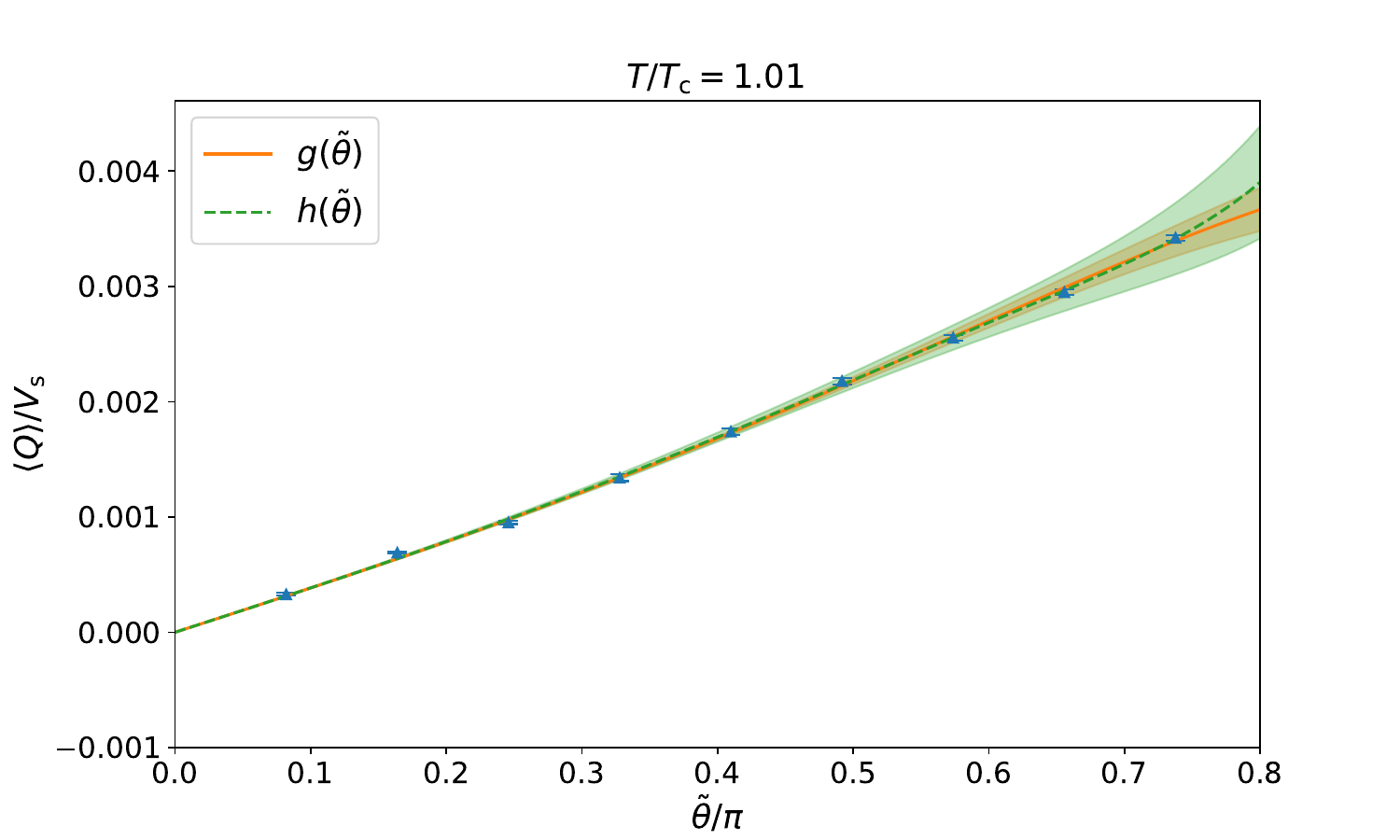}\\
    \includegraphics[width=0.475\hsize]{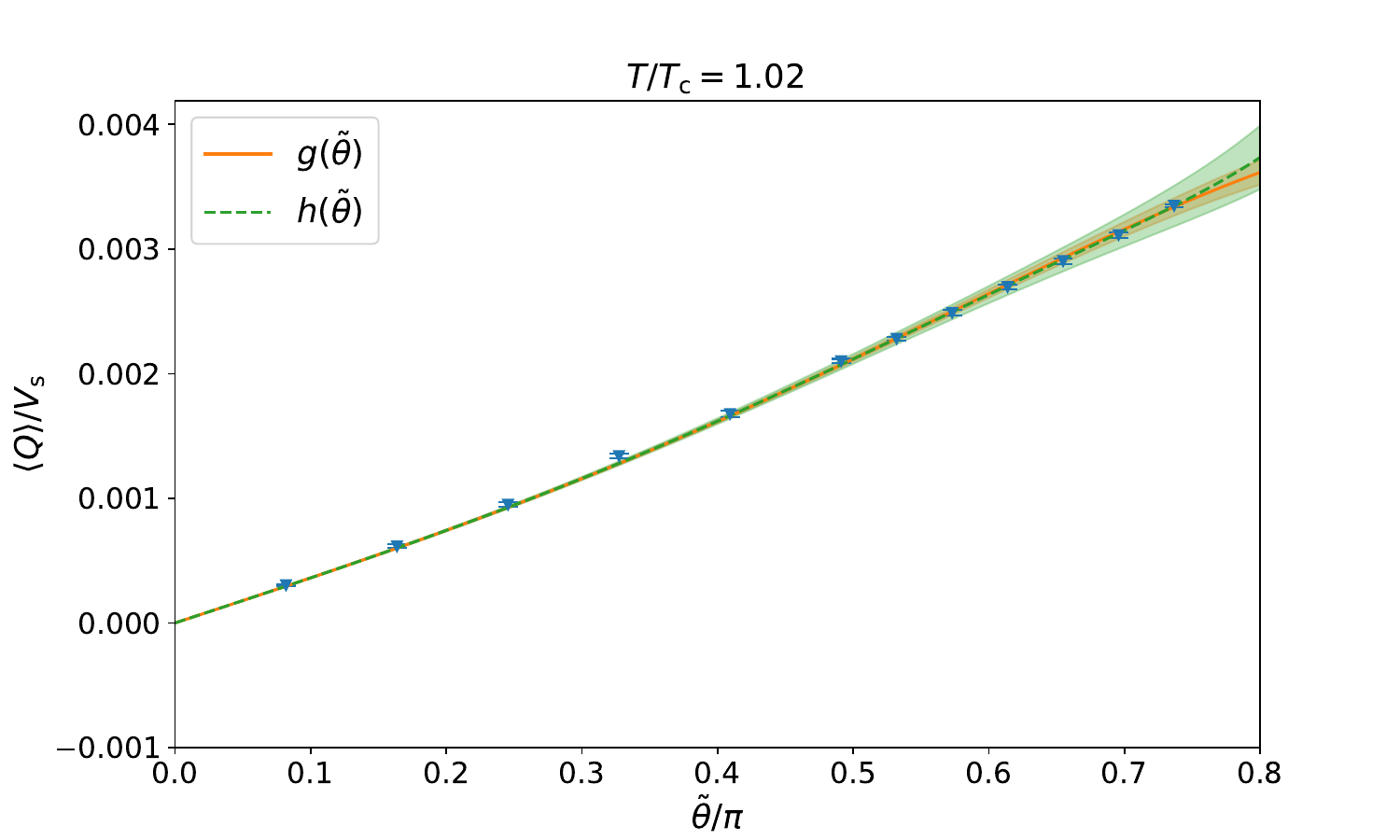}
    \includegraphics[width=0.475\hsize]{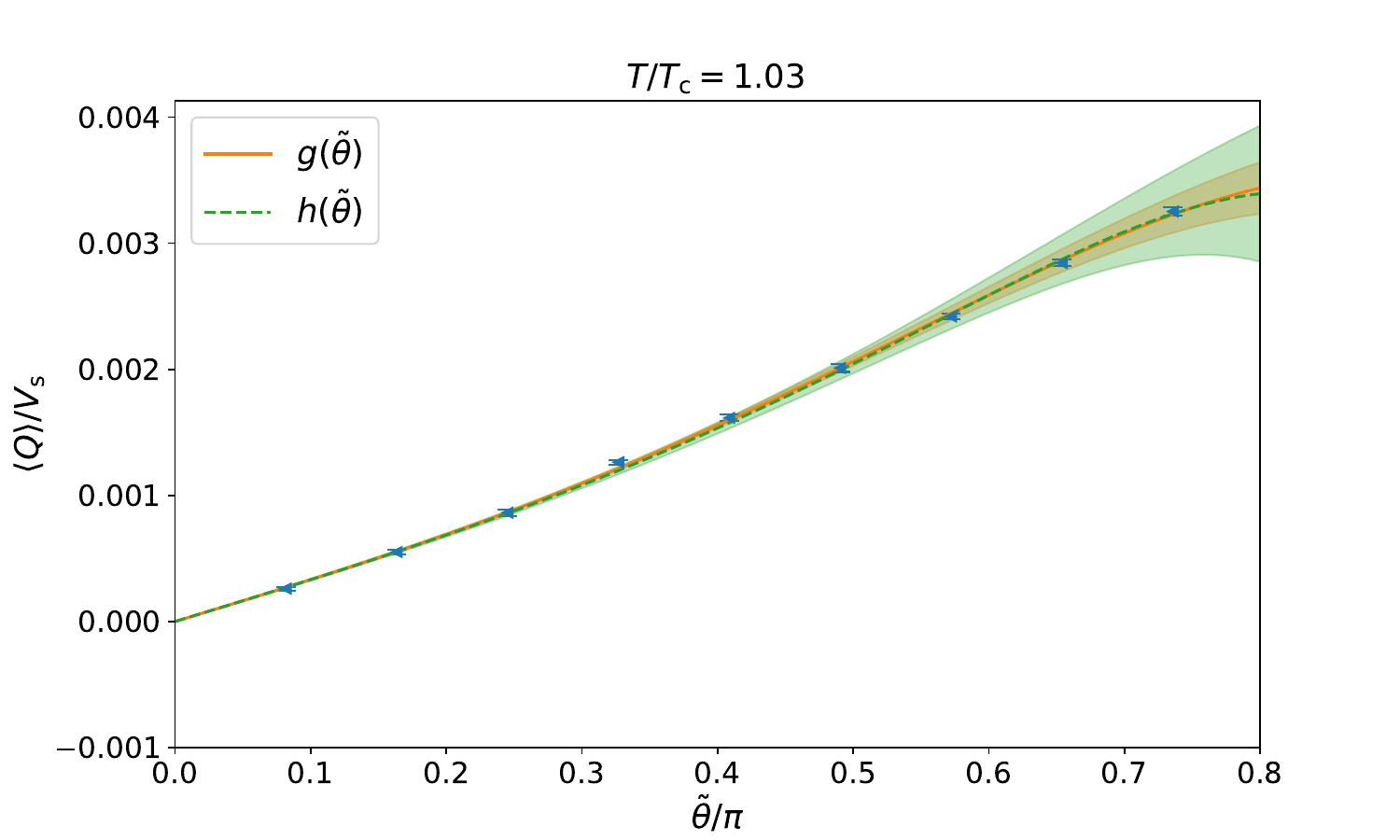}\\
    \includegraphics[width=0.475\hsize]{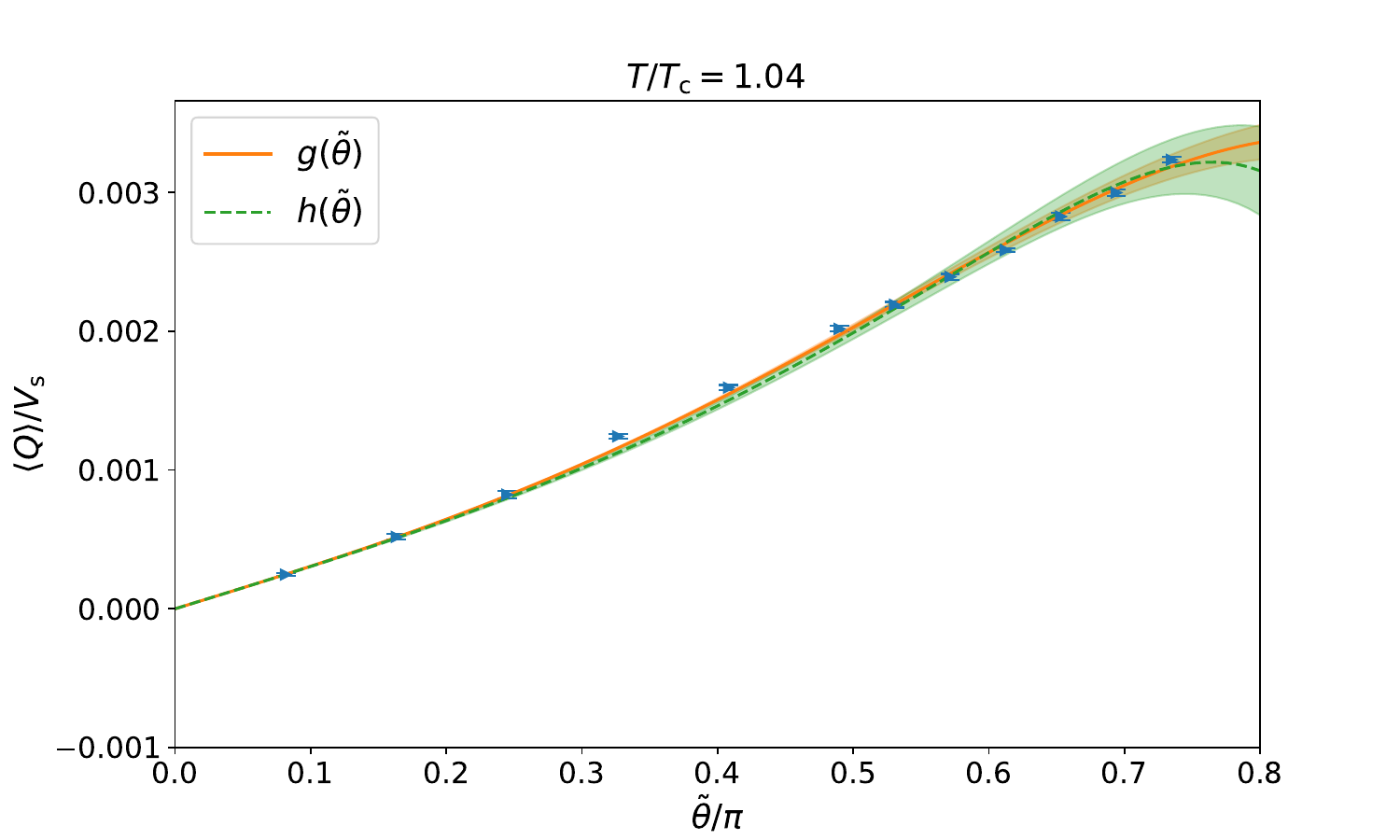}
    \includegraphics[width=0.475\hsize]{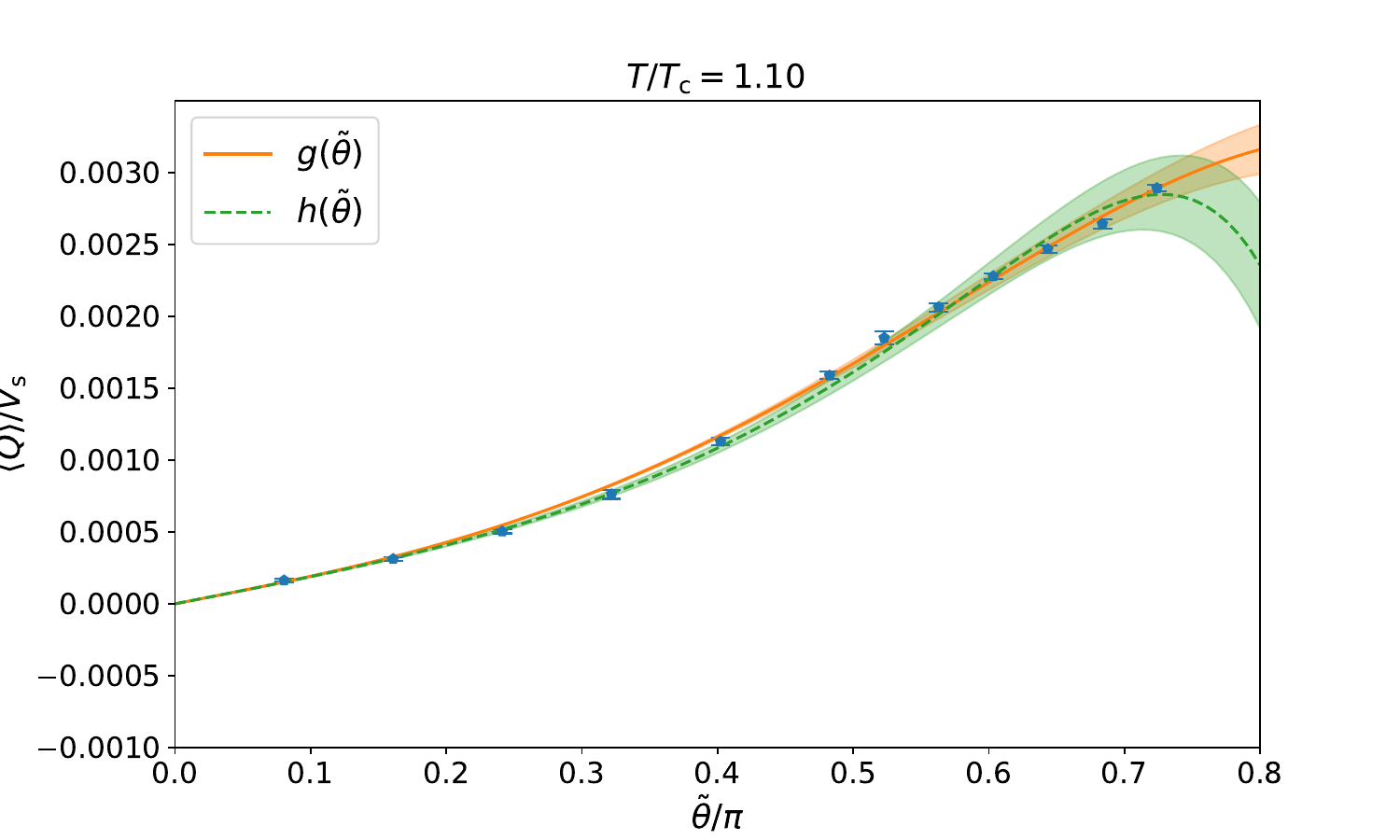}\\
    \caption{The topological charge density
      $\Braket{Q}_{i\tilde{\theta}}/V_{\rm s}$ 
      after the infinite volume extrapolation is plotted against $\tilde{\theta}/\pi$
      at various temperature within $0.9 \le T/T_{\mathrm{c}} \le 1.1$.
      We fit the data points to the functions \eqref{g-fitting} (orange curve) and
      \eqref{h-fitting} (green curve) at each temperature.
      The values of the fitting parameters are given in table~\ref{tab:fit_parm}.
      The error bands are also shown with light colors.}
    \label{fig:fitting}
\end{figure}

\begin{figure}[H]
    \centering
    \includegraphics[width=0.475\hsize]{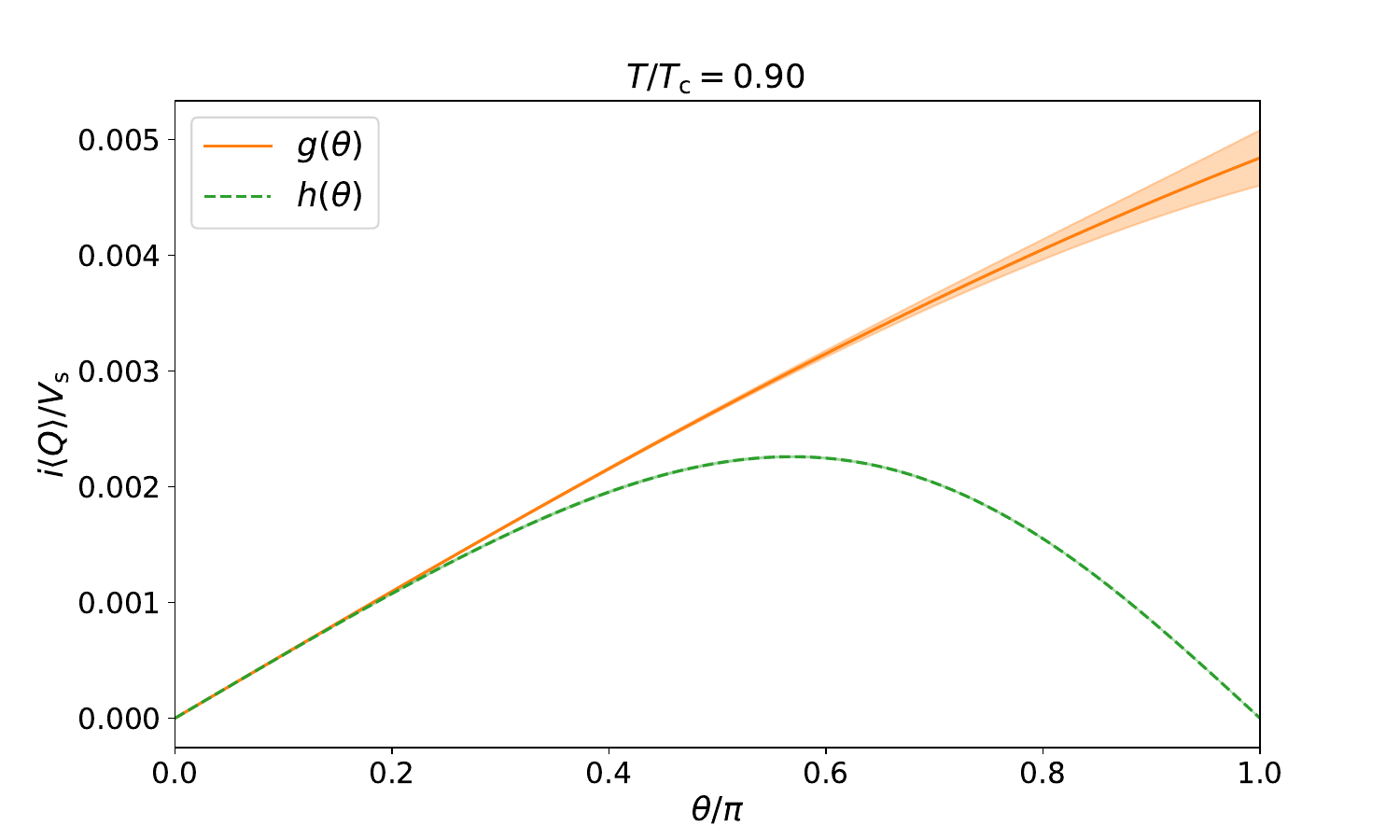}
    \includegraphics[width=0.475\hsize]{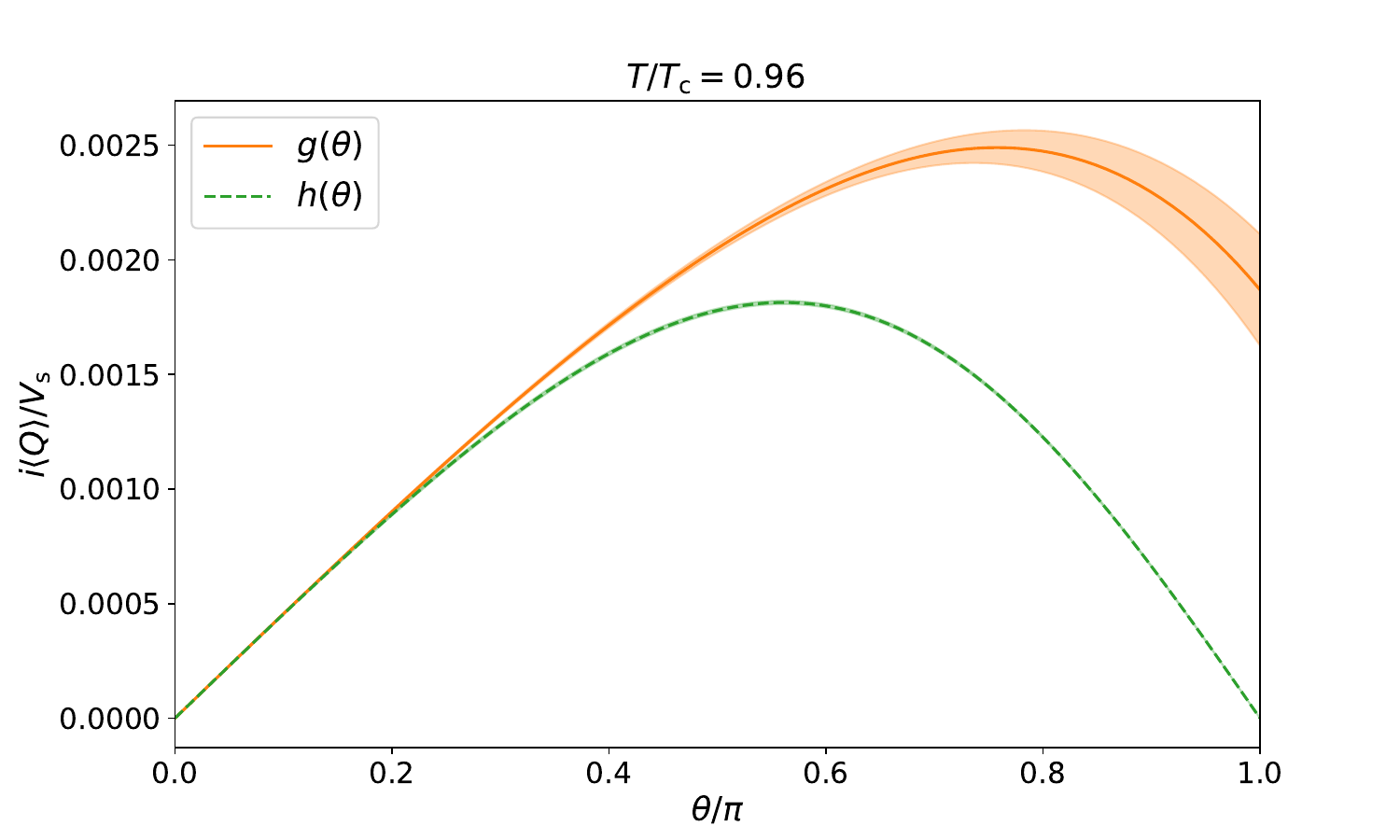}\\
    \includegraphics[width=0.475\hsize]{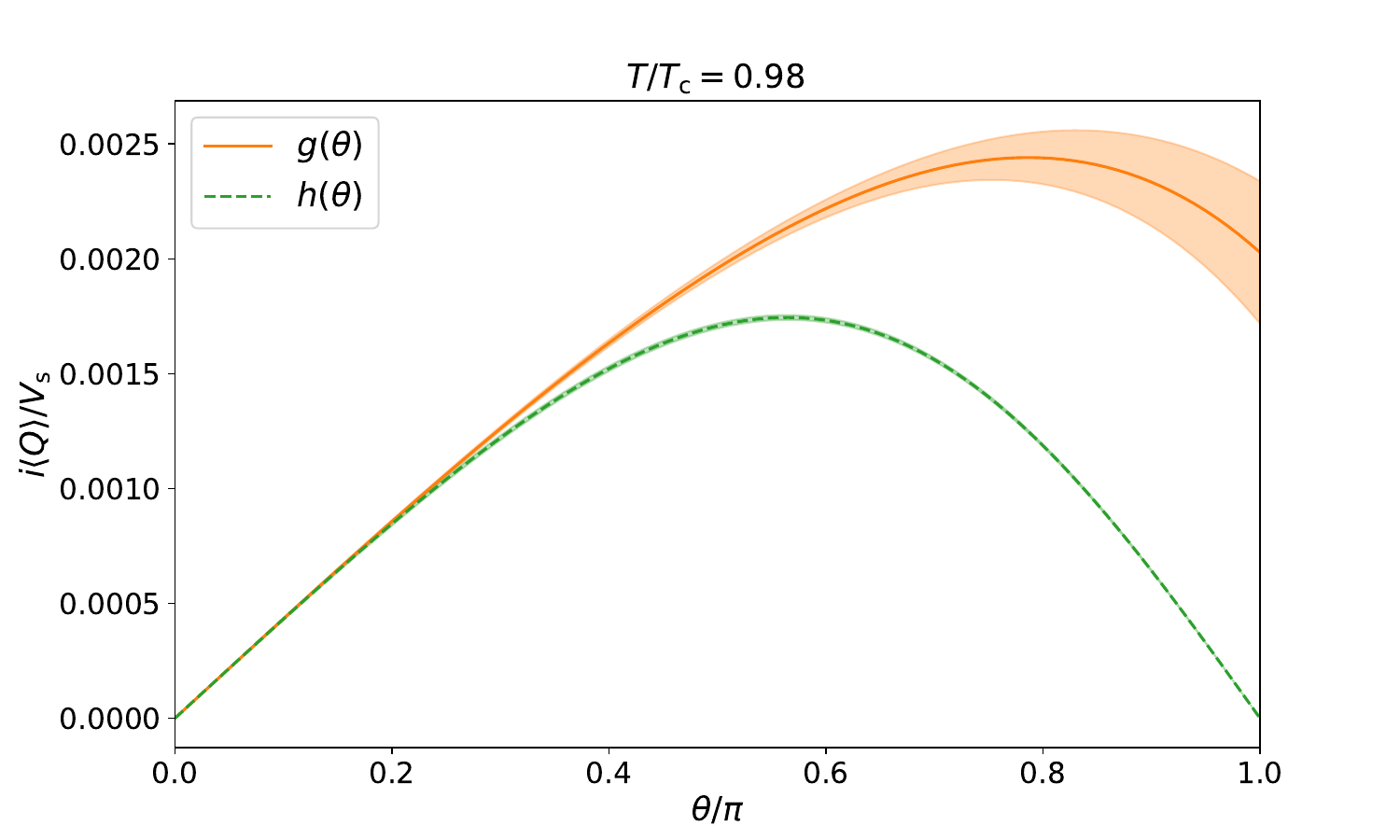}
    \includegraphics[width=0.475\hsize]{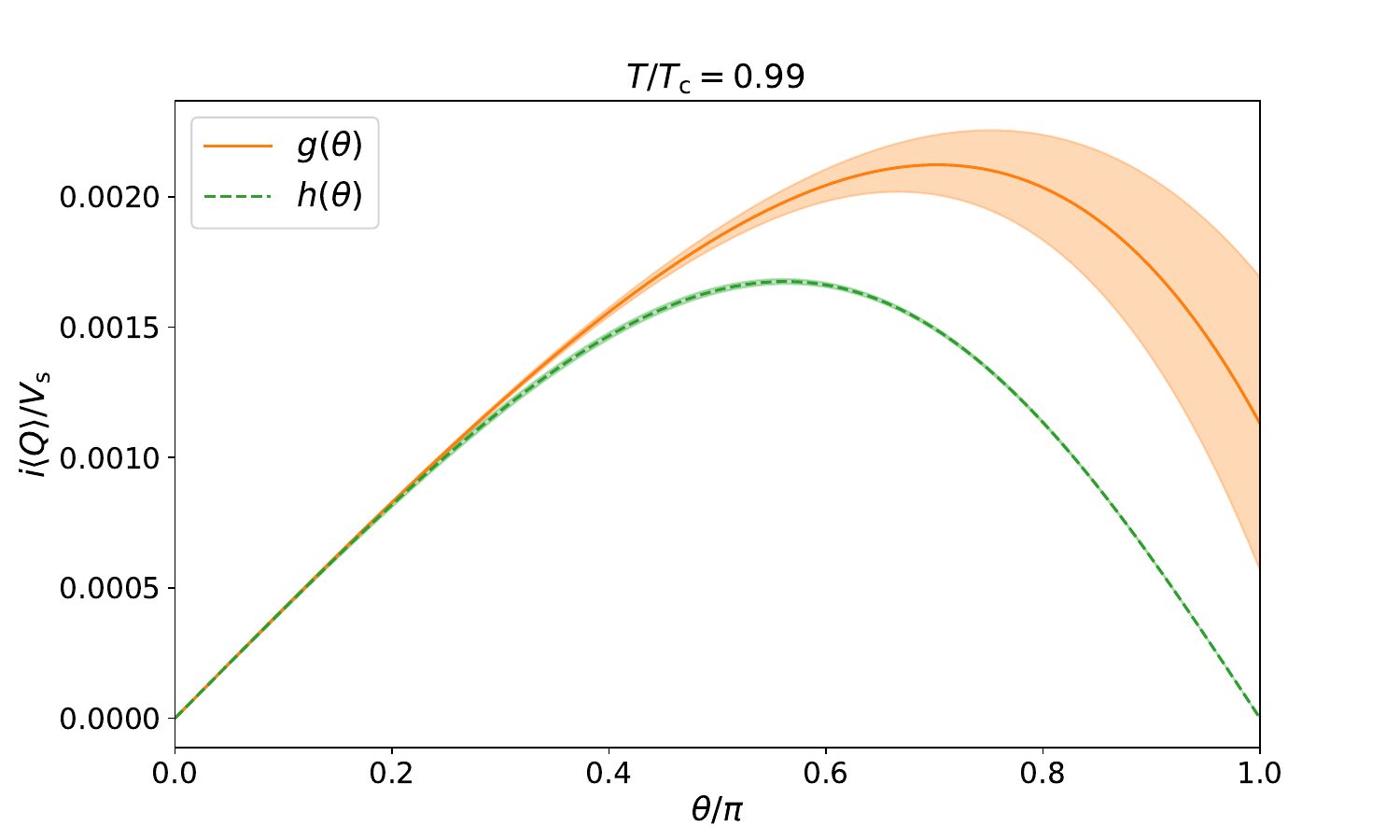}\\
    \includegraphics[width=0.475\hsize]{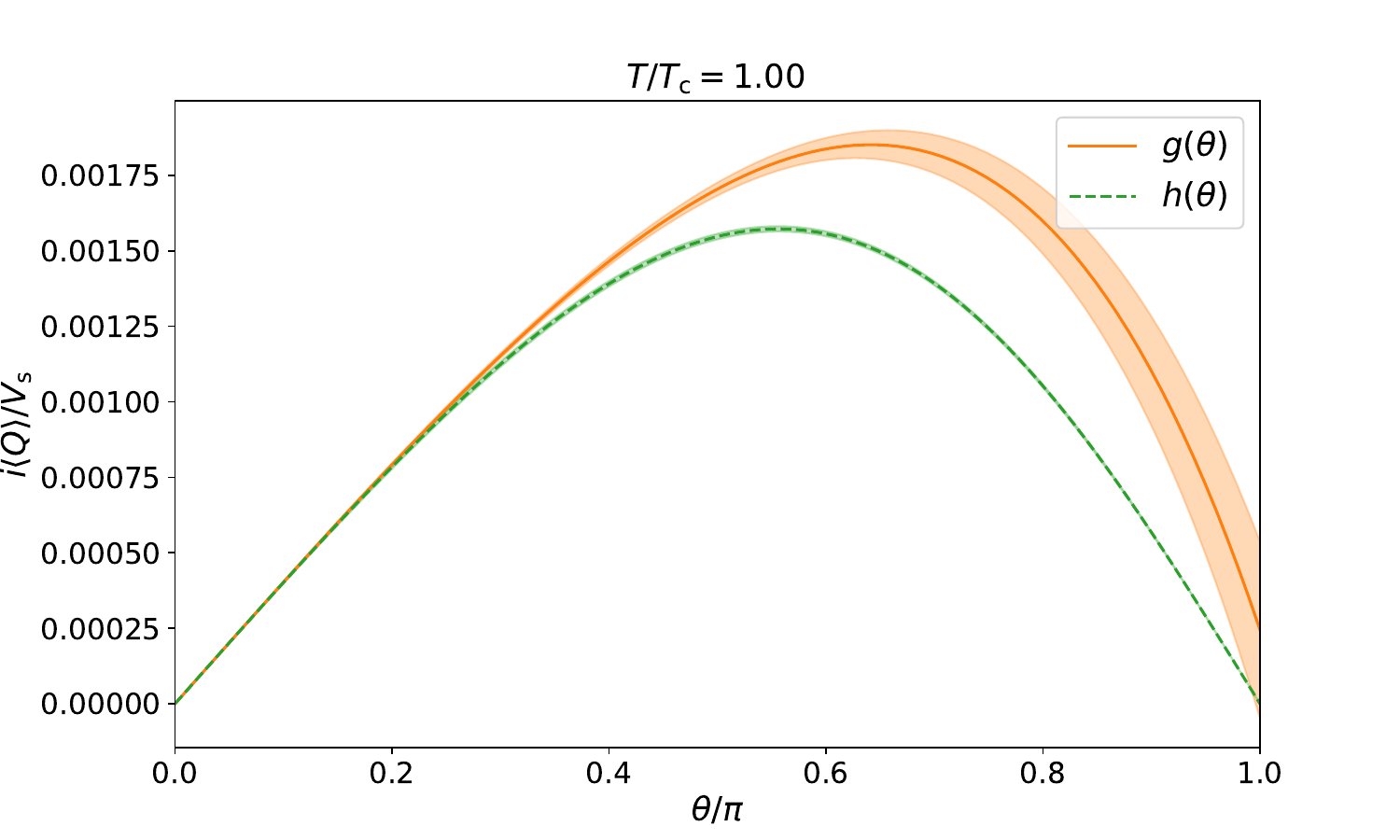}
    \includegraphics[width=0.475\hsize]{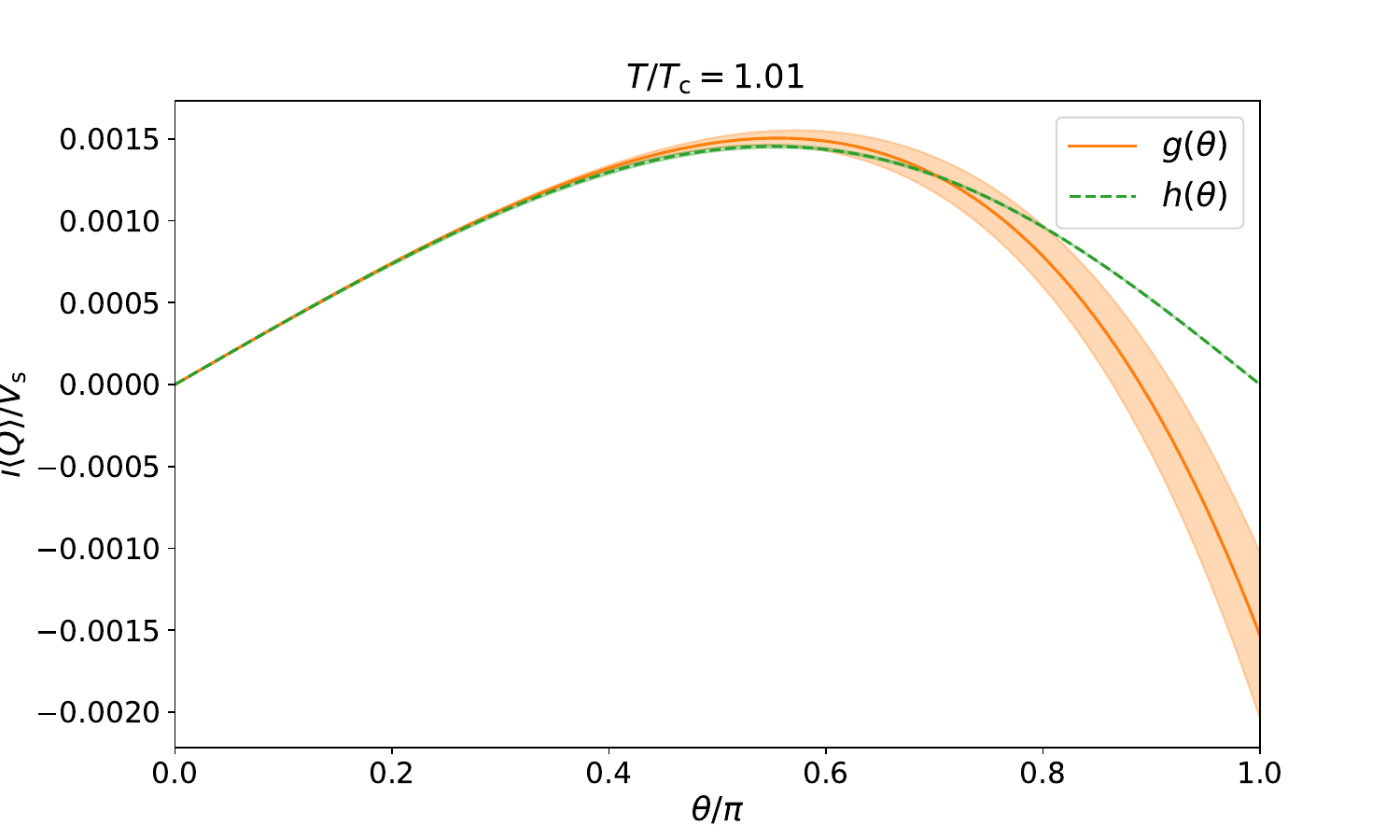}\\
    \includegraphics[width=0.475\hsize]{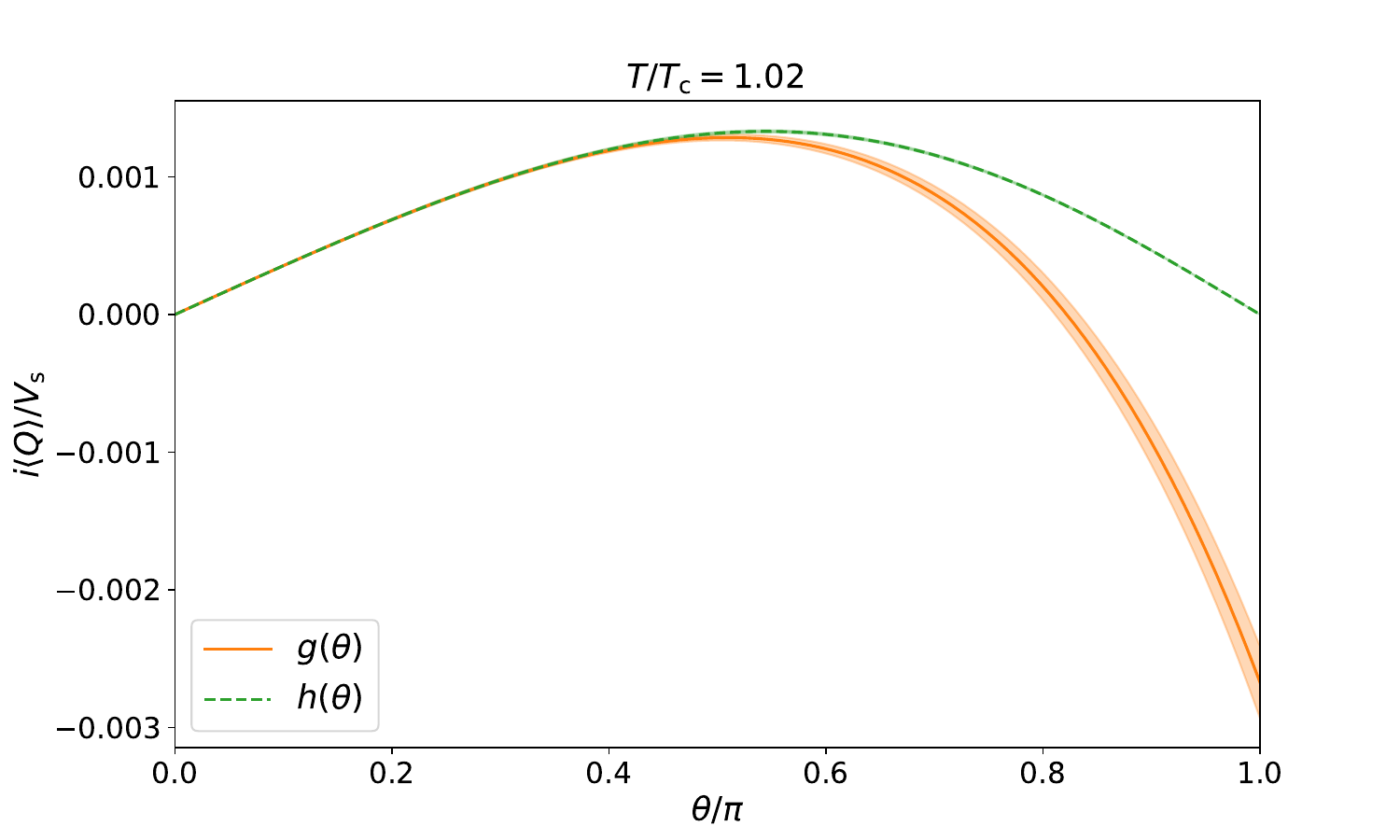}
    \includegraphics[width=0.475\hsize]{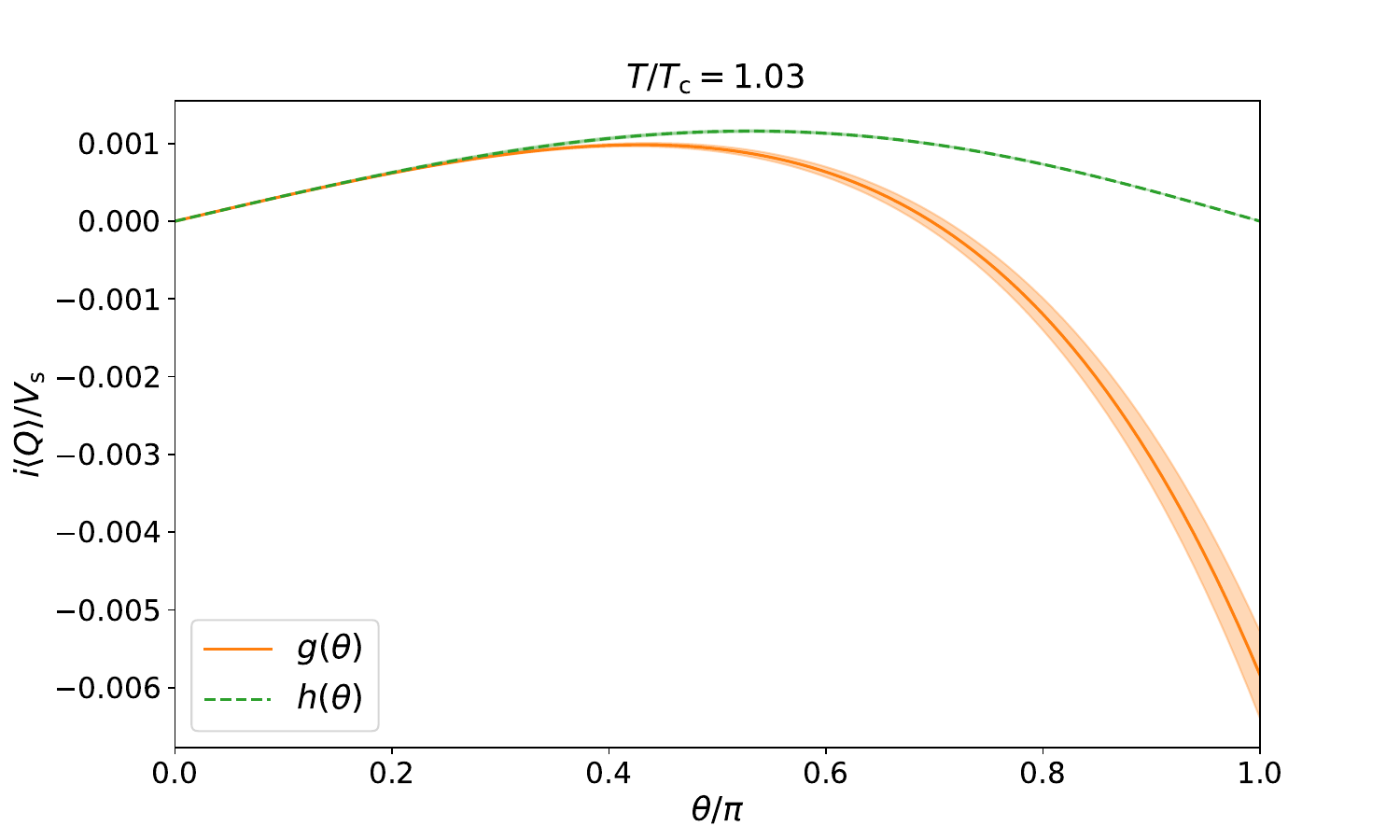}\\
    \includegraphics[width=0.475\hsize]{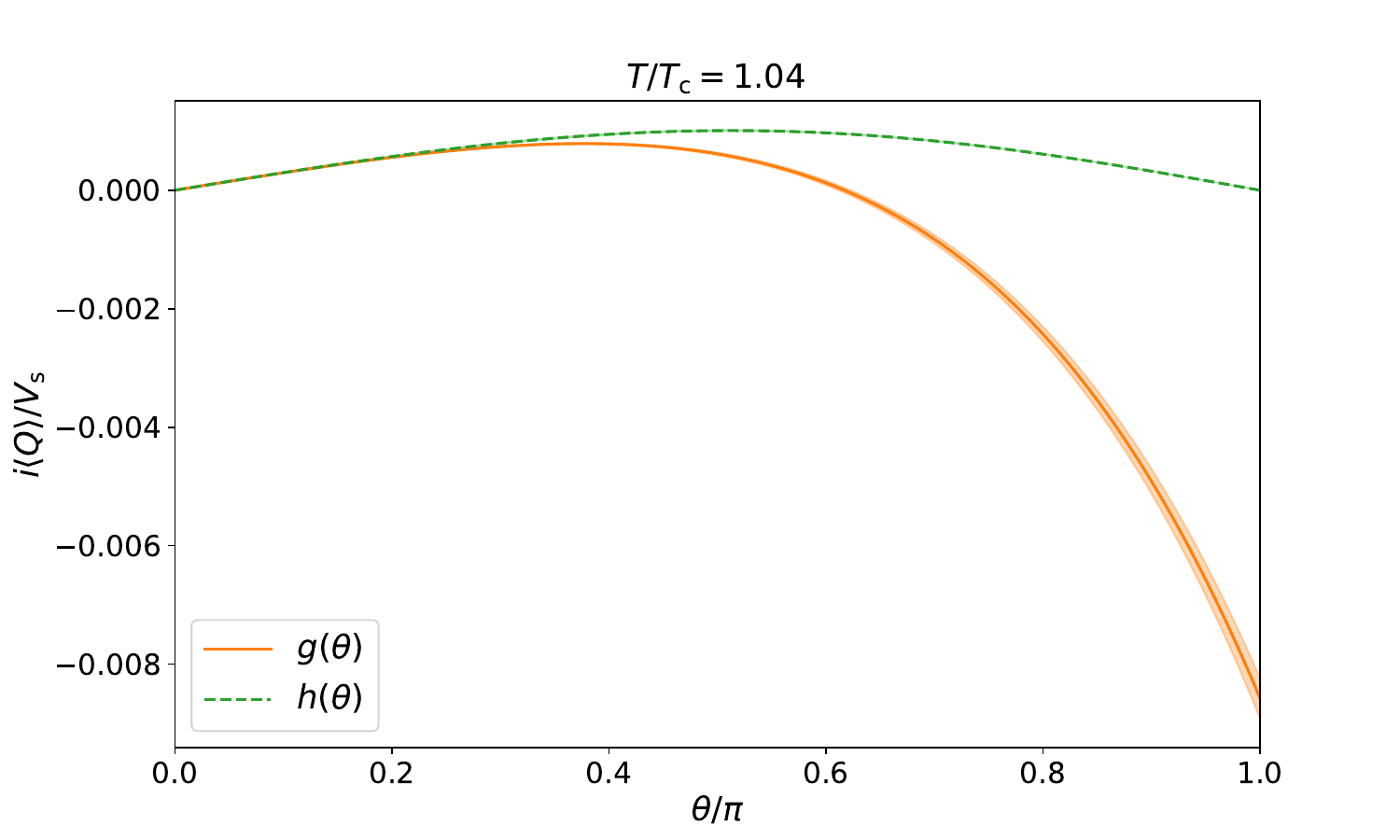}
    \includegraphics[width=0.475\hsize]{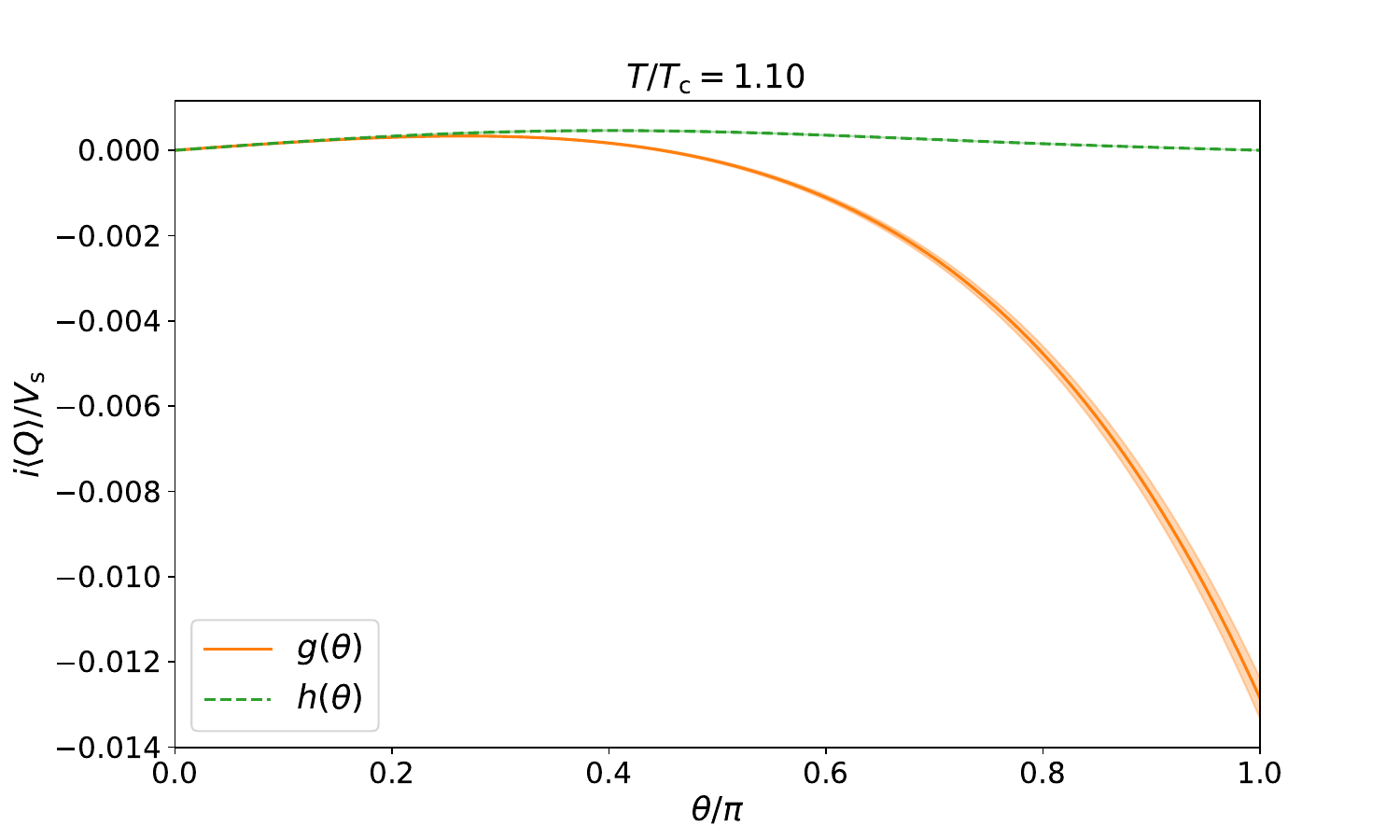}\\
    \caption{The topological charge density
      $\lim_{V_s \rightarrow \infty}  i\Braket{Q}_{\theta}/V_{\rm s}$
      for real $\theta$
%%      $\frac{1}{V_{\rm s}} \Braket{Q}_{\theta}}$ after the infinite volume extrapolation
      obtained by analytic continuation of the fitting functions
      is plotted against $\theta/\pi$ at various temperature
      within $0.9 \le T/T_{\mathrm{c}} \le 1.1$.
      The orange and green curves correspond to the functions
      $g(\theta)$ and $h(\theta)$ in \eqref{gh-theta}, respectively.
%%      \ref{g-theta} and \ref{h-theta}, respectively.
      The error bands are also shown with light colors.}
    \label{fig:analytic_cont}
\end{figure}

Using the results for the topological charge density and the topological susceptibility
after the infinite volume extrapolation,
we determine the holomorphic function that describes
the $\tilde{\theta}$ dependence of the topological charge density
as follows.
According to our discussions in section \ref{sec:imaginary},
a natural ansatz for $\frac{1}{V_{\rm s}} \Braket{Q}_\theta$ in the real $\theta$ region
within $|\theta|<\pi$ is\footnote{Note, in particular,
that the function $h(\theta)$ includes the behavior
$\Braket{Q}\propto  \sin{\theta}$ obtained by
the dilute instanton gas approximation, which is expected to be valid
at sufficiently high temperature \cite{Gross:1980br,Weiss:1980rj}.}
%%in the CP broken and restored phases as
\begin{equation}
  \begin{array}{ll}
    g(\theta) = -i (a_1 \theta + a_3 \theta^3 + a_5 \theta^5)
    \quad    & \mbox{in the CP broken phase} \ ,
%%    \label{g-theta}
    \\
    h(\theta) = -i(b_1 \sin{\theta} + b_2 \sin{2\theta} + b_3 \sin{3\theta}) \quad 
    & \mbox{in the CP restored phase} \ ,
%%    \label{h-theta}
  \end{array}
  \label{gh-theta}
\end{equation}
respectively, where we have taken into account the fact that
the topological charge is a CP-odd quantity.
%% only the odd powers of ${\theta}$ appear in $g(\theta)$.
We find that adding more terms in the expansion causes overfitting
and hence the truncation at this order seems to be optimal\footnote{This
applies to the fitting ansatz \eqref{eq:fit_Tdec} as well.}.

%%Finally, we perform the analytic continuation.

Using the topological susceptibility $\chi_0$
%%$\chi_0=\frac{1}{i} \left.\frac{\partial_{\theta} \Braket{Q}_\theta}{V}\right|_{\theta=0}$,
at $\theta=0$ after the infinite volume extrapolation,
we get the constraints $a_1 = \chi_0$ and $b_1 + 2 b_2 + 3 b_3 = \chi_0$,
which can be used to reduce the number of fitting parameters.
%%
%[Atis] Here the coefficients of the leading terms are determined by $\left.\frac{\partial_{\theta} \Braket{Q}_\theta}{V}\right|_{\theta=0} = \chi_0$.
%[Atis] Since the topological charge is a CP-odd quantity, only the odd powers of $\tilde{\theta}$ appear in $g(\theta)$.
%%
Thus, we obtain the ansatz for imaginary $\theta=i \tilde{\theta}$
%% We therefore fit our data points obtained
%% for imaginary $\theta= - i \tilde{\theta}$ at each temperature
%% to the functions
%%
%%use the following two functions for fitting the data points obtained by simulations with
\begin{align}
%%\begin{split}
  g(i \tilde{\theta}) &= \chi_0 \tilde{\theta}
  - a_3 \tilde{\theta}^3 + a_5 \tilde{\theta}^5 \ ,
\label{g-fitting}
  \\
    h(i \tilde{\theta})   &= (\chi_0 - 2b_2 - 3b_3) \sinh{\tilde{\theta}} +
    b_2 \sinh{2\tilde{\theta}} + b_3 \sinh{3\tilde{\theta}}  \ .
    \label{h-fitting}
%%\end{split}
\end{align}
%%\subsection{Fitting}

In figure~\ref{fig:fitting}, we fit our data points to
\eqref{g-fitting} (orange curve) and
\eqref{h-fitting} (green curve) at various temperature
within $0.9\le T/T_{\mathrm{c}} \le 1.1$.
The values of the fitting parameters and the normalized chi-square $\chi^2/N_{\rm DF}$
are given in table~\ref{tab:fit_parm}.
We find at lower temperature
$T/T_{\mathrm{c}} \le 1.0$
%% $T/T_{\mathrm{c}} \lesssim 1.0$
that the fitting to \eqref{h-fitting}
%%the hyperbolic sine series are not
is not good, which suggests that
the system is in the CP broken phase.
%[Atis] This result suggests that the CP breaking occurs at such temperature.
%%This suggests the CP broken phase.
On the other hand, at higher temperature
$T/T_{\mathrm{c}} \ge 1.01$,
%% $T/T_{\mathrm{c}} \gtrsim 1.01$,
the normalized chi-square $\chi^2/N_{\rm DF}$ turns out to be large
for both ansatzes, which may be due to a few data points around
the deconfining phase transition at $\tilde{\theta}= \tilde{\theta}_{\rm c}$.
(See the last paragraph of Section \ref{sec:deconf-temp-theta-dep}.)
Therefore, in the following analysis, we focus on the fitting curve
obtained at lower temperature $T/T_{\mathrm{c}} \le 1.0$
%% $T/T_{\mathrm{c}} \lesssim 1.0$
with the fitting ansatz \eqref{g-fitting}.

\begin{figure}[H]
    \centering
    \includegraphics[width=0.75\hsize]{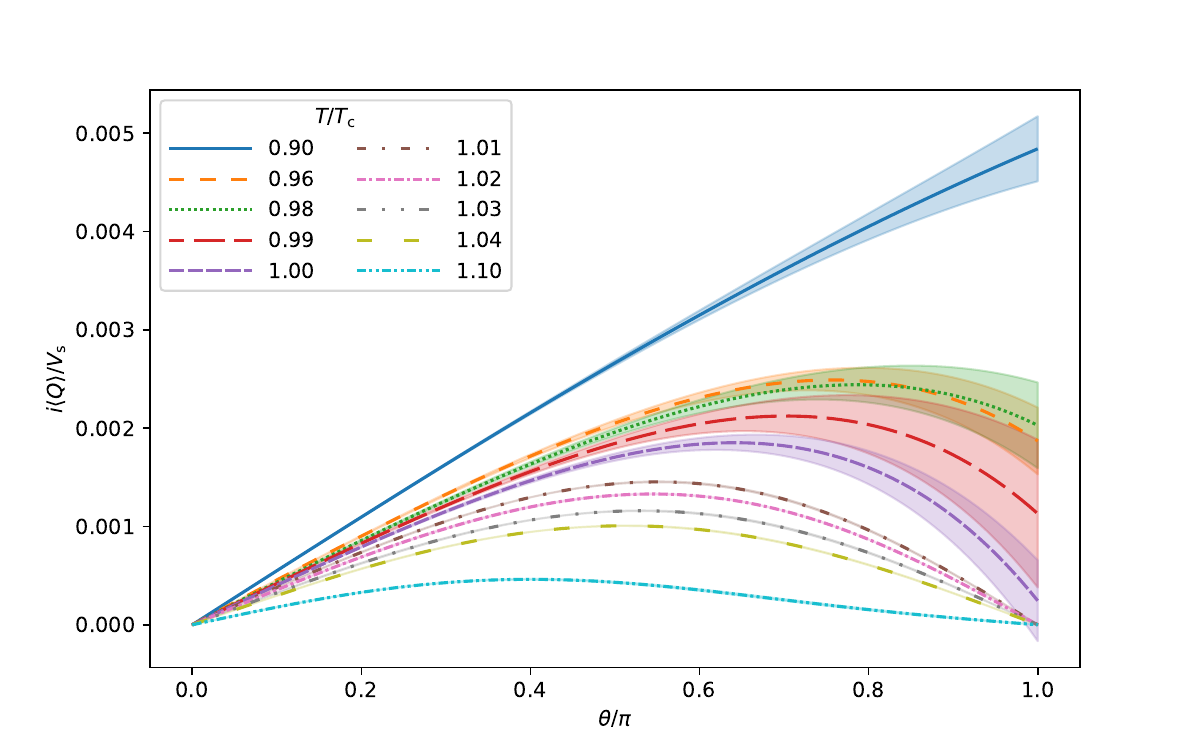}
    \caption{Our prediction for
      $\lim_{V_{\rm s} \rightarrow \infty} i \Braket{{Q}}_{\theta}/V_{\rm s}$
      %%       obtained by the analytic continuation of the fitting functions
      is plotted against $\theta/\pi$ at various temperature
      within $0.9 \le T/T_{\mathrm{c}} \le 1.1$.
      The gap at $\theta = \pi$ disappears
      at some $T$
      %%some temperature
      within $1.0 \lesssim T/T_{\mathrm{c}} \lesssim 1.01$.}
    \label{fig:ana_con_poly_det_Tcp}
\end{figure}

In figure~\ref{fig:analytic_cont}, we plot the fitting functions
\eqref{gh-theta}
after the analytic continuation.
%[Atis] The orange and green curves are corresponding to the function $g(\theta)$ and $f(\theta)$, respectively.
The orange and green curves correspond to
%%the functions
$g(\theta)$ and $h(\theta)$, respectively.
To determine the CP restoration temperature, we focus
%% on the value of the function $g(\theta)$ at $\theta = \pi$
on the curve corresponding to the function $g(\theta)$ in the low temperature region.
As the temperature is increased,
the curve starts to bend and the gap at $\theta = \pi$
disappears around $T/T_{\mathrm{c}} \sim 1.0$,
which we identify as the CP restoration temperature $T_{\rm CP}$.
%%We identify the temperature where the gap disappears with the CP restoration temperature.
Based on this, we arrive at the prediction for $\frac{1}{V_s} \Braket{Q}_\theta$
%% summarize our results
in figure~\ref{fig:ana_con_poly_det_Tcp},
where we plot the function
$g(\theta)$ for $T/T_{\mathrm{c}} \leq 1.00$
and the function $h(\theta)$ for $T/T_{\mathrm{c}} \geq 1.01$.
%%in the same figure.
%[Atis] We plot $g(\theta)$ up to $\le T/T_{\mathrm{c}} = 0.99$ and $h(\theta)$ above that temperature

\begin{table}[H]
    \centering
    \begin{tabular}{|c||c|c|c||c|c|c|}\hline
         &  \multicolumn{3}{c||}{polynomial} & \multicolumn{3}{c|}{hyperbolic sine} \tabularnewline\hline
        $T/T_{\mathrm{c}}$ & $a_3$ & $a_5$ & $\chi^2/N_{\rm DF}$ & $b_2$ & $b_3$ & $\chi^2/N_{\rm DF}$  \tabularnewline\hline 
        \hline 
        0.90 & 0.0007(1) & 0.0001(2) & 1.92 & -0.000249(2) & 0.0000117(2) & 16.14 \tabularnewline\hline 
        0.96 & 0.00162(9) & -0.0011(2) & 2.63 & -0.000177(2) & 0.0000076(2) & 5.12 \tabularnewline\hline 
        0.98 & 0.0015(1) & -0.0008(3) & 0.92 & -0.000179(3) & 0.0000084(3) & 1.74 \tabularnewline\hline 
        0.99 & 0.0018(2) & -0.0013(5) & 1.42 & -0.000166(4) & 0.0000075(4) & 1.92 \tabularnewline\hline 
        1.00 & 0.0021(1) & -0.0017(3) & 1.55 & -0.000141(2) & 0.0000056(2) & 2.33 \tabularnewline\hline 
        1.01 & 0.0028(2) & -0.0025(5) & 7.73 & -0.000118(4) & 0.0000043(3) & 6.92 \tabularnewline\hline 
        1.02 & 0.0033(1) & -0.0029(2) & 2.18 & -0.000094(2) & 0.0000030(2) & 1.68 \tabularnewline\hline 
        1.03 & 0.0045(2) & -0.0046(5) & 1.46 & -0.000056(4) & 0.0000006(4) & 2.48 \tabularnewline\hline 
        1.04 & 0.0057(1) & -0.0059(3) & 5.10 & -0.000021(3) & -0.0000015(2) & 11.29 \tabularnewline\hline 
        1.10 & 0.0077(2) & -0.0069(4) & 2.67 & 0.000092(3) & -0.0000077(3) & 4.26 \tabularnewline\hline 
    \end{tabular}
    \caption{The values of the fitting parameters obtained by
      fitting $\Braket{Q}_{i\tilde{\theta}}/V_{\rm s}$
      after the infinite volume extrapolation
      at various temperature within $0.9 \le T/T_{\mathrm{c}} \le 1.1$.}
    \label{tab:fit_parm}
\end{table}

%%\begin{figure}[tb]
\begin{figure}[H]
    \centering
    \includegraphics[width=0.75\hsize]{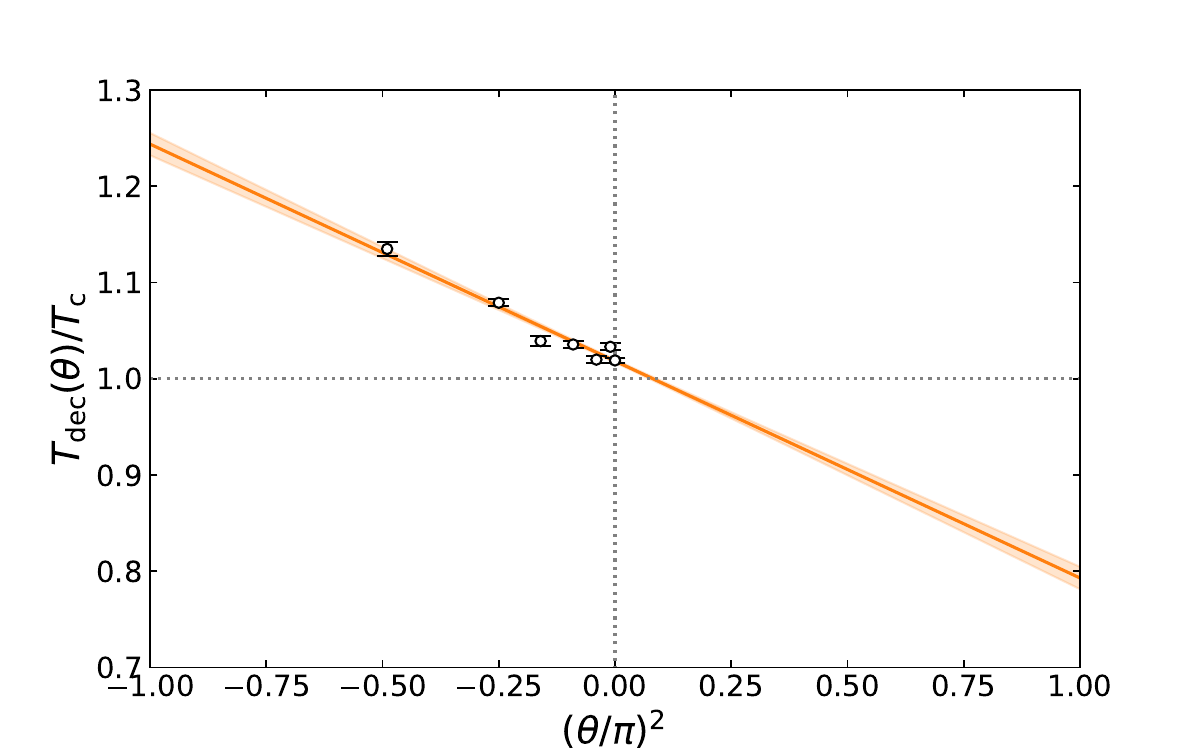}
    \caption{The deconfining temperature after the infinite volume extrapolation
      is plotted against $(\theta/\pi)^2$.
      The curves represent the fit to the function \eqref{eq:fit_Tdec}.}
    \label{fig:T_dec}
\end{figure}

\subsection{$\theta$ dependence of the deconfining temperature}
\label{sec:deconf-temp-theta-dep}

In order to investigate the inequality \eqref{inequality-CP-dec},
we still
have to determine $T_{\mathrm{dec}}(\pi)$ that appears on the right-hand side.
For that, we
%% In this section, we
discuss the $\theta$ dependence of the deconfining temperature,\footnote{
When this paper was about to be completed, we encountered a preprint~\cite{Yamada:2024pjy},
which addresses this issue using the subvolume method.
The deconfinement temperatures obtained at real $\theta$ in that paper lie close to the fitting curve in our figure~\ref{fig:T_dec}.
%The results are in rough agreement with ours.
}
again by analytic continuation from the results obtained at imaginary $\theta$.
Note that in pure SU($N$) Yang-Mills theory
that we are dealing with, the Polyakov loop
%% $\mathcal{P}$
is an order parameter for the spontaneous $\mathbb{Z}_{N}$
center symmetry breaking,
which corresponds to the deconfining phase transition. Therefore we expect
to observe a peak in the corresponding susceptibility at the deconfining temperature.

In order to determine $T_{\mathrm{dec}}$ at imaginary $\theta$,
we measure the Polyakov loop susceptibility $\chi_{\rm P}(T)$
at various temperature
and fit our results
to the Lorentzian function
\begin{equation}
    \chi_{\rm P}(T) = \frac{A}{(T-T_{\mathrm{peak}})^2+w^2} \ ,
\end{equation}
where $A$, $T_{\mathrm{peak}}$ and $w$ are the fitting parameters.
By extrapolating $T_{\mathrm{peak}}$ to the infinite volume,
we estimate the deconfining temperature $T_{\mathrm{dec}}(i\tilde{\theta})$
at each $\tilde{\theta}$.

The result of $T_{\mathrm{dec}}(i\tilde{\theta})/T_{\mathrm{c}}$
is plotted in figure~\ref{fig:T_dec} against $(\theta/\pi)^2=-(\tilde{\theta}/\pi)^2$.
%% on the left half of the figure.
Since the deconfining temperature is invariant 
under the CP transformation $\theta\rightarrow-\theta$,
it should be an even function of $\theta$.
%% Thus, the leading term for expanding
%% $T_{\mathrm{dec}}(\tilde{\theta})/T_{\mathrm{c}}$ 
%% around $\tilde{\theta}=0$ will be $\tilde{\theta}^2$.
Here we fit our results for the deconfining temperature to
%%for the $\theta$-dependence of the deconfining temperature, 
\begin{equation}
  \frac{T_{\mathrm{dec}}(\theta)}{T_{\mathrm{c}}}
  %%  = c_0 - c_2 \theta^2 \ ,
    = c_0 - c_2 \left( \frac{\theta}{\pi} \right) ^2 \ ,
%%   =\left\{
%% \begin{array}{l}
%% %% 1+R_1\tilde{\theta}^2 \\
%%   1+R_2\tilde{\theta}^2 + c_2 \tilde{\theta}^4 \\
%%   1 - c_1 \theta^2 - c_2 \theta^4  \ ,
%% %% 1-c_3(1-\cos{\theta})  \ ,
%% \end{array}
%% \right.
\label{eq:fit_Tdec}
\end{equation}
%%including the polynomial and the hyperbolic cosine.
%% where the second one is motivated by the $2\pi$ periodicity of $\theta$ 
%% in the real $\theta$ region although 
%% $T_{\mathrm{dec}}(\tilde{\theta})/T_{\mathrm{c}}$
%% is not necessarily smooth at $\theta=\pi$.
%%
%%We fitted our results 
%%to the two functions \eqref{eq:fit_Tdec}.
which yields\footnote{In the SU(3) case, it is found that
the coefficient $c_2$ in the first ansatz is positive as
well \cite{Anber:2013sga, DElia:2013uaf, Otake:2022bcq, Borsanyi:2022fub}.}
%% at finite lattice spacing and
%% may due to the small discrepancy 
%%  between the deconfining temperatures at finite lattice spacing and
$c_0=1.0183(16)$, $c_2=0.225(12)$.
%% \red{}
%If we set $c_2=0$, we obtain $c_1=0.0303(9)$.
%% For the second ansatz, we obtain $c_3=0.045(1)$.
The result of the fit is shown in figure~\ref{fig:T_dec}, % as well,
where we also extend the fitting function to 
%%show the behaviors of the fitting curve in
the real $\theta$ region.
%% \begin{equation}
%%     \frac{T_{\mathrm{dec}}(\theta)}{T_{\mathrm{c}}}=\left\{
%% \begin{array}{l}
%% 1-R_1\theta^2 \\
%% 1-R_2\theta^2 + c_2 \theta^4 \\
%% 1+c_3(\cos{\theta}-1)
%% \end{array}
%% \right. .
%% \label{eq:fit_Tdec_ana_cont}
%% \end{equation}
%%
%% Thus we conclude that $T_{\mathrm{dec}}(\theta) < T_{\mathrm{c}}$.
%% We have tried various fitting ansatz, but our conclusion
%% $T_{\mathrm{dec}}(\theta) < T_{\mathrm{c}}$ seems to be robust.
Note that the vertical intercept of the fitting line being $c_0 \neq 1$ is not a contradiction
since $T_{\rm c}$ represents the deconfining temperature in the continuum limit,
whereas $T_{\rm dec}(\theta)$ is obtained at finite lattice spacing.
While the prediction for
%%the $\theta$-dependence
the deconfining temperature $T_{\mathrm{dec}}(\theta)/T_{\mathrm{c}}$
in the real $\theta$ region depends much on the ansatz,
we may safely conclude that
%%$T_{\mathrm{dec}}(\theta)/T_{\mathrm{c}} \lesssim 0.91T_{\mathrm{c}}$.
$T_{\mathrm{dec}}(\theta)
%% /T_{\mathrm{c}}
< T_{\mathrm{c}}$.
%%<1$.
%The coefficient RR of the quadratic term of
%Tdec(˜θ)/TcT_{\mathrm{dec}}(\tilde{\theta})/T_{\mathrm{c}}
%turned out to be negative as in the case of the SU(3) YM
%\cite{DElia:2012pvq, DElia:2013uaf, Otake:2022bcq, Borsanyi:2022fub}.
%%
%%The natural expectation is that the deconfining temperature at $\theta=\pi$
%%is lower than that of $\theta=0$.

Having identified the deconfining phase transition line in the imaginary $\theta$ region,
%%Using our results in the imaginary $\theta$ region,
we can discuss
how it
%the deconfining phase transition
affects the $\tilde{\theta}$ dependence of
the topological charge density $\Braket{Q}_{i\tilde{\theta}}/V_{\rm s}$
after the infinite volume extrapolation.
Let us focus on the bottom two plots in figure~\ref{fig:fitting}.
From figure~\ref{fig:T_dec}, we find that the critical $\tilde{\theta}$
at which the deconfining phase transition occurs is
$\tilde{\theta}_{\rm c}/\pi= 0.32(2)$ and $\tilde{\theta}_{\rm c}/\pi= 0.60(2)$,
respectively, for $T/T_{\rm c}=1.04$ and $T/T_{\rm c}=1.10$.
%% $\tilde{\theta}_{\rm c}/\pi= 0.3 \sim 0.4$ and $\tilde{\theta}_{\rm c}/\pi= 0.5\sim 0.6$,
%% respectively, for $T/T_{\rm c}=1.04$ and $T/T_{\rm c}=1.10$.
%%
%% ----------------
%%    \tilde{\theta}_{c} / \pi = 0.32(2) at T_{dec}(\theta)/T_c = 1.04
%%    \tilde{\theta}_{c} / \pi = 0.60(2) at T_{dec}(\theta)/T_c = 1.1
%% ----------------
%%
In fact, we do see some deviation from the fitting curve
around $\tilde{\theta} = \tilde{\theta}_{\rm c}$, which might be due to the
deconfining phase transition.
Despite this subtlety,
we find that the data points are nicely fitted to the analytic functions
\eqref{g-fitting} and \eqref{h-fitting}.
Therefore our basic assumption that
$\Braket{Q}_{\theta}$ is not only continuous but also analytic across
the deconfining phase transition seems to be valid to good accuracy
at least in the imaginary $\theta$ region.

%\clearpage
%%%%%%%%%%%%%%%%%%%%%%%%%%%%%%%
%%%%%%%%%%%%%%%%%%%%%%%%%%%%%%%
%%%%%%%%%%%%%%%%%%%%%%%%%%%%%%%
%\section{Summary}
\section{Summary and discussions}
\label{sec:discussion}
%%%%%%%%%%%%%%%%%%%%%%%%%%%%%%%
%%%%%%%%%%%%%%%%%%%%%%%%%%%%%%%
%%%%%%%%%%%%%%%%%%%%%%%%%%%%%%%

In this paper, we have investigated the spontaneous breaking of
CP symmetry in four-dimensional pure Yang-Mills theory at $\theta=\pi$,
which has
%The phase structure at $\theta=\pi$ attracts
attracted a lot of attention in recent years
since the 't Hooft anomaly matching condition suggested
a nontrivial phase structure.
%% recent studies have suggested that it should be nontrivial
%% as a consequence of 
In fact, at large $N$, it is known that the CP symmetry
at $\theta=\pi$ is spontaneously broken in the confined phase
and it gets restored in the deconfined phase.
However, a CP broken deconfined phase may appear at small $N$
since it is allowed by the anomaly matching condition.
The aim of this work is to investigate this possibility by numerical simulations.

In order to circumvent the sign problem that occurs in 
ordinary Monte Carlo methods due to the $\theta$ term,
we perform simulations at imaginary $\theta$,
%%which is possible since
where the sign problem is absent,
and make an analytic continuation to real $\theta$.
As the order parameter of the spontaneous CP symmetry breaking,
we calculate the expectation value of the topological charge $\langle Q \rangle$
at $\theta=\pi$ in this way.
Our results suggest that the CP restoration temperature $T_{\mathrm{CP}}$
is very close to $T_{\mathrm{c}}$, which is the deconfining temperature at $\theta=0$.
We have also estimated the $\theta$ dependence of
the deconfining temperature by analytic continuation and found
that
%%$T_{\mathrm{dec}}(\theta=\pi) \lesssim 0.91T_{\mathrm{c}}$.
$T_{\mathrm{dec}}(\theta=\pi)  < T_{\mathrm{c}}$.
Combining these two results, we have concluded
%% namely $0.99\lesssim T_{\mathrm{CP}}/T_{\mathrm{c}} \lesssim 1$.
%%Thus our results suggest
the existence of a CP broken deconfined phase at $\theta=\pi$,
in striking contrast to the situation at large $N$.
%% that the CP symmetry at $\theta=\pi$ is
%% restored at temperature higher than the deconfinement temperature---unlike the situation at large $N$.

%% While our result is consistent with that of
%% the subvolume method \cite{Kitano:2020mfk, Kitano:2021jho, Yamada:2024vsk},
%% %[Atis] which suggests that the CP is broken at zero temperature and
%% which suggested that the CP symmetry is spontaneously broken at zero temperature and
%% restored at $T_{{\rm CP}} \le 1.2T_{\mathrm{c}}$,
%% This result is consistent with our more stringent 

%% Let us mention that previous studies based on
%% %%Our result is consistent with that of
%% the subvolume method \cite{Kitano:2020mfk, Kitano:2021jho, Yamada:2024vsk}
%% %[Atis] which suggests that the CP is broken at zero temperature and
%% suggested that the CP symmetry is spontaneously broken at zero temperature and
%% restored at $T_{{\rm CP}} \le 1.2T_{\mathrm{c}}$.

Let us note that it is not straightforward to 
take the continuum limit by increasing the inverse coupling $\beta$
since that will cause the so-called topology freezing problem.
(See, however, Ref.~\cite{Eichhorn:2023uge} and references therein for
recent developments in solving this problem.)
Instead, it would be worthwhile to improve the gauge action
by including $1 \times 2$ Wilson loops,
which will enable us to reduce the number of smearing steps.
This is useful not only in reducing the possible artifacts of the stout smearing
but also in reducing the computational cost, 
which will be important in increasing the lattice size.

%[Atis] For this purpose, the gauge action should be promoted to
%% For this purpose, the gauge action should be replaced by
%% an improved action including rectangular Wilson loops.
%% The number of smearing steps will be reduced by using
%% such an action, which is also crucial in increasing the lattice size.

%%It would be also interesting
We are currently trying
to extend our work to the SU(3) YM theory, 
where the phase structure at $\theta=\pi$ is expected \cite{Gaiotto:2017yup}
to be like figure~\ref{fig:phase_diag} (Left), similarly to the large-$N$ case.
Considering that $T_{\mathrm{dec}}(\pi) < T_{\mathrm{c}}$ is expected 
in SU(3) \cite{Poppitz:2012nz,Anber:2013sga, DElia:2013uaf, Otake:2022bcq, Borsanyi:2022fub} as well,
%% similarly to what we have found in SU(2),
we may find that $T_{\rm CP}$ is significantly lower than $T_{\mathrm{c}}$
unlike in the SU(2) case, where we have found $T_{\rm CP} \simeq T_{\mathrm{c}}$.

%% might be able to see that $T_{{\rm CP}}  < T_{\mathrm{c}}$
%% unlike what we have found in SU(2).
%%
%%should be able to find a possible qualitative difference
%%of the topological charge distribution between $N=2$ and $N=3$.

% Our method will be useful to investigate the CP symmetry
% of any other models with a θ\theta term,
% if the Monte Carlo simulation for imaginary θ\theta is possible.
% For example, the phase structure of CP(N-1) model
% at θ=π\theta = \pi is interesting in terms of Haldane's conjecture
% \cite{Azcoiti:2007cg, Nakayama:2021iyp}.

%%%%%%%%%%%%%%%%%%%%%%%%%%
\acknowledgments
We would like to thank Kohta Hatakeyama for his participation at the earlier stage
of this work.
The authors are also grateful to
%%We would like to thank
Yuta Ito and Yuya Tanizaki for valuable discussions and comments.
The computations were carried out on Yukawa-21 at YITP in Kyoto University 
and the PC clusters at KEK Computing Research Center and KEK Theory Center.
This work also used computational resources of supercomputer NEC
SX-Aurora TSUBASA provided by the Particle, Nuclear, and Astro Physics
Simulation Program No.2020-009 (FY2020), No.2021-005 (FY2021),
No.2022-004 (FY2022), and 2023-002(FY2023) of Institute of Particle and
Nuclear Studies, High Energy Accelerator Research Organization (KEK).
%% This work is supported by the Particle, Nuclear, and Astro Physics Simulation Program 
%% No.2020-009 (FY2020), No.2021-005 (FY2021), No.2022-004 (FY2022), and 2023-002(FY2023) 
%% of Institute of Particle and Nuclear Studies, High Energy Accelerator Research Organization (KEK).
%%
%% M.~Honda is supported by MEXT Q-LEAP, JST PRESTO Grant Number JPMJPR2117,
%% JSPS Grant-in-Aid for Transformative Research Areas (A) JP21H05190 and JSPS KAKENHI Grant Number 22H01222.
M.~Honda is supported by JST PRESTO Grant Number JPMJPR2117, JST CREST
Grant Number JPMJCR24I3, JSPS Grant-in-Aid for Transformative Research
Areas (A) ``Extreme Universe" JP21H05190 [D01] and JSPS KAKENHI Grant
Number 22H01222.
A.~M. is supported by JSPS Grant-in-Aid for Transformative Research Areas (A) JP21H05190.
A.~Y. is supported by JSPS Grant-in-Aid for Transformative Research Areas (A) JP21H05191.

%%%%%%%%%%%%%%%%%%%%%%%%%%
%%%%%%%%%%%%%%%%%%%%%%%%%%
%%%%%%%%%%%%%%%%%%%%%%%%%%
\appendix
\section{Derivation of the HMC force term with the stout smearing}
\label{app:stout}
%%%%%%%%%%%%%%%%%%%%%%%%%%
%%%%%%%%%%%%%%%%%%%%%%%%%%
%%%%%%%%%%%%%%%%%%%%%%%%%%

%%In this appendix, we derive the HMC force term for the stout smearing
%%used in our simulation.
The stout smearing used in our simulation
has been proposed in Ref.~\cite{Morningstar:2003gk}
originally in the SU(3) case, and it has also been applied to
the SU(2) case, for instance in Refs.~ \cite{Hasenfratz:2007rf,Catterall:2013koa}.
%%In this appendix,
Here we present the derivation of the HMC force term 
for the reader's convenience.
%%using slightly different notations from the one given in \cite{Catterall:2013koa}.

Let us consider the first smearing step 
\begin{equation}
U_{n,\mu}\rightarrow U_{n,\mu}^{(1)} \equiv U_{n,\mu}^{\prime}=e^{iY_{n,\mu}}U_{n,\mu} \ ,
\end{equation}
where $e^{iY_{n,\mu}}$ is defined by \eqref{exp-Y},
and calculate the drift term $D_{n,\mu}^{a}S[U^{\prime}(U)]$
%%with respect to
for the original link variable, where the derivative is defined as
\begin{align}
  D_{n,\mu}^{a}F(U_{n,\mu}) \equiv
  \lim_{\epsilon \rightarrow 0} \frac{F(e^{i\epsilon\tau^{a}}U_{n,\mu}) - F(U_{n,\mu})}{\epsilon} \ ,
  \label{eq:def-D-derivative}
\end{align}
where $\tau^a$ are the SU(2) generators  with the normalization
$\tr (\tau^a \tau^b) = \frac{1}{2} \delta_{ab}$.
Note first that
\begin{align}
  U_{m,\nu}^{\prime}\left(e^{i\alpha^{a}\tau^{a}}U_{n,\mu}\right)
  %% & \approx U_{m,\nu}^{\prime}
  %% +\alpha^{a}\left[\frac{\partial}{\partial\epsilon}
  %%   U_{m,\nu}^{\prime}\left(e^{i\epsilon\tau^{a}}U_{n,\mu}\right)\right]_{\epsilon=0}\nonumber \\
  & \approx
  \left\{\mathbbm{1}+\alpha^{a}\left(D_{n,\mu}^{a}U_{m,\nu}^{\prime}\right) (U_{m,\nu}^{\prime})^\dagger \right\} U_{m,\nu}^{\prime} \ ,
  \label{eq:Uprime-U-rel}
\end{align}
up to the first order in $\alpha^{a}$.
The quantity in the parenthesis on the right-hand side can be rewritten as 
\begin{align}
  \alpha^{a}\left(D_{n,\mu}^{a}U_{m,\nu}^{\prime}\right) (U_{m,\nu}^{\prime})^\dagger
  & =2\alpha^{a}\mathrm{Tr}\left\{\left(D_{n,\mu}^{a}U_{m,\nu}^{\prime}\right)(U_{m,\nu}^{\prime})^\dagger\tau^{b}\right\}\tau^{b}  \nn \\
  &=  i\alpha^{a}\Delta_{n,\mu;m,\nu}^{a;b}\tau^{b} \ ,
\label{eq:DU-Delta-U}
\end{align}
where we have defined 
\begin{equation}
  \Delta_{n,\mu;m,\nu}^{a;b} \equiv
  - 2i\mathrm{Tr}\left\{\left(D_{n,\mu}^{a}U_{m,\nu}^{\prime}\right)
    (U_{m,\nu}^{\prime})^\dagger\tau^{b}\right\} \ .
\end{equation}
In deriving \eqref{eq:DU-Delta-U}, 
we have used the identity
%(???\ref{eq:formula_tata}) is used,
\begin{equation}
  \mathrm{Tr} \left( W\tau^{a}\right) \tau^{a}
  =\frac{1}{2}\left\{ W-\frac{1}{N}\mathrm{Tr}\left(W\right) \mathbbm{1}\right\} \ , %.
\end{equation}
in which the trace part on the right-hand side vanishes in the present case as
\begin{align}
\mathrm{Tr}\left\{\left(D_{n,\mu}^{a}U_{m,\nu}^{\prime}\right) (U_{m,\nu}^{\prime})^\dagger\right\}
& =\mathrm{Tr}\left\{D_{n,\mu}^{a}\left(e^{iY_{m,\nu}}U_{m,\nu}\right)U_{m,\nu}^\dagger e^{-iY_{m,\nu}}\right\} \nonumber \\
& =\mathrm{Tr}\left(
   e^{-iY_{m,\nu}}D_{n,\mu}^{a}e^{iY_{m,\nu}}+
  U_{m,\nu}^\dagger D_{n,\mu}^{a}U_{m,\nu}\right)\nonumber \\
  & =
i \mathrm{Tr}\left(D_{n,\mu}^{a}Y_{m,\nu}\right) +
  i\delta_{nm}\delta_{\mu\nu} \mathrm{Tr}\left(
%   e^{-iY_{m,\nu}}\left(D_{n,\mu}^{a}iY_{m,\nu}\right)e^{iY_{m,\nu}}
  U_{m,\nu}^\dagger \tau^{a}U_{m,\nu}\right)\nonumber \\
  %% & =i D_{n,\mu}^{a}\mathrm{Tr}\left(Y_{m,\nu}\right)
  %% +i\delta_{n,m}\delta_{\mu,\nu}\mathrm{Tr}\left(\tau^{a}\right)\nonumber \\
 & =0 \ .
\end{align}
%%The relation \eqref{eq:Delta-D-rel}
Plugging \eqref{eq:DU-Delta-U} in \eqref{eq:Uprime-U-rel},
%%enables us to rewrite \eqref{eq:Uprime-U-rel} as
we obtain
\begin{align}
    U_{m,\nu}^{\prime}\left(e^{i\alpha^{a}\tau^{a}}U_{n,\mu}\right)
 & =\left[\mathbbm{1}+i\alpha^{a}\Delta_{n,\mu;m,\nu}^{a;b}\tau^{b}\right]U_{m,\nu}^{\prime}\nonumber \\
 & \approx\exp\left(i\alpha^{a}\Delta_{n,\mu;m,\nu}^{a;b}\tau^{b}\right)U_{m,\nu}^{\prime} \ .
\end{align}
%%Thus, up to the first order in $\alpha^{a}$,
%% implies that the small change in $U_{n,\mu}$
%% can be replaced by that in $U_{n,\mu}^{\prime}$.
Thus we obtain
%%enables us to write
the chain rule
\begin{align}
  D_{n,\mu}^{a}S\left[U^{\prime}(U)\right]
 %% & =\sum_{m,\nu}\left.\frac{d}{d\epsilon}S\left[U_{m,\nu}^{\prime}\left(e^{i\epsilon\tau^{a}}U_{n,\mu}\right)\right]\right|_{\epsilon\rightarrow0}\nonumber \\
 %% & =\sum_{m,\nu}\left.\frac{d}{d\epsilon}S\left[e^{i\epsilon\Delta_{n,\mu;m,\nu}^{a;b}\tau^{b}}U_{m,\nu}^{\prime}\right]\right|_{\epsilon\rightarrow0}\nonumber \\
 %% & =\sum_{m,\nu}\left.\Delta_{n,\mu;m,\nu}^{a;b}\frac{d}{d\epsilon^{\prime}}S\left[e^{i\epsilon^{\prime}\tau^{b}}U_{m,\nu}^{\prime}\right]\right|_{\epsilon^{\prime}\rightarrow0}\nonumber \\
 & =\sum_{m,\nu}\Delta_{n,\mu;m,\nu}^{a;b}D_{m,\nu}^{\prime b}S\left[U^{\prime}\right] \ ,
\end{align}
where the primed derivative $D_{m,\nu}^{\prime b}$ is assumed to act on $U^{\prime}$.
%% In the third line, $\epsilon$ is rescaled to
%% $\epsilon^{\prime}=\Delta_{n,\mu;m,\nu}^{a;b}\epsilon$.
The relation between the drift
\begin{equation}
F_{n,\mu}^{\prime} \equiv  i\tau^{a}D_{n,\mu}^{\prime a}S\left[U^{\prime}(U)\right]
\end{equation}
for the smeared link and the drift $F_{n,\mu}$ for the original link
is obtained as
\begin{align}
  F_{n,\mu} & \equiv
  i\tau^{a}D_{n,\mu}^{a}S\left[U^{\prime}(U)\right]\nonumber \\
  & =i\tau^{a}\sum_{m,\nu}\Delta_{n,\mu;m,\nu}^{a;b}
  D_{m,\nu}^{\prime b}S\left[U^{\prime}\right]\nonumber \\
 %& =i\tau^{a}\sum_{m,\nu}(-2i)\mathrm{Tr}\left[\left(D_{n,\mu}^{a}U_{m,\nu}^{\prime}\right)U_{m,\nu}^{\prime-1}\tau^{b}\right]D_{m,\nu}^{\prime b}S\left[U^{\prime}\right]\nonumber \\
 & =-2i\tau^{a}\sum_{m,\nu}\mathrm{Tr}\left[\left\{ \left(D_{n,\mu}^{a}e^{iY_{m,\nu}}\right)e^{-iY_{m,\nu}}+e^{iY_{m,\nu}}\left(D_{n,\mu}^{a}U_{m,\nu}\right)U_{m,\nu}^\dagger e^{-iY_{m,\nu}}\right\} F_{m,\nu}^{\prime}\right]\nonumber \\
 & =-2i\tau^{a}\sum_{m,\nu}\left\{ \mathrm{Tr}\left[e^{-iY_{m,\nu}}F_{m,\nu}^{\prime}D_{n,\mu}^{a}e^{iY_{m,\nu}}\right]+i\delta_{nm}\delta_{\mu\nu}\mathrm{Tr}\left[e^{iY_{m,\nu}}\tau^{a}e^{-iY_{m,\nu}}F_{m,\nu}^{\prime}\right]\right\} \nonumber \\
  & =-2i\tau^{a}\sum_{m,\nu}\mathrm{Tr}\left[e^{-iY_{m,\nu}}F_{m,\nu}^{\prime}D_{n,\mu}^{a}e^{iY_{m,\nu}}\right]+e^{-iY_{n,\mu}}F_{n,\mu}^{\prime}e^{iY_{n,\mu}} \ .
  \label{drift-F}
\end{align}
%% In the last line, the trace part vanishes since $F_{n,\mu}^{\prime}$
%% is traceless
%% \begin{equation}
%% \mathrm{Tr}\left[e^{-iY_{n,\mu}}F_{n,\mu}^{\prime}e^{iY_{n,\mu}}\right]=\mathrm{Tr}F_{n,\mu}^{\prime}=0 \ .
%% \end{equation}

Let us recall that
$e^{iY_{n,\mu}}$ can be written as \eqref{def-exp-Y} using \eqref{def-kappa}.
We rewrite \eqref{def-exp-Y} as
\begin{equation}
  \exp(iY_{n,\mu}) =(\cos \kappa_{n,\mu}) \mathbbm{1}
  + i \, f(\kappa_{n,\mu})    \, Y_{n,\mu} \ ,
  \label{def-exp-Y-2}
\end{equation}
% where we have defined 
using the analytic function $f(x)$ defined as\footnote{The fact that
  the function $f(x)$ is analytic is important for using the stout smearing
  in the HMC algorithm.}
%% which has no singularity and is smooth for any $x\in\mathbb{R}$.
\begin{equation}
  f(x)=
  \frac{\sin x}{x}
  =1-x^{2}\left[\frac{1}{6}-x^{2}\left(\frac{1}{120}-\frac{x^{2}}{5040}\right)\right]
  +\mathrm{O}(x^{8})\ .
\end{equation}
% with $x\in\mathbb{C}$.
Using this function, the derivative in the first term in \eqref{drift-F}
is decomposed as
\begin{align}
&\quad \mathrm{Tr}\left(e^{-iY_{m,\nu}}F_{m,\nu}^{\prime}D_{n,\mu}^{a}e^{iY_{m,\nu}}\right)\nonumber \\
% & =\mathrm{Tr}\left[e^{-iY_{m,\nu}}F_{m,\nu}^{\prime}D_{n,\mu}^{a}\left(\cos\kappa_{m,\nu}\mathbbm{1}+f(\kappa_{m,\nu})iY_{m,\nu}\right)\right]\nonumber \\
 & =\mathrm{Tr}\left[ e^{-iY_{m,\nu}}F_{m,\nu}^{\prime}\left\{-\sin\kappa_{m,\nu}\mathbbm{1}+f^{\prime}(\kappa_{m,\nu}) iY_{m,\nu}\right\}\right]D_{n,\mu}^{a}\kappa_{m,\nu}\nonumber \\
  & \qquad+f(\kappa_{m,\nu})\mathrm{Tr}\left(e^{-iY_{m,\nu}}F_{m,\nu}^{\prime}iD_{n,\mu}^{a}Y_{m,\nu}\right) \ .
  \label{eq:derivative-decomposition}
\end{align}
The first trace in \eqref{eq:derivative-decomposition}
can be simplified by using the property
$Y_{m,\nu}^{2}=\kappa_{m,\nu}^{2}\mathbbm{1}$ of the $2\times2$
traceless matrix and $\mathrm{Tr}F_{n,\mu}^{\prime}=0$ as
\begin{align}
\mathrm{Tr}\left[e^{-iY_{m,\nu}}F_{m,\nu}^{\prime}\left\{-\sin\kappa_{m,\nu}\mathbbm{1}+f^{\prime}(\kappa_{m,\nu})iY_{m,\nu}\right\}\right]
=\frac{1-f(2\kappa_{m,\nu})}{\kappa_{m,\nu}}\mathrm{Tr}\left(F_{m,\nu}^{\prime}iY_{m,\nu}\right) \ .
\end{align}
Converting the derivative of $\kappa_{m,\nu}$ to the derivative of
$Y_{m,\nu}$ as
\begin{equation}
D_{n,\mu}^{a}\kappa_{m,\nu}=D_{n,\mu}^{a}\sqrt{\frac{1}{2}\mathrm{Tr}Y_{m,\nu}^{2}}=-\frac{1}{2\kappa_{m,\nu}}\mathrm{Tr}\left(iY_{m,\nu}iD_{n,\mu}^{a}Y_{m,\nu}\right) \ ,
\end{equation}
%%\eqref{eq:derivative-decomposition} can be written as
all the derivatives in \eqref{eq:derivative-decomposition}
%%the trace
can be written in terms of $D_{n,\mu}^{a}Y_{m,\nu}$ as
\begin{align}
 & \mathrm{Tr}\left(e^{-iY_{m,\nu}}F_{m,\nu}^{\prime}D_{n,\mu}^{a}e^{iY_{m,\nu}}\right)\nonumber \\
 & =-\frac{1-f(2\kappa_{m,\nu})}{2\kappa_{m,\nu}^{2}}\mathrm{Tr}\left(F_{m,\nu}^{\prime}iY_{m,\nu}\right)\mathrm{Tr}\left(iY_{m,\nu}iD_{n,\mu}^{a}Y_{m,\nu}\right)+f(\kappa_{m,\nu})\mathrm{Tr}\left(e^{-iY_{m,\nu}}F_{m,\nu}^{\prime}iD_{n,\mu}^{a}Y_{m,\nu}\right)\nonumber \\
  & =\mathrm{Tr}\left[\left\{ -\frac{1-f(2\kappa_{m,\nu})}{2\kappa_{m,\nu}^{2}}\mathrm{Tr}\left(F_{m,\nu}^{\prime}iY_{m,\nu}\right)iY_{m,\nu}+f(\kappa_{m,\nu})e^{-iY_{m,\nu}}F_{m,\nu}^{\prime}\right\} iD_{n,\mu}^{a}Y_{m,\nu}\right] \ .
 %%  \nonumber \\
 %% & =:\mathrm{Tr}\left[\hat{\Lambda}_{m,\nu}iD_{n,\mu}^{a}Y_{m,\nu}\right]\ .
\end{align}
Plugging this in \eqref{drift-F},
we obtain the form of the drift term
\begin{align}
  F_{n,\mu}&=-2i\tau^{a}\sum_{m,\nu}\mathrm{Tr}\left(\hat{\Lambda}_{m,\nu}iD_{n,\mu}^{a}Y_{m,\nu}\right)+e^{-iY_{n,\mu}}F_{n,\mu}^{\prime}e^{iY_{n,\mu}} \ ,
\label{eq:form-of-drift-F}
  \\
\hat{\Lambda}_{m,\nu}&\equiv
-\frac{1-f(2\kappa_{m,\nu})}{2\kappa_{m,\nu}^{2}}\mathrm{Tr}\left(F_{m,\nu}^{\prime}iY_{m,\nu}\right)iY_{m,\nu}+f(\kappa_{m,\nu})e^{-iY_{m,\nu}}F_{m,\nu}^{\prime}\ .
\end{align}

Let us recall here that $Y_{m,\nu}$ is defined in terms of $J_{m,\nu}$
as \eqref{smearing-param}.
The first term of the left-hand side of \eqref{eq:form-of-drift-F}
can be rewritten as
%%$F_{n,\mu}$,
%% the trace part of the matrix $\hat{\Lambda}_{m,\nu}$
%% does not contribute since
\begin{align}
-2i\tau^{a}\sum_{m,\nu}\mathrm{Tr}\left(\hat{\Lambda}_{m,\nu}iD_{n,\mu}^{a}Y_{m,\nu}\right) & =2i\rho\tau^{a}\sum_{m,\nu}\mathrm{Tr}\left\{\hat{\Lambda}_{m,\nu}\mathrm{Tr}\left( D_{n,\mu}^{a}J_{m,\nu}\tau^{b}\right) \tau^{b}\right\} \nonumber \\
 & =2i\rho\tau^{a}\sum_{m,\nu}\mathrm{Tr}\left\{ \mathrm{Tr}\left(\hat{\Lambda}_{m,\nu}\tau^{b}\right)\tau^{b}D_{n,\mu}^{a}J_{m,\nu}\right\} \nonumber \\
 & =2i\rho\tau^{a}\sum_{m,\nu}\mathrm{Tr}\left(\Lambda_{m,\nu}D_{n,\mu}^{a}J_{m,\nu}\right) \ ,
\end{align}
where we have defined the traceless matrices
%%$\Lambda_{m,\nu}$ by subtracting the trace part from $\hat{\Lambda}_{m,\nu}$
\begin{equation}
  \Lambda_{m,\nu} \equiv
  \mathrm{Tr}\left(\hat{\Lambda}_{m,\nu}\tau^{b}\right)\tau^{b}
  =\frac{1}{2}\left\{\hat{\Lambda}_{m,\nu}-\frac{1}{2}\mathrm{Tr}\left(\hat{\Lambda}_{m,\nu}\right) \right\} \ .
\end{equation}
Therefore the drift term 
\begin{equation}
  F_{n,\mu}=e^{-iY_{n,\mu}}F_{n,\mu}^{\prime}e^{iY_{n,\mu}}+2i\rho\tau^{a}\sum_{m,\nu}\mathrm{Tr}\left[\Lambda_{m,\nu}D_{n,\mu}^{a}J_{m,\nu}\right]
  \label{drift-after-each-smearing}
\end{equation}
is obtained
%%completely determined
by calculating the derivative of the $J_{m,\nu}$,
which is given by
\begin{align}
 & D_{n,\mu}^{a}J_{m,\nu}\nonumber \\
 & =D_{n,\mu}^{a}(U_{m,\nu}\Omega_{m,\nu}-\Omega^\dagger_{m,\nu}U_{m,\nu}^\dagger)\nonumber \\
 & =i\delta_{n,m}\delta_{\mu\nu}(\tau^{a}U_{m,\nu}\Omega_{m,\nu}+\Omega^\dagger_{m,\nu}U_{m,\nu}^\dagger\tau^{a})+U_{m,\nu}(D_{n,\mu}^{a}\Omega_{m,\nu})-(D_{n,\mu}^{a}\Omega^\dagger_{m,\nu})U_{m,\nu}^\dagger \ ,
\end{align}
\begin{align}
 & D_{n,\mu}^{a}\Omega_{m,\nu}\nonumber \\
 & =i\left(1-\delta_{\mu\nu}\right)\left(\delta_{n,m+\nu}\tau^{a}U_{n,\mu}U_{n+\mu-\nu,\nu}^\dagger U_{n-\nu,\mu}^\dagger-\delta_{n,m}U_{n+\nu,\mu}U_{n+\mu,\nu}^\dagger U_{n,\mu}^\dagger\tau^{a}\right.\nonumber \\
 & \hphantom{=i\left(1-\delta_{\mu\nu}\right)}\left.-\delta_{n,m+\nu-\mu}U_{n,\mu}^\dagger\tau^{a}U_{n-\nu,\nu}^\dagger U_{n-\nu,\mu}+\delta_{n,m-\mu}U_{n+\nu,\mu}^\dagger U_{n,\nu}^\dagger\tau^{a}U_{n,\mu}\right)\nonumber \\
 & \quad -i\delta_{\mu\nu}\sum_{\sigma(\neq\mu)}\left(\delta_{n,m+\sigma}U_{n+\mu-\sigma,\sigma}U_{n,\mu}^\dagger \tau^{a}U_{n-\sigma,\sigma}^\dagger+\delta_{n,m-\sigma}U_{n+\mu,\sigma}^\dagger U_{n,\mu}^\dagger\tau^{a}U_{n,\sigma}\right) \ ,
\end{align}
\begin{align}
 & D_{n,\mu}^{a}\Omega^\dagger_{m,\nu}\nonumber \\
 & =i\left(1-\delta_{\mu\nu}\right)\left(\delta_{n,m}\tau^{a}U_{n,\mu}U_{n+\mu,\nu}U_{n+\nu,\mu}^\dagger-\delta_{n,m+\nu}U_{n-\nu,\mu}U_{n+\mu-\nu,\nu}U_{n,\mu}^\dagger\tau^{a}\right.\nonumber \\
 & \hphantom{=i\left(1-\delta_{\mu\nu}\right)}\left.-\delta_{n,m-\mu}U_{n,\mu}^\dagger\tau^{a}U_{n,\nu}U_{n+\nu,\mu}+\delta_{n,m+\nu-\mu}U_{n-\nu,\mu}^\dagger U_{n-\nu,\nu}\tau^{a}U_{n,\mu}\right)\nonumber \\
 & \quad +i\delta_{\mu\nu}\sum_{\sigma(\neq\mu)}\left(\delta_{n,m+\sigma}U_{n-\sigma,\sigma}\tau^{a}U_{n,\mu}U_{n+\mu-\sigma,\sigma}^\dagger+\delta_{n,m-\sigma}U_{n,\sigma}^\dagger \tau^{a}U_{n,\mu}U_{n+\mu,\sigma}\right) \ .
\end{align}

\bibliographystyle{JHEP}
\bibliography{ref}
%% ref.bib is the same as that for the Lattice proceedings.

\end{document}